\documentclass[letterpaper,twocolumn,10pt]{article}
\usepackage{zhanggroup}

\usepackage{xspace}
\usepackage{amsmath}
\usepackage{multirow} 
\usepackage{booktabs}
\usepackage{graphicx}
\usepackage{subcaption}
\usepackage[most]{tcolorbox}
\usepackage{CJK}
\usepackage{tabularx}
\usepackage{enumitem}
\usepackage[table]{xcolor}
\usepackage{colortbl}
\usepackage{pifont}
\usepackage[dvipsnames]{xcolor}
\usepackage{amsthm}
\usepackage{wasysym}
\newcommand{\cmark}{{\color{blue!60!green}\ding{51}}}
\newcommand{\xmark}{{\color{red}\ding{55}}}
\definecolor{myred}{RGB}{227, 23, 13} 
\definecolor{myyellow}{RGB}{255,255,200} 
\definecolor{mygreen}{RGB} {220,255,220}   
\definecolor{myblue}{RGB} {0,0,255}   
\newcommand{\mypara}[1]{\noindent{\bf {#1}.} \xspace}

\usepackage[nameinlink]{cleveref}
\crefname{section}{Section}{Sections}
\crefname{appendix}{Appendix}{Appendices}

\begin{document}
%-------------------------------------------------------------------------------

\date{}

\title{Real Money, Fake Models: Deceptive Model Claims in Shadow APIs}

\newcommand{\authorthanksmark}[1]{\ifcase\value{#1}\or\ensuremath{\dagger}\or *\else\fnsymbol{#1}\fi}
\thanksmarkseries{authorthanksmark}

\author{
Yage Zhang\textsuperscript{1}\thanks{Equal contribution.}\ \ \
Yukun Jiang\textsuperscript{1}\footnotemark[1]\ \ \
Zeyuan Chen\textsuperscript{1}\ \ \
Michael Backes\textsuperscript{1}\ \ \
Xinyue Shen\textsuperscript{1,2}\ \ \
Yang Zhang\textsuperscript{1}\thanks{Corresponding author.}
\\
\\
\textsuperscript{1}\textit{CISPA Helmholtz Center for Information Security} \ \ \
\textsuperscript{2}\textit{University of Waterloo}
}

\maketitle

%-------------------------------------------------------------------------------
\begin{abstract}
Access to frontier large language models (LLMs), such as GPT-5 and Gemini-2.5, is often hindered by high pricing, payment barriers, and regional restrictions.
These limitations drive the proliferation of \textit{shadow APIs}, third-party services that claim to provide access to official model services without regional limitations via indirect access.
Despite their widespread use, it remains unclear whether shadow APIs deliver outputs consistent with those of the official APIs, raising concerns about the reliability of downstream applications and the validity of research findings that depend on them.
In this paper, we present the first systematic audit between official LLM APIs and corresponding shadow APIs.
We first identify 17 shadow APIs that have been utilized in 187 academic papers, with the most popular one reaching more than 5{,}900 citations and 58{,}000 GitHub stars by December 6, 2025.
Through multidimensional auditing of three representative shadow APIs across utility, safety, and model verification, we uncover widespread behavioral inconsistency and fingerprint-based evidence consistent with deceptive model claims in a subset of audited endpoints.
Specifically, we reveal performance divergence reaching up to $47.21\%$, significant unpredictability in safety behaviors, and identity verification failures in $45.83\%$ of fingerprint tests.
These practices critically undermine the reproducibility and validity of scientific research, harm the interests of shadow API users, and damage the reputation of official model providers.
By the time of writing, 4 of the 17 providers have already ceased operations, underscoring the operational volatility of this market.
Meanwhile, unverifiable compliance claims and independent model-substitution testing platforms have emerged in the ecosystem, reflecting growing community awareness of this risk.\footnote{The evaluation code is available at \url{https://github.com/TrustAIRLab/ShadowAPI}.}
\end{abstract}
%-------------------------------------------------------------------------------

%-------------------------------------------------------------------------------
\section{Introduction}
%-------------------------------------------------------------------------------

\begin{figure}[t!]
    \centering
    \includegraphics[width=\linewidth]{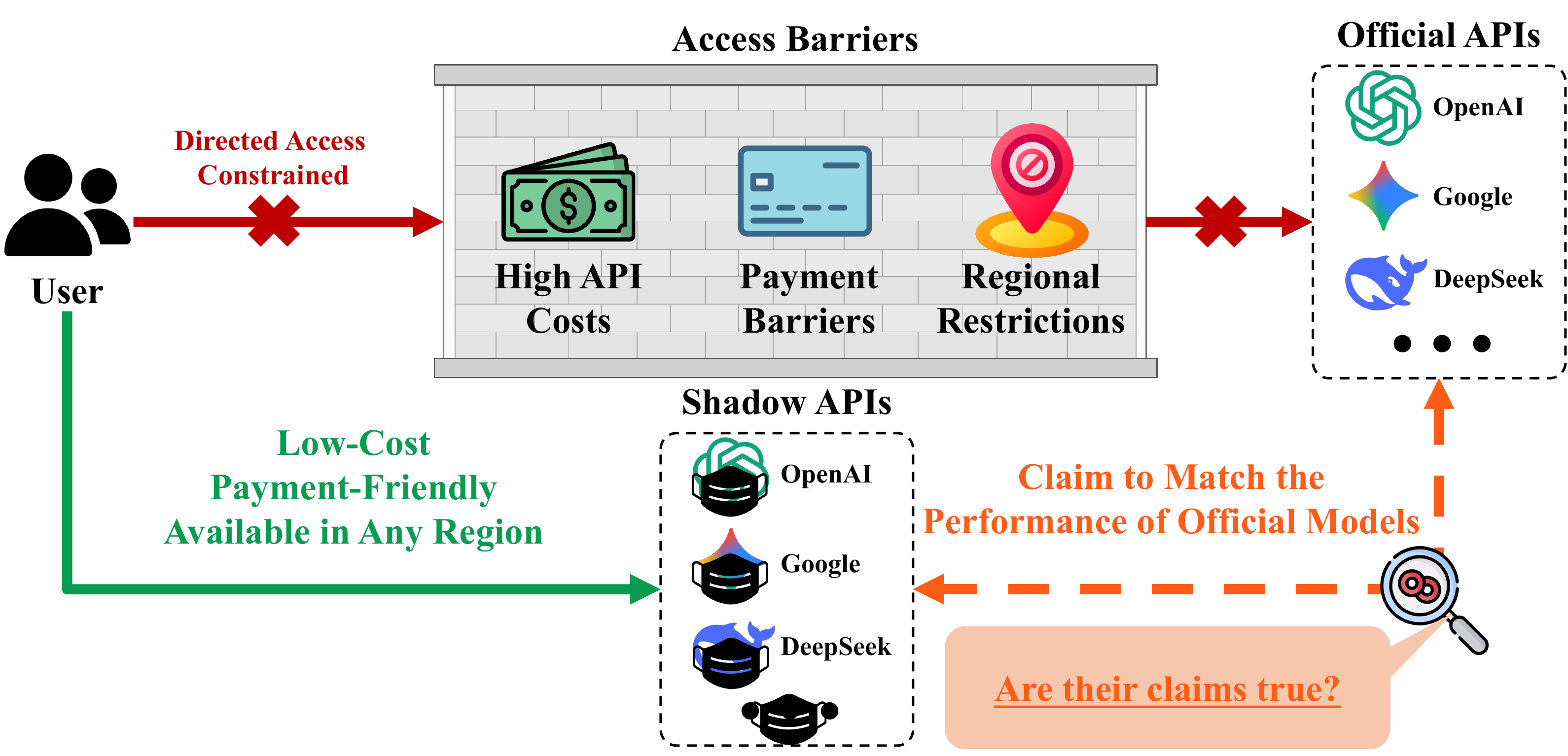}
    \caption{
        Overview of the shadow API ecosystem.
    }
    \label{figure:intro}
\end{figure}

Frontier large language models (LLMs) have transformed numerous domains, becoming essential infrastructure for scientific research and daily applications~\cite{ECDGGGHKLLLMZM25,ZZLTWHMZZDDYCCJRLTLLNW23,BCEGHKLLLLNPRZ23}. 
By integrating advanced reasoning and domain knowledge, these models demonstrate enhanced effectiveness in complex tasks~\cite{WWSBIXCLZ22,O23,SATMWCSTCPPSGKSCMAWCMCGTLRBSKN23}.
Given that many frontier LLMs are proprietary or require substantial computational resources, local deployment is often infeasible. 
In recent years, access to these models has been mediated through commercial APIs provided by companies like OpenAI and Google~\cite{XWGZJ26,LBLTSYZNWKNYYZCMRAHZDLRRYWSOZYSKGCKHHCXSGHIZCWLMZK22}.
However, official APIs often impose high pricing, payment barriers, and strict geographical restrictions~\cite{CZZ23}.
For instance, the official OpenAI API cannot be accessed directly from several regions, such as China, Russia, and Iran~\cite{OpenAI_Supported_Countries,OpenAI_China_Access}.
As a result, a vast third-party provider market has emerged, offering compatible endpoints at a lower cost and without regional limitations.
We refer to these services as \textbf{\textit{shadow APIs}} (illustrated in~\autoref{figure:intro}), platforms that claim to generate the same output as official LLMs via indirect access.
By December 6, 2025, the shadow APIs we identify have been utilized in 187 academic papers, and the most widely used one alone accounts for more than 5{,}900 citations and 58{,}000 GitHub stars. 
Most of these papers have been accepted by top venues, such as ACL, CVPR, and ICLR (see details in~\autoref{section:academia}).

Shadow API use arises within a broader access ecosystem: official frontier-model APIs can be costly or geographically unavailable, while researchers may face resource constraints and publication pressure~\cite{XZLCLLWL25,LLM_API_Economics}.
We therefore do not treat the use of these services alone as individual misconduct.
The security concern arises when results are attributed to a named model without verifiable endpoint provenance, because neither authors nor readers can determine whether the claimed model actually produced the outputs.
This motivates our audit of whether shadow endpoints serve the models they claim and our recommendation that endpoint provenance be disclosed and, where feasible, verified.

Recent studies have highlighted the unreliability of API access for open-source models~\cite{CSZS25,GLG25,SLZJZ24}, raising serious concerns about whether shadow APIs can faithfully replace official ones.
Users may treat shadow APIs as interchangeable substitutes for official APIs, considering the access problem mainly from a utility perspective, without conducting further verification.
From a security standpoint, the relevant concern is \emph{model provenance integrity}: users cannot directly verify the binding between the advertised model identity and the backend serving their requests.
Undocumented backend substitution would violate this binding, whereas operational instability can produce behavioral divergence without establishing an identity mismatch.
Both forms of uncertainty can undermine reproducibility and user trust.

In this paper, we address the fundamental opacity of the shadow API market by answering three research questions:

\begin{enumerate}[label=\textbf{RQ\arabic*:}, leftmargin=*, labelsep=0.5em, itemsep=0.5em]
    \item What shadow APIs currently exist, and to what extent are they used?
    \item Do shadow APIs perform consistently with official ones for any given request?
    \item What evidence can model verification methods provide?
\end{enumerate}

To address \textbf{RQ1}, we identify 17 shadow APIs and survey 187 academic papers to quantify the prevalence of these services~(\autoref{section:academia}).
For \textbf{RQ2}, we conduct a multi-dimensional benchmarking across both utility and safety perspectives~(\autoref{section:utility}).
Alarmingly, our experiments reveal significant performance inconsistencies between shadow APIs and official APIs.
On high-risk medical benchmarks like MedQA, the accuracy of the Gemini-2.5-flash model drops precipitously, from $83.82\%$ with the official API to approximately $37.00\%$ across all examined shadow APIs.
Shadow APIs also exhibit unpredictable safety divergence from official baselines, with measured harmfulness scores either falling below the official baseline by about $0.23$ points or rising to nearly twice the official value.
Both directions undermine reproducibility: a lower shadow-endpoint score may understate the risk posed by the official model, whereas a higher score may inflate attack success rates that do not replicate on the official endpoint.
Finally, for \textbf{RQ3}, we utilize model fingerprinting and output metadata analysis to audit these shadow APIs. 
We find fingerprint-based evidence consistent with model substitution in a subset of audited endpoints, alongside broader behavioral and metadata inconsistencies.
Specifically, across 24 evaluated endpoints, $45.83\%$ fail fingerprint verification, and an additional $12.50\%$ exhibit substantial Euclidean distance deviations from the official models~(\autoref{section:fingerprinting}).

Our contributions are summarized as follows.
\begin{itemize}[leftmargin=*]
    \item We identify 17 widely used shadow APIs and present the first systematic audit on them compared with the official baselines.
    \item We demonstrate frequent failures and unreliability of shadow APIs in utility and safety evaluation, distinguishing inconsistency, substitution, and deceptive model claims.
    \item Verification based on model fingerprint and meta information provides supportive evidence of the differences between shadow and official APIs.
    \item Furthermore, we provide suggestions to enforce provenance awareness and reduce reproducibility risks when using LLM APIs.
\end{itemize}

\mypara{Disclosure}
We are committed to responsibly reporting our findings to official model providers as well as the authors of papers that use shadow APIs. 
We have received partial acknowledgment from them.
We did not contact the shadow API providers during the audit or follow-up period because doing so could have altered endpoint behavior or availability.

%-------------------------------------------------------------------------------
\section{Preliminary}
%-------------------------------------------------------------------------------

%-------------------------------------------------------------------------------
\subsection{Background}
%-------------------------------------------------------------------------------

LLMs have rapidly become the backbone of modern natural language processing (NLP), machine learning (ML), as well as even computer vision (CV) research, and the base for building personal agents such as OpenClaw~\cite{JZSBZ26}.
A recent analysis of 16{,}193 papers shows that by 2024, over $60\%$ of NLP papers, around $20\%$ of ML papers, and nearly $10\%$ of CV papers are related to LLMs~\cite{XZLCLLWL25}. 
The number of LLM papers has exploded from a few hundred in 2019 to more than 7{,}000 in 2024.
At the same time, access to frontier proprietary models is tightly controlled. 
Providers restrict API access to certain regions for legal and security reasons~\cite{Gemini_Available_Regions}. 
For instance, Anthropic explicitly disallows sales in unsupported regions~\cite{Anthropic_Regional_Restrictions}.
OpenAI likewise warns that accessing or reselling its API from unsupported countries may lead to account suspension~\cite{OpenAI_Unsupported_Regions}.
These restrictions collide with the geographic barriers of artificial intelligence (AI) research, given the reality that major AI conferences such as AAAI and CVPR attract large numbers of submissions and accepted papers from regions with restricted access to frontier APIs (e.g., China)~\cite{AAAI26_Review_Process,CVPR24_in_Numbers}.

In addition to geographic barriers, expensive bills put pressure on users.
Frontier LLM APIs are typically priced for enterprise customers, which can be prohibitive for individual researchers or students. 
Analysis of LLM API economics suggests that many current usage patterns are unsustainable for small actors without cross-subsidies~\cite{LLM_API_Economics}.
This drives the rise of a commercial shadow market offering discounted access to frontier LLMs without geographical restrictions~\cite{LLM_Grey_Market,GenAI_Credential_Theft}.

Moreover, official terms of service prohibit any form of API key resale or redistribution~\cite{openai_business_license2025}.
In addition, Chinese government regulations require AI services to operate in compliance with applicable laws and administrative requirements.\footnote{\url{http://www.cac.gov.cn/2023-07/13/c_1690898327029107.htm}}
As a result, these sellers operate in violation of both the service contract and applicable regulatory requirements.
These trends suggest a growing disconnect between academic demand for frontier models and the constraints of official distribution channels, which drives the emergence of a substantial market of unofficial third-party services.

%-------------------------------------------------------------------------------
\subsection{Definition}
\label{section:definition}
%-------------------------------------------------------------------------------

For clarity, we introduce the term \textbf{\textit{shadow APIs}} to describe third-party LLM API services characterized by
\begin{enumerate}
    \item \textbf{indirect access} to frontier proprietary models through intermediate providers.
    \item \textbf{access provision in restricted regions}, where the service bypasses the official provider's geographic access controls.
\end{enumerate}

To apply these criteria consistently, we classify a service as a shadow API only when it also sells access using credentials or credits issued by the intermediate provider and does not publicly document its upstream route~(\autoref{table:shadow_vs_aggregator}).
We therefore exclude provider-operated endpoints, publicly documented distribution channels, routing platforms that disclose their upstream routing, and self-hosted gateways that use the user's own credentials.
No service was included or excluded based on its measured outputs, and the three providers audited in depth were chosen for popularity, accessibility, and model coverage.

%-------------------------------------------------------------------------------
\begin{table}[t!]
\centering
\caption{Operational distinction between documented distribution channels and shadow APIs. }
\label{table:shadow_vs_aggregator}
\scalebox{0.65}{
\begin{tabular}{@{}lcc@{}}
\toprule
\textbf{Criterion} & \textbf{Documented Channel} & \textbf{Shadow API} \\
\midrule
Upstream route       & \cmark~Publicly documented      & \xmark~Not documented \\
Credential source    & \cmark~User's own upstream key & \xmark~Intermediate provider's key \\
Legal entity disclosed  & \cmark~Verifiable corporation   & \xmark~Typically individual \\
Registration identifier & \cmark~Business / ICP registry  & \xmark~Rarely disclosed \\
Legal documentation     & \cmark~ToS, SLA, privacy policy & \xmark~Often absent \\
Regional compliance     & \cmark~Authorized regions only  & \xmark~Bypasses restrictions \\
Payment accountability  & \cmark~Corporate invoicing      & \xmark~Individual Alipay / WeChat / USDT \\
\bottomrule
\end{tabular}
}
\end{table}
%-------------------------------------------------------------------------------

%-------------------------------------------------------------------------------
\subsection{\texorpdfstring{Security Framing and Scope}{Security Framing and Scope}}
\label{section:threat_model}
%-------------------------------------------------------------------------------

\mypara{Security Property}
The relevant security property is \emph{model provenance integrity}: an endpoint sold as access to a named model should preserve the binding between that model identity and the backend serving the request.
The risk arises when users cannot verify this binding, causing downstream evaluations to be attributed to a model that may not have produced the outputs.

\mypara{Interpreting the Evidence}
We distinguish measured behavioral divergence from evidence about backend identity.
Differences in utility, safety, latency, metadata, or MET outcomes show that an endpoint does not reproduce the official baseline on the measured dimension, but they do not identify the cause or the backend.
We report fingerprint-based evidence consistent with model substitution only when the fingerprint matches a specific alternative model.
Because black-box fingerprinting is probabilistic, such a match supports but does not prove backend identity.
We use \emph{deceptive model claim} only for the subset of endpoints that are commercially offered under one model identity while producing fingerprint evidence consistent with another.
This term characterizes the mismatch between the commercial claim and the observed evidence and does not establish operator intent.
This claim-centered usage is analogous to regulatory concerns about unsupported AI capability claims~\cite{FTC_Deceptive_AI_Claims}.

\mypara{Implications for Security Research}
Safety and alignment measurements obtained through an unverified endpoint may characterize a backend other than the model named in the study.
A lower measured harmfulness score may understate the risk posed by the named official model, whereas a higher score may overstate attack effectiveness and yield conclusions that do not replicate on the official endpoint.
More generally, results attributed to a named model are difficult to reproduce and reliably advance subsequent security/safety research when endpoint provenance is not documented.

\mypara{Scope of Claims}
Ecosystem-level findings concern all 17 identified services and rest on collection, compliance, and metadata evidence~(\autoref{section:academia}).
Behavioral and identity findings concern only shadow APIs A, E, and H and the 24 endpoints audited directly~(\autoref{section:utility},~\autoref{section:fingerprinting}).
We do not generalize these endpoint-level measurements to the remaining 14 services.

%-------------------------------------------------------------------------------
\subsection{Related Work}
%-------------------------------------------------------------------------------

LLMs exhibit distinct linguistic patterns and features that function as fingerprints, enabling the identification of the LLM that generated a given piece of content~\cite{MSSA24, YCWPWC24, BZTYZ24}.
Pasquini et al.~\cite{PKA25} introduce LLMmap, an active fingerprinting method that queries models with carefully crafted inputs to estimate the likelihood of the response under different reference models.

Cai et al.~\cite{CSZS25} audit model substitution in open-source LLM APIs and evaluate Trusted Execution Environments (TEEs) as a hardware-level solution for verifiable model integrity.
In contrast, we focus on the shadow APIs, indirect services that operate as opaque black boxes. 
Relatedly, research on model extraction and imitation~\cite{KTPPI20} shows that surrogate models can be trained to mimic the outputs of BERT-based NLP APIs, making it difficult for users to distinguish between original APIs and extracted surrogates based on surface-level interactions alone.
Recent work examines adjacent risks in the broader LLM deployment ecosystem.
Hou et al.~\cite{HHZW25} conduct an Internet-wide measurement of public-facing LLM deployments, characterizing their prevalence, configurations, and exposure-related security risks.
Puthal et al.~\cite{PRM25} examine the security and governance risks arising from unsanctioned ``shadow AI'' use within organizations.
These studies focus respectively on insecure public deployments and organizationally ungoverned AI use. 
Our audit instead asks whether commercially sold indirect API endpoints actually serve the models they claim.
However, systematic audits of indirect API services operating in this unregulated shadow market remain largely absent.

%-------------------------------------------------------------------------------
\section{Landscape of Shadow APIs}
\label{section:academia}
%-------------------------------------------------------------------------------
For \textbf{RQ1}, we investigate the prevalence of shadow APIs by quantifying their widespread adoption across both the research and the open-source community.

%-------------------------------------------------------------------------------
\subsection{Shadow APIs Collection}
%-------------------------------------------------------------------------------
We start from screening the accepted papers from ICLR 2024 (2{,}260 total, 1{,}108 with code) and papers from ACL 2024 (1{,}923 total, 1{,}005 with code)~\cite{ICLR2024_Accepted_Papers,ACL2024_Accepted_Papers}.
We use these two venues as seeds because both concentrate LLM-related papers and are recent yet old enough for most associated code repositories to have been publicly released.
The collection is not restricted to these two venues, as the endpoint-based expansion below covers repositories from any source. 
To retrieve the associated code repositories, we systematically parsed the abstracts and footnotes of these papers to extract URLs matching standard repositories.
From this combined pool of 2,113 code-available papers, we identify 92 projects that use LLM APIs via their associated GitHub repositories.
Among these, 21 papers ($22.82\%$) use at least one shadow API endpoint, and we search GitHub for these URLs to find additional repositories that call the same endpoint URL, collecting the venue, institution, country, citation counts, and the GitHub stars of associated repositories, with all metrics updated as of December 6, 2025. 
We iterate this process by harvesting new shadow API endpoint URLs from these repositories, repeating until no new ones are discovered.

We include an endpoint when (i) a repository replaces an official provider's base URL with that endpoint while retaining an official model identifier, and (ii) the service satisfies the structural criteria in~\autoref{section:definition} and~\autoref{table:shadow_vs_aggregator}.
We exclude publicly documented cloud distribution channels such as Azure OpenAI, routing platforms that disclose their upstream routing such as OpenRouter, and self-hosted relays that forward a user's own provider credential.

%-------------------------------------------------------------------------------
\subsection{Prevalence and Impact of Shadow APIs}
%-------------------------------------------------------------------------------

\begin{figure*}[!t]
\centering
\begin{subfigure}{0.32\linewidth}
    \centering
    \includegraphics[width=\linewidth]{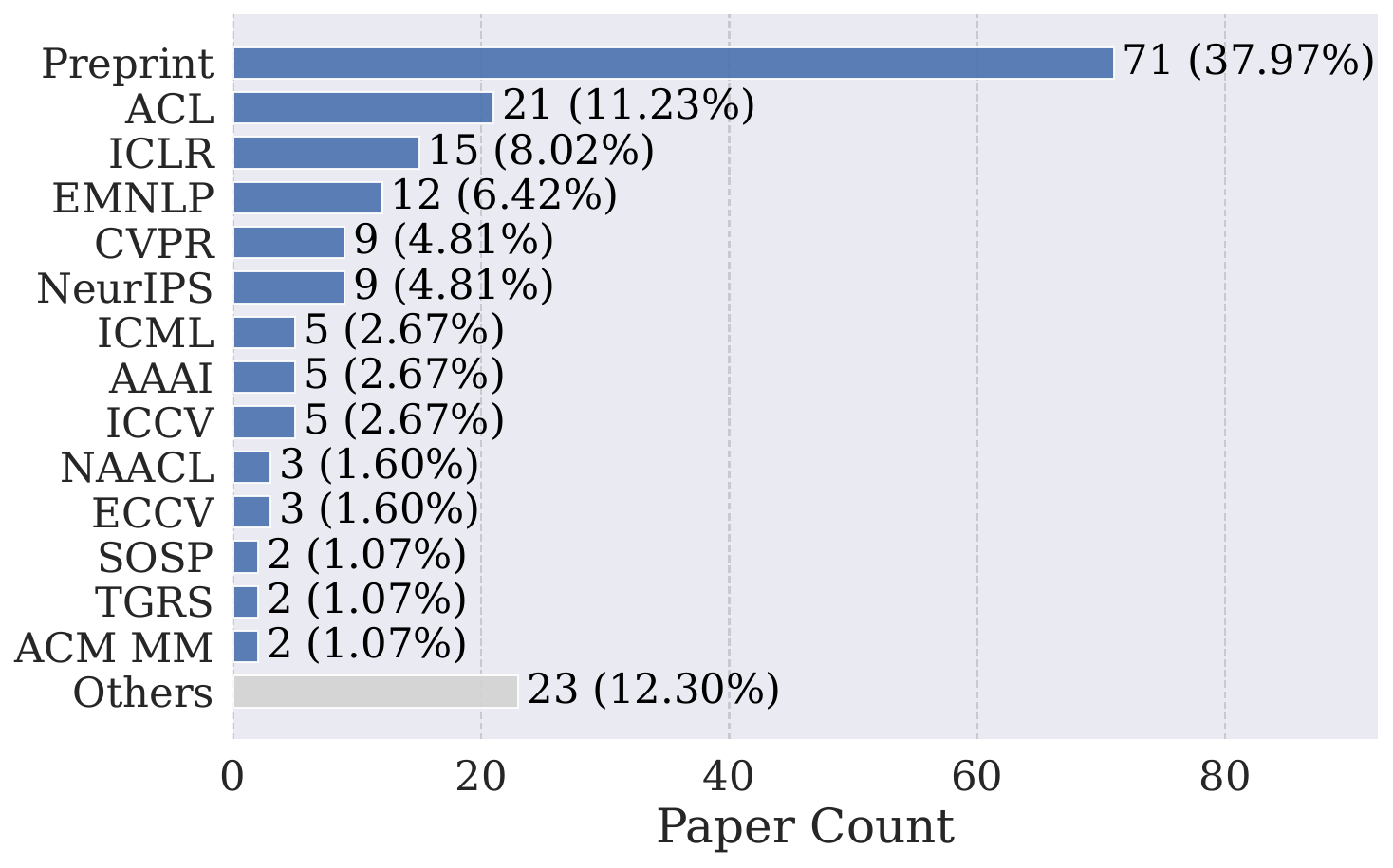}
    \caption{Paper counts by research venue.}
    \label{figure:publish_counts}
\end{subfigure}
\begin{subfigure}{0.30\linewidth}
    \centering
    \includegraphics[width=\linewidth, trim={3.3cm 0 0 0}, clip]{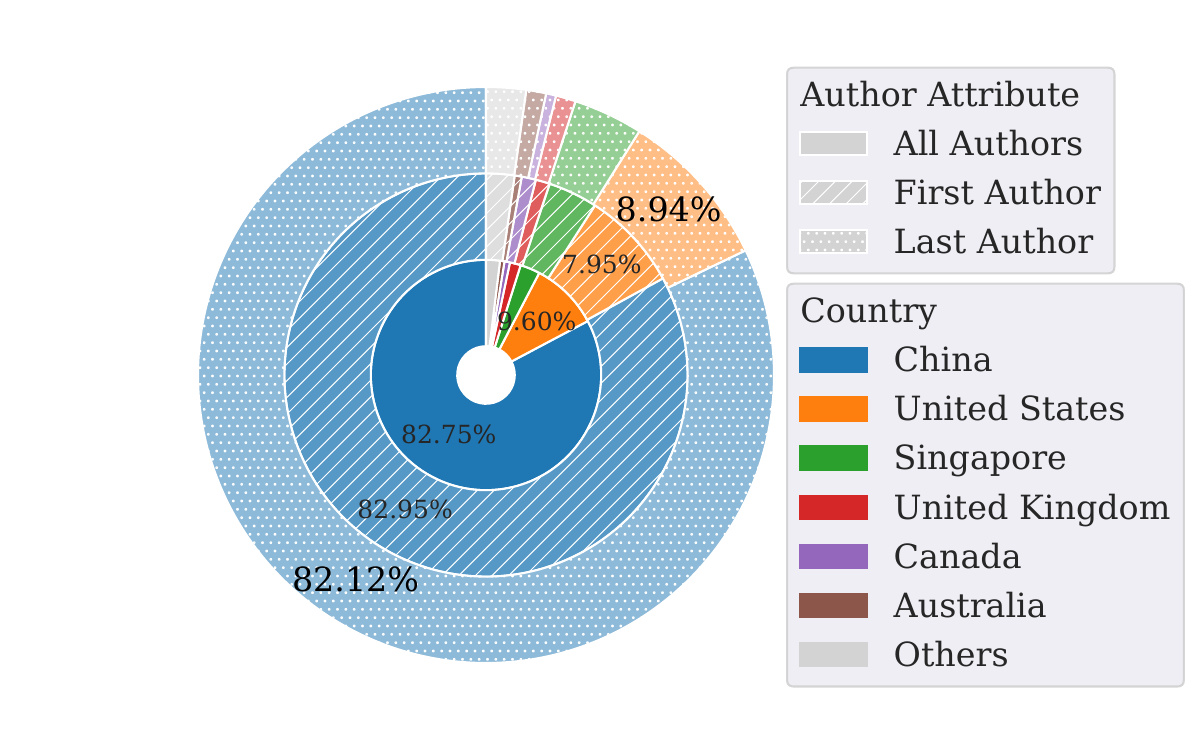}
    \caption{Geographic distribution of authors.}
    \label{figure:author_regions}
\end{subfigure}
\begin{subfigure}{0.34\linewidth}
    \centering
    \includegraphics[width=\linewidth]{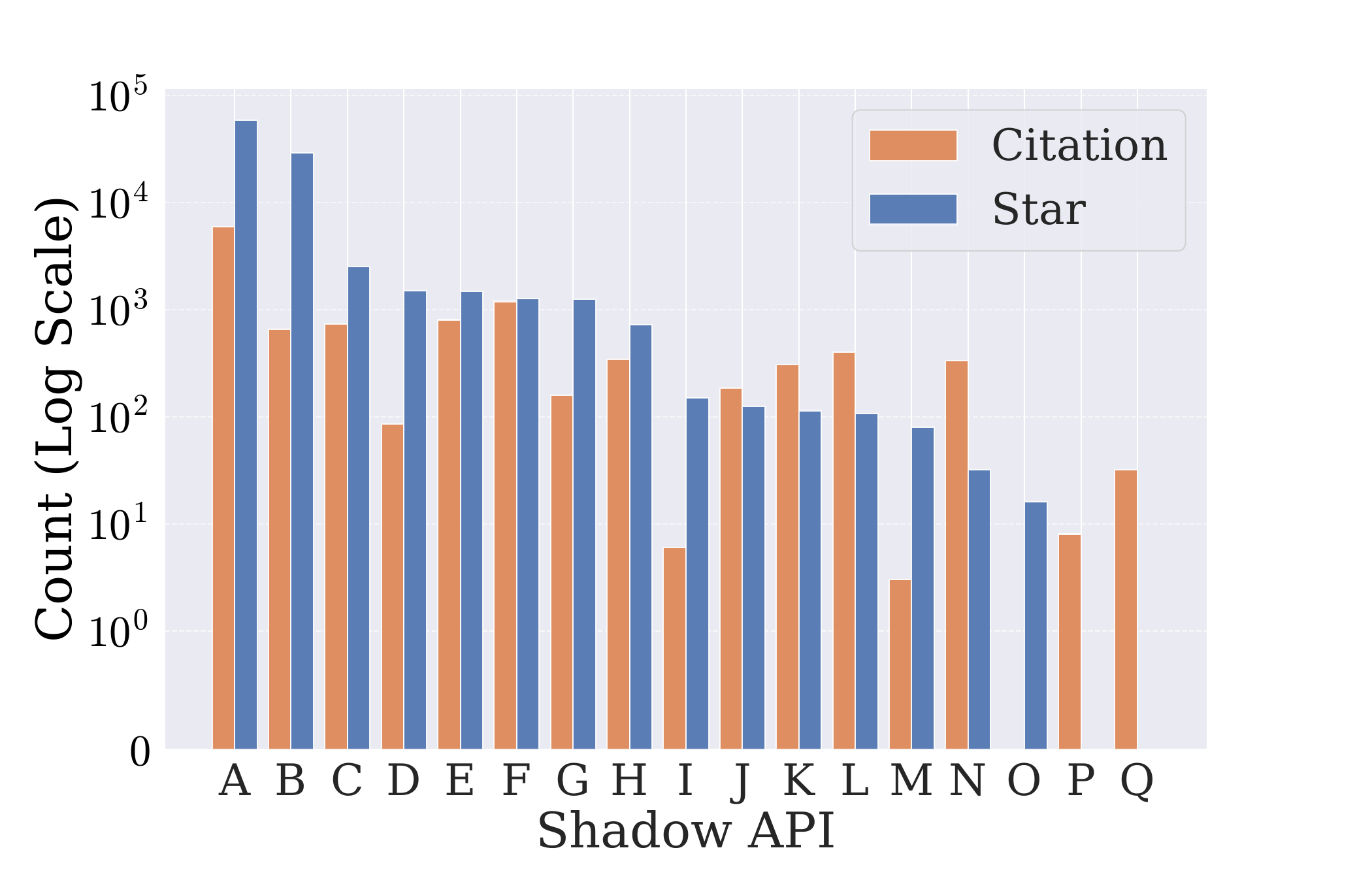}
    \caption{Citations and GitHub stars.}
    \label{figure:github_cites}
\end{subfigure}
\caption{Landscape of the shadow APIs.}
\label{figure:overall_impact}
\end{figure*}

\mypara{Usage}
We ultimately identify 17 shadow APIs whose endpoints appear in 187 research papers. Among these, 116 papers ($62.03\%$) have been accepted at peer-reviewed conferences\footnote{For *ACL conferences, our statistics also encompass papers published in Findings.} or journals, such as ACL, CVPR, and ICLR.
The most widely used shadow API alone has accumulated more than 5{,}900 citations, and the associated GitHub repositories have received more than 58{,}000 stars in total, as shown in~\autoref{figure:github_cites}.
These results indicate that shadow API usage is already widespread in research.
As shown in~\autoref{figure:author_regions}, most authors are affiliated with institutions in regions such as China, where access to certain proprietary models is restricted.

\mypara{Infrastructure} 
Among the 17 identified shadow API services, our analysis reveals that 11 are built upon open-source AI model aggregation and redistribution systems, primarily OneAPI~\cite{OneAPI} and its derivative, NewAPI~\cite{NewAPI}. 
OneAPI is an open-source tool designed for self-hosted deployment. 
It unifies interfaces from various commercial LLM providers into a standard OpenAI-compatible format. 
The infrastructure system supports key features such as API key management, secondary redistribution, request routing, and automatic retries, thereby increasing the potential for exploitation, resale, and abuse more than the official API.

\mypara{Compliance and Transparency}
The compliance and transparency of third-party API services are critical, as they determine whether users' rights can receive legal protection, including whether the service operates as described and whether it can ensure continuous and stable availability.
To assess the compliance status of shadow APIs, we examine publicly available information on provider identity, corporate registration, and service-related disclosures.
Our analysis shows that 15 of the 17 identified services are operated by individuals without transparent identity information or verifiable provenance.  
Only one provider holds a valid corporate registration through an Internet Content Provider filing in China.  
As a result, the provider ecosystem exhibits high operational volatility: two services had already ceased operations during our measurements, and four in total had ceased by the time of writing~(\autoref{section:post_disclosure}).  
In addition, because upstream routing is not transparently documented, our measurements within a bounded observation window cannot establish whether the audited endpoints change backend models over time. 
Passing our verification checks during one measurement window therefore does not establish continued consistency in subsequent requests or periods. 
This limited observability constrains users’ ability to verify model provenance and assess operational risk.
Full compliance metadata for all 17 providers is provided in~\autoref{appendix:anonymized_apis}.

\begin{tcolorbox}[colback=gray!25!white, size=title,breakable,boxsep=1mm,colframe=white,before={\vskip1mm}, after={\vskip0mm}]
\textbf{RQ1 Take-Aways:}
Shadow APIs are already widely popular in usage, yet they operate with minimal transparency and governance. 
Most providers lack verifiable identity, stable infrastructure, or disclosure of upstream models.
\end{tcolorbox}

%-------------------------------------------------------------------------------
\section{Performance Evaluation}
\label{section:utility}
%-------------------------------------------------------------------------------
To address \textbf{RQ2}, we evaluate the performance of shadow APIs via utility and safety benchmarks. 
We opt for this controlled approach over reproducing specific prior studies to mitigate the instability of shadow services and preserve the anonymity of affected researchers.

%-------------------------------------------------------------------------------
\subsection{Experimental Setups}
%-------------------------------------------------------------------------------

\mypara{Models}
To ensure a comprehensive evaluation across different providers and models, we select target models based on token usage statistics. 
We refer to the November 2025 usage ranking of LLMs on the OpenRouter public leaderboard, with the detailed chart provided in~\autoref{appendix:ranking}.
From the ranking, we identify three primary model families (F-A, F-B, and F-C) covering both proprietary models and open-source model frontiers:
\begin{itemize}
    \item \textbf{F-A (OpenAI)}: GPT-4o-mini, GPT-5, and GPT-5-mini~\cite{gpt5}.
    \item \textbf{F-B (Google)}: Gemini-2.0-flash, Gemini-2.5-flash, and Gemini-2.5-pro.
    \item \textbf{F-C (DeepSeek)}: DeepSeek-Chat and DeepSeek-Reasoner.
\end{itemize}
For the science domain evaluation, we evaluate all the above eight LLMs.
For the sensitive domain and safety evaluations, we employ a filtered subset to focus on representative behaviors.
We select one distinct, widely deployed model from each family:
GPT-5-mini (representing F-A), Gemini-2.5-flash (representing F-B), and DeepSeek-Chat (representing F-C).

\mypara{Shadow API Selection}
To ensure the representativeness of our study, we select shadow APIs based on two popularity proxies, i.e., project citation counts and GitHub stars~\cite{BHV16}.
We assign identifiers to providers (shadow APIs A, B, etc.) based on a descending sort of their associated GitHub star counts, which are summarized in~\autoref{figure:github_cites}.
The corresponding service names are provided in~\autoref{table:compliance_full}.
Specifically, we select shadow APIs A, E, and H based on three criteria: popular with regard to citations and GitHub stars, publicly accessible, and comprehensively covered for models in the three identified model families.
Brief profiles for these selected APIs are provided in~\autoref{appendix:profiles}.
All official baselines are queried directly through the corresponding official APIs.

%-------------------------------------------------------------------------------
\mypara{Experimental Details}
%-------------------------------------------------------------------------------
To reduce variance, we average over three trials for all experiments and report accuracy with its standard deviation.
We perform utility and safety evaluations via API queries, which do not require local GPU acceleration. 
For the LLMmap method, we utilize an NVIDIA DGX A100 with GPU acceleration to train the model fingerprint database.
Detailed model configurations and API parameters are provided in~\autoref{appendix:model_configurations}.

%-------------------------------------------------------------------------------
\subsection{Utility Evaluation}
%-------------------------------------------------------------------------------

\mypara{Benchmarks and Methodology}
We assess model utility across scientific and sensitive domains for the evaluated model families, including Gemini 2.5, DeepSeek-V3.2, and Qwen3~\cite{T255,D252,T25}.
For the science domain, we employ the AIME 2025 benchmark~\cite{AIME_Problems}, which tests competition-level mathematics, and the GPQA (Diamond) benchmark~\cite{RHSPPDMB23}, targeting PhD-level scientific questions.
For the sensitive domain, we focus on high-risk fields, i.e., medical and legal fields, where ``high-risk'' refers to the severity of real-world consequences that incorrect outputs carry for patients and legal parties, rather than to any regulatory classification.
In the medical field, following prior work~\cite{STSCLZXWZWSCG25}, we use the MedQA (USMLE) dataset~\cite{JPOWFS20}, covering diagnosis, treatment, and medical concepts.
In the legal field, we select LegalBench~\cite{GNHRCNCPWRZTHSFSDPHWNNNCTHMLRHKHRGGWGZIL23}, specifically the SCALR task for evaluating legal reasoning and reading comprehension, consistent with~\cite{CTXWML25, HYGWKW25}.

\mypara{Implementation Details}
To ensure fair comparison, we adopt EvalScope prompt templates~\cite{EvalScope_Datasets} for the AIME 2025 and GPQA benchmarks, and for the MedQA (USMLE) benchmark, we use the Hulu-Med multiple-choice template~\cite{JWSHZPZYFZHCWTLXWXFZLXSWL25}, and the original task-specific prompts for LegalBench (Scalr)~\cite{GNHRCNCPWRZTHSFSDPHWNNNCTHMLRHKHRGGWGZIL23}.
We obtain all results using the model configurations detailed in~\autoref{table:api-capability-comparison}.

\begin{figure}[!t]
\centering
\begin{subfigure}{\linewidth}
    \centering
    \includegraphics[width=\linewidth]{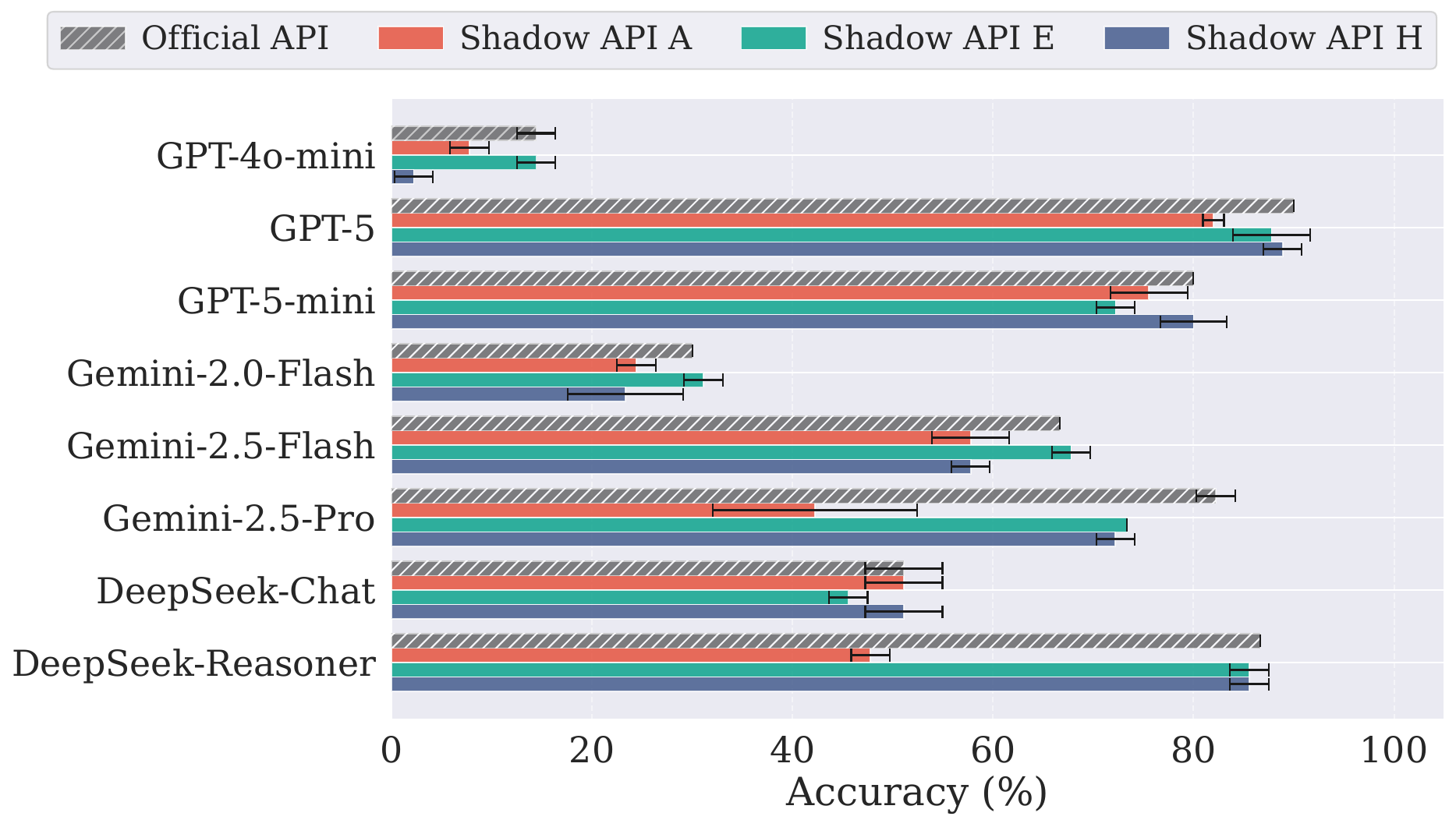}
    \caption{AIME 2025}
    \label{figure:AIME}
\end{subfigure}
\begin{subfigure}{\linewidth}
    \centering
    \includegraphics[width=\linewidth]{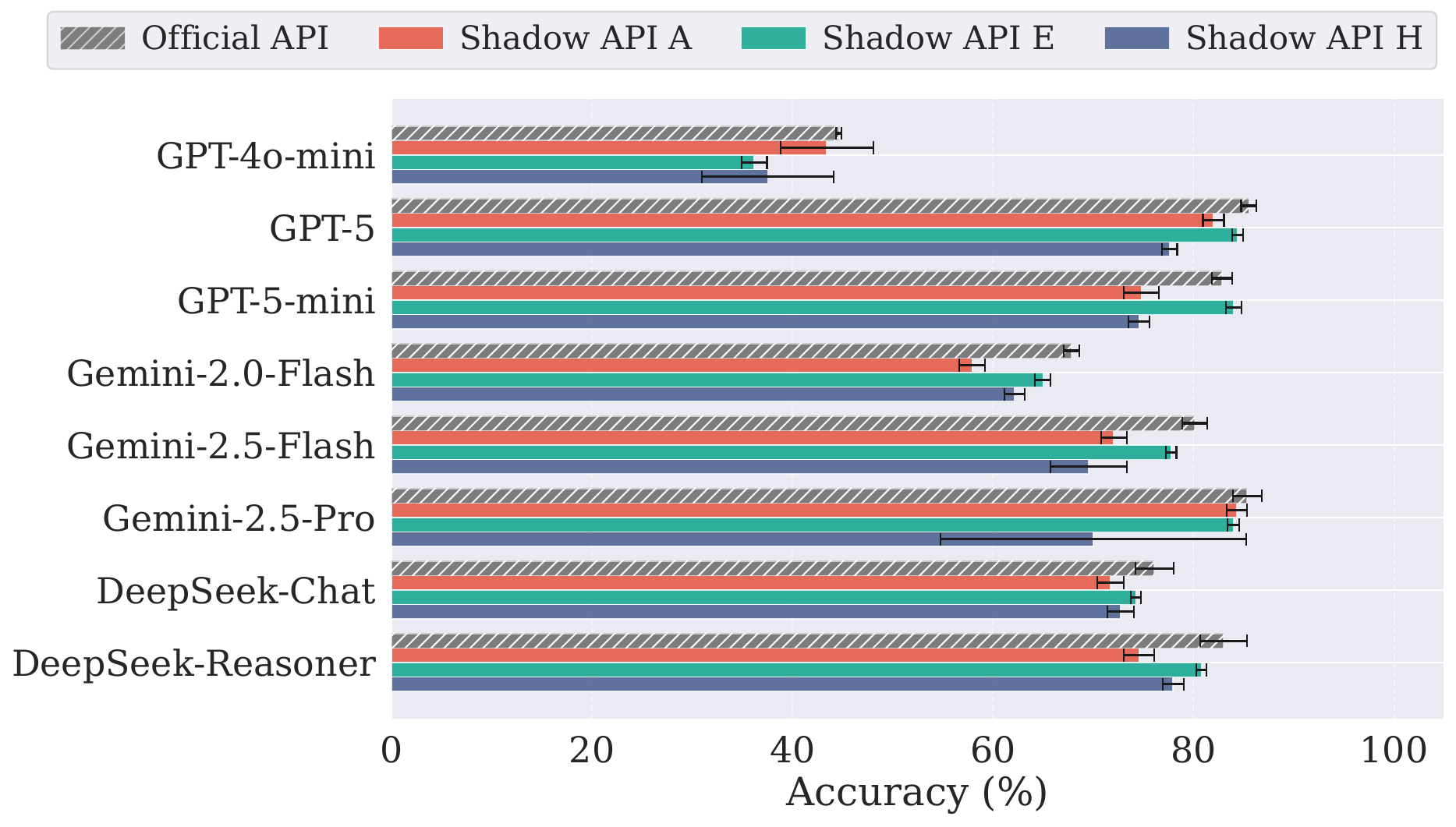}
    \caption{GPQA (Diamond)}
    \label{figure:GPQA}
\end{subfigure}
\caption{Performance comparison on (a) AIME 2025 and (b) GPQA benchmarks across official and shadow APIs.}
\label{figure:main_results}
\end{figure}

\mypara{Science Domain Performance}
The official API serves as the reference baseline rather than a strict performance upper bound, as illustrated in~\autoref{figure:main_results}.
Throughout, the quantity of interest is divergence from the official endpoint under identical prompts and decoding configurations, not absolute benchmark accuracy: an endpoint faithfully serving the claimed model should track the official baseline regardless of how high that baseline is.
Among the shadow APIs, shadow API E exhibits the strongest consistency, maintaining a minimal average accuracy drop of $2.64\%$ from the official API.
In some instances, such as GPT-5-mini on GPQA, it even marginally outperforms the official API by $1.18\%$, likely reflecting small evaluation or decoding variance.
In contrast, official APIs exhibit minimal performance variance, whereas shadow APIs show substantially higher variability.
In particular, shadow APIs A and H display pronounced divergences from official APIs, with average accuracy gaps of $9.81\%$ and $6.46\%$, respectively.
Although these shadow APIs perform comparably on non-reasoning tasks, their performance degrades markedly on reasoning-oriented models.
Notably, on the AIME 2025 benchmark, shadow API A suffers severe accuracy drops, with deficits $40.00\%$ for Gemini-2.5-pro and $38.89\%$ for DeepSeek-Reasoner.
These results indicate that advanced reasoning capabilities are significantly compromised in shadow APIs A and H, resulting in substantial deviations from their claimed parity with official API performance.

\begin{figure}[!t]
\centering
\begin{subfigure}{\linewidth}
    \centering
    \includegraphics[width=\linewidth]{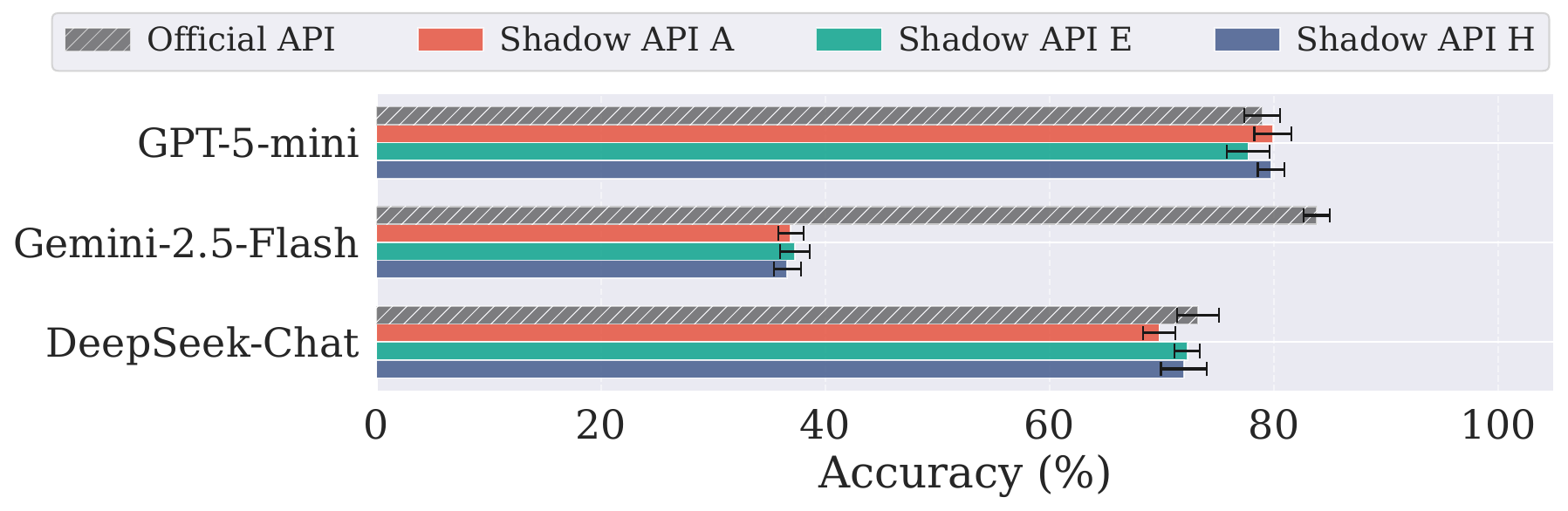}
    \caption{MedQA (USMLE)}
    \label{figure:MedQA}
\end{subfigure}
\begin{subfigure}{\linewidth}
    \centering
    \includegraphics[width=\linewidth]{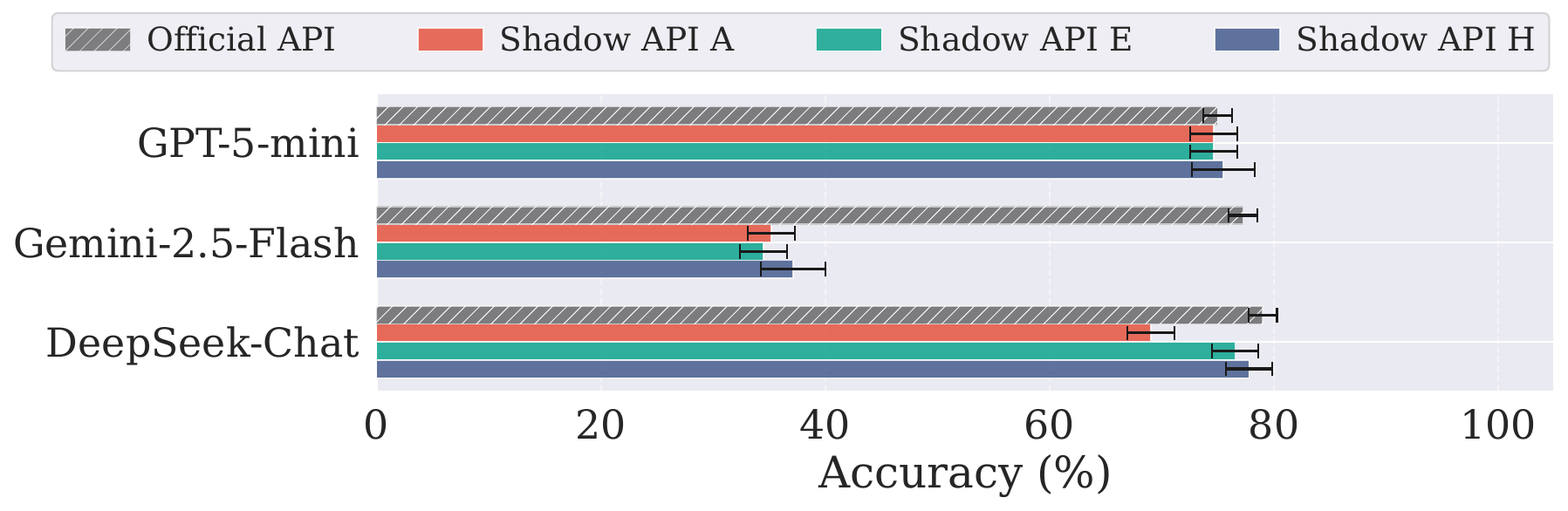}
    \caption{LegalBench (Scalr)}
    \label{figure:LegalBench}
\end{subfigure}
\caption{Performance comparison on (a) MedQA (USMLE) and (b) LegalBench (Scalr) benchmarks across official and shadow APIs.}
\end{figure}

\mypara{Sensitive Domain Performance}
For sensitive domain, including medicine field (MedQA (USMLE)) and law field (LegalBench (Scalr)), performance results are shown in~\autoref{figure:MedQA} and~\autoref{figure:LegalBench}.
Shadow APIs A, E, and H exhibit average accuracy drops of $16.96\%$, $15.71\%$, and $14.75\%$, respectively.
Performance on Gemini-2.5-flash degrades across all shadow APIs. 
In particular, its accuracy on MedQA decreases sharply from an official score of $83.82\%$ to an average of $36.95\%$, corresponding to a substantial performance deficit of $46.51\%$–$47.21\%$.
A consistent collapse is observed on LegalBench, where all shadow APIs lag behind the official endpoints by $40.10\%$ to $42.73\%$.
DeepSeek-Chat also demonstrates instability, particularly in the legal field.
While shadow APIs E and H remain relatively stable, shadow API A exhibits a significant accuracy drop of $9.98\%$ on LegalBench.
\autoref{table:safety_case} presents two cases in which all three shadow endpoints return incorrect answers in either all three runs or two of the three runs.
\autoref{table:medqa_discrepancy} and~\autoref{table:legalbench_discrepancy} (in~\autoref{appendix:detailed_discrepancy}) quantify this unreliability at the per-query level.
For Gemini-2.5-flash on MedQA, the three shadow APIs produce answers consistent with the official API in only $49.57\%$ to $49.65\%$ of cases, and the corresponding three-way consistency range on LegalBench is $52.36\%$ to $54.64\%$.
The tightness of this range across three independently operated providers is itself informative.
The shadow APIs exhibit nearly identical consistency rates on the same query set, a recurring pattern whose cause cannot be established from performance evidence alone~(\autoref{section:threat_model}).
GPT-5-mini and DeepSeek-Chat exhibit smaller but still non-trivial consistency gaps, ranging from $80.56\%$ to $96.70\%$.
Reliance on shadow APIs for professional guidance therefore poses severe safety risks even when headline accuracy gaps appear moderate.

%-------------------------------------------------------------------------------
\begin{table}[t]
\centering
\renewcommand{\arraystretch}{1.3}
\caption{Medical and legal fields failure response examples.}
\label{table:safety_case}
\footnotesize
\begin{tabularx}{\linewidth}{@{}p{1.75cm} X p{1.7cm} p{2.15cm}@{}}
\toprule
\textbf{Benchmark} &
\textbf{Question (abridged)} &
\textbf{Official API}
\newline\textbf{(R1/R2/R3)} &
\textbf{Shadow APIs}
\newline\textbf{(R1/R2/R3)} \\
\midrule

MedQA (USMLE)
\newline{\scriptsize Gemini-2.5-Flash}
\newline{\scriptsize CSV ID 764; Gold: D} &
A patient in active labor with major bleeding requests transfer from a
hospital not equipped for high-risk pregnancy. What is the appropriate
next step? &
D~\cmark{} / D~\cmark{} / D~\cmark{}
\newline
{\scriptsize Treat and transfer after a written request.} &
\textbf{A, E, H:}
E~\xmark{} / E~\xmark{} / E~\xmark{}
\newline
{\scriptsize Transfer only if the benefits outweigh the risks.} \\
\midrule

LegalBench (Scalr)
\newline{\scriptsize DeepSeek-Chat}
\newline{\scriptsize CSV ID 147; Gold: 0} &
Whether nondisclosure of drug adverse-event reports can support a
Rule 10b-5 claim when the reports are not statistically significant. &
0~\cmark{} / 3~\xmark{} / 3~\xmark{} &
\textbf{A:}
0~\cmark{} / 1~\xmark{} / 1~\xmark{}
\newline
\textbf{E:}
0~\cmark{} / 3~\xmark{} / 3~\xmark{}
\newline
\textbf{H:}
0~\cmark{} / 4~\xmark{} / 4~\xmark{}
\newline\\
\bottomrule
\end{tabularx}

\vspace{1mm}
\parbox{\linewidth}{\scriptsize
Entries are ordered as Runs 1/2/3. A checkmark denotes agreement
with the benchmark gold answer; a cross denotes disagreement.}
\end{table}
%-------------------------------------------------------------------------------

%-------------------------------------------------------------------------------
\subsection{Safety Evaluation}
%-------------------------------------------------------------------------------

\mypara{Datasets}
We consider two widely used benchmarks, JailbreakBench~\cite{CDRACSDFPTHW24} and AdvBench~\cite{ZWCNKF23}, which respectively contain 100 and 520 harmful requests covering different categories (such as deception, discrimination, and physical harm) to evaluate the safety capabilities of LLMs in the face of unsafe user requests.
For AdvBench, we use the 50-item subset provided by~\cite{CRDHPW25}.
We choose these benchmarks to quantify one dimension of endpoint safety (i.e., response to harmful requests) through black-box API queries without access to model internals.
JailbreakBench~\cite{CDRACSDFPTHW24} offers balanced, policy-aligned coverage across ten misuse categories, whereas the representative, redundancy-reduced AdvBench-50~\cite{ZWCNKF23,CRDHPW25} provides a second established inventory and spans 32 harmful categories; using both reduces reliance on any single benchmark's curation.
Their wide use in evaluating various jailbreak methods and LLMs~\cite{MZKNASK23,CRDHPW25,RSTLLASLFSAABDMDGSLVRXMCIBHMJSCRMAGMTYMASAIKLHSSCRSSo24,LHXFDH24} supports their relevance as established safety probes.

\mypara{Implementation Details}
We employ four distinct jailbreak attacks to evaluate the safety: GCG~\cite{ZWCNKF23}, Base64~\cite{WHS23}, Combination~\cite{WHS23}, and FlipAttack~\cite{LHXFDH24}.
For GCG, we use LLaMA3-8B to generate a sample-level suffix and then transfer it to other LLMs.
For Base64 and Combination, we follow the settings for Base64 and \texttt{combination\_1} in~\cite{WHS23}. 
For FlipAttack, we use its well-performed ``flip char in sentence'' mode.

\mypara{Metrics}
Following prior work~\cite{CRDHPW25, MZKNASK23, JLBZ25, JHLZBZ26} such as PAIR, TAP, and JAIL-CON, we introduce a lightweight judge model to output the \textbf{\emph{harmfulness score}} and verify whether the generated answer is valid and harmful.
Specifically, based on~\cite{SLBTHPASEWT24}, we build the evaluator with GPT-4o-mini and the rubric-based prompt template from StrongREJECT. 
A higher harmfulness score indicates a more harmful answer, i.e., lower safety.

\begin{figure*}[!t]
\centering
\begin{subfigure}{0.32\linewidth}
    \centering
    \includegraphics[width=\linewidth]{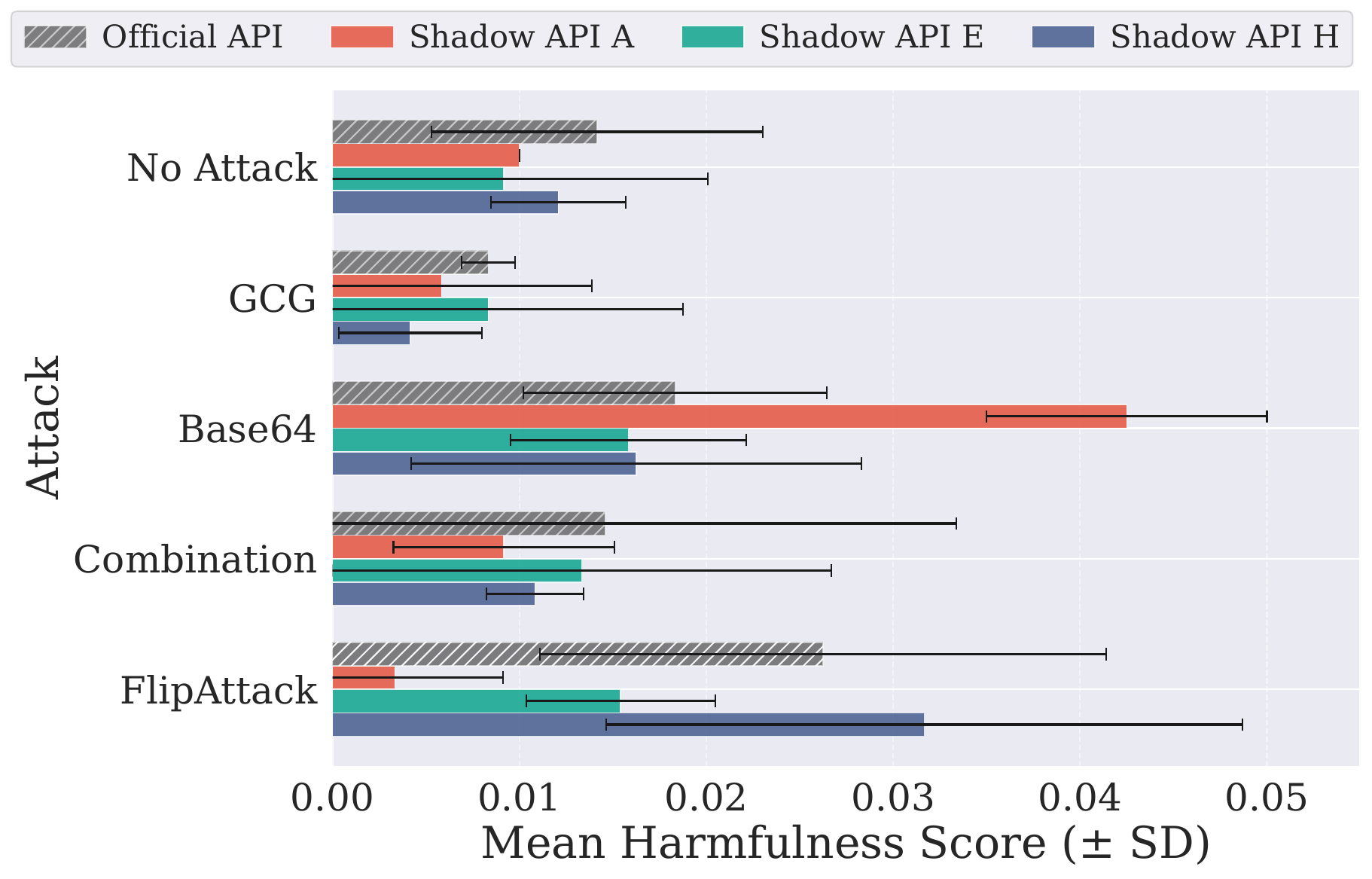}
    \caption{GPT-5-mini}
    \label{figure:GPT_safety}
\end{subfigure}
\begin{subfigure}{0.323\linewidth}
    \centering
    \includegraphics[width=\linewidth]{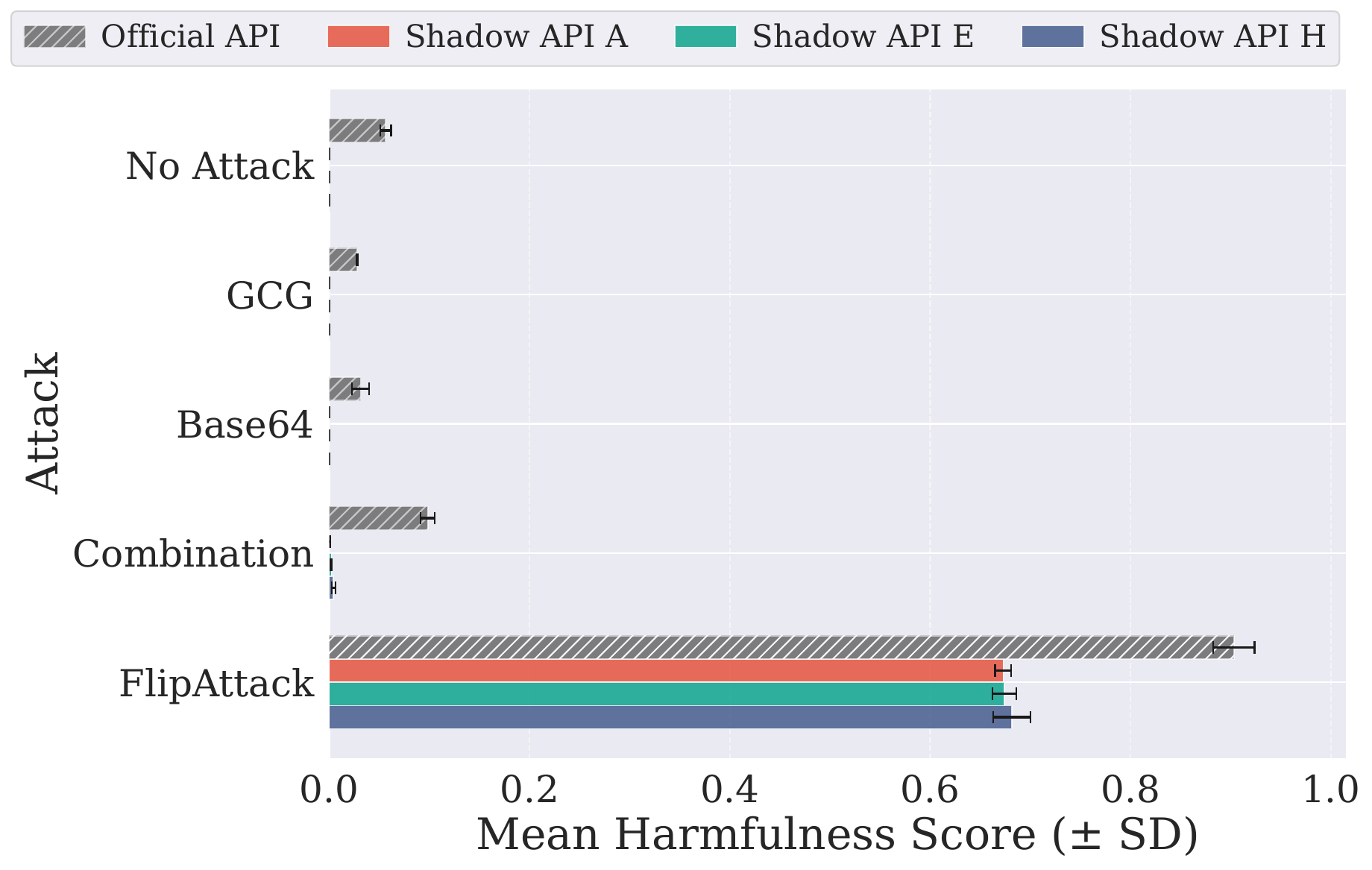}
    \caption{Gemini-2.5-flash}
    \label{figure:gemini_safety}
\end{subfigure}
\begin{subfigure}{0.322\linewidth}
    \centering
    \includegraphics[width=\linewidth]{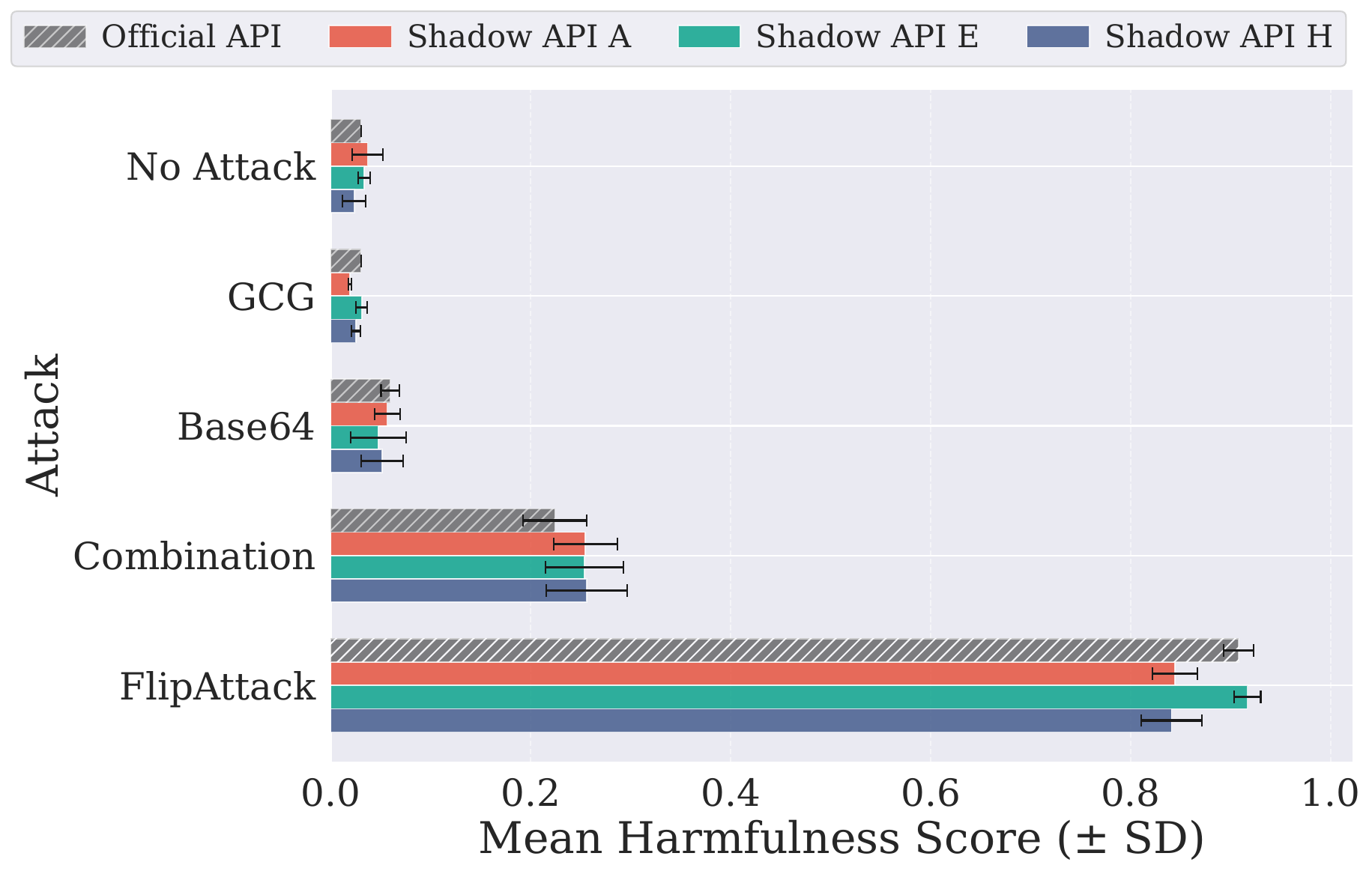}
    \caption{DeepSeek-Chat}
    \label{figure:deepseek_safety}
\end{subfigure}
\caption{Safety performance comparison on the JailbreakBench dataset.}
\label{figure:safety_JBB}
\end{figure*}

\begin{figure*}[!t]
\centering
\begin{subfigure}{0.32\linewidth}
    \centering
    \includegraphics[width=\linewidth]{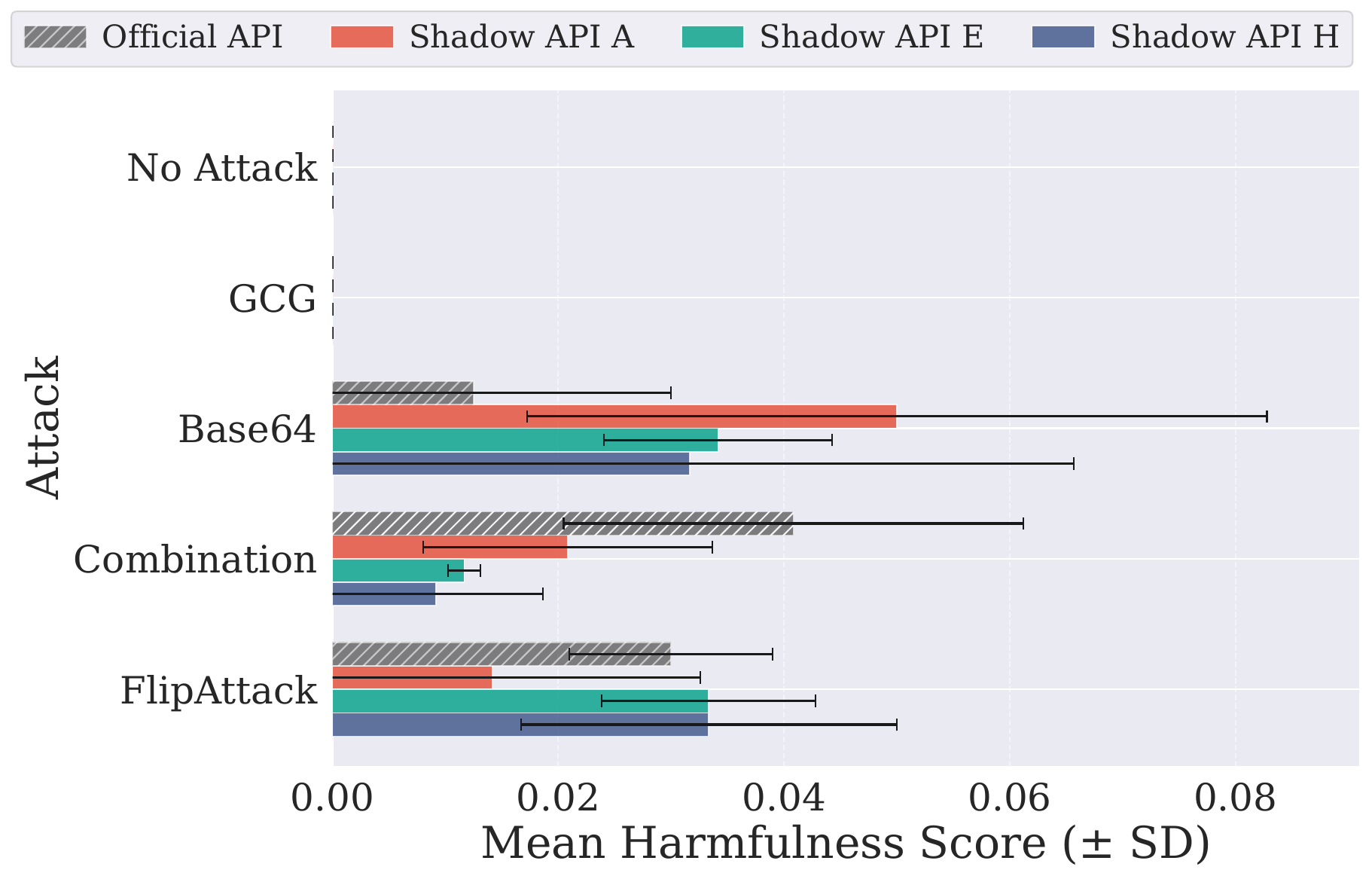}
    \caption{GPT-5-mini}
    \label{figure:GPT_safety_adv}
\end{subfigure}
\begin{subfigure}{0.32\linewidth}
    \centering
    \includegraphics[width=\linewidth]{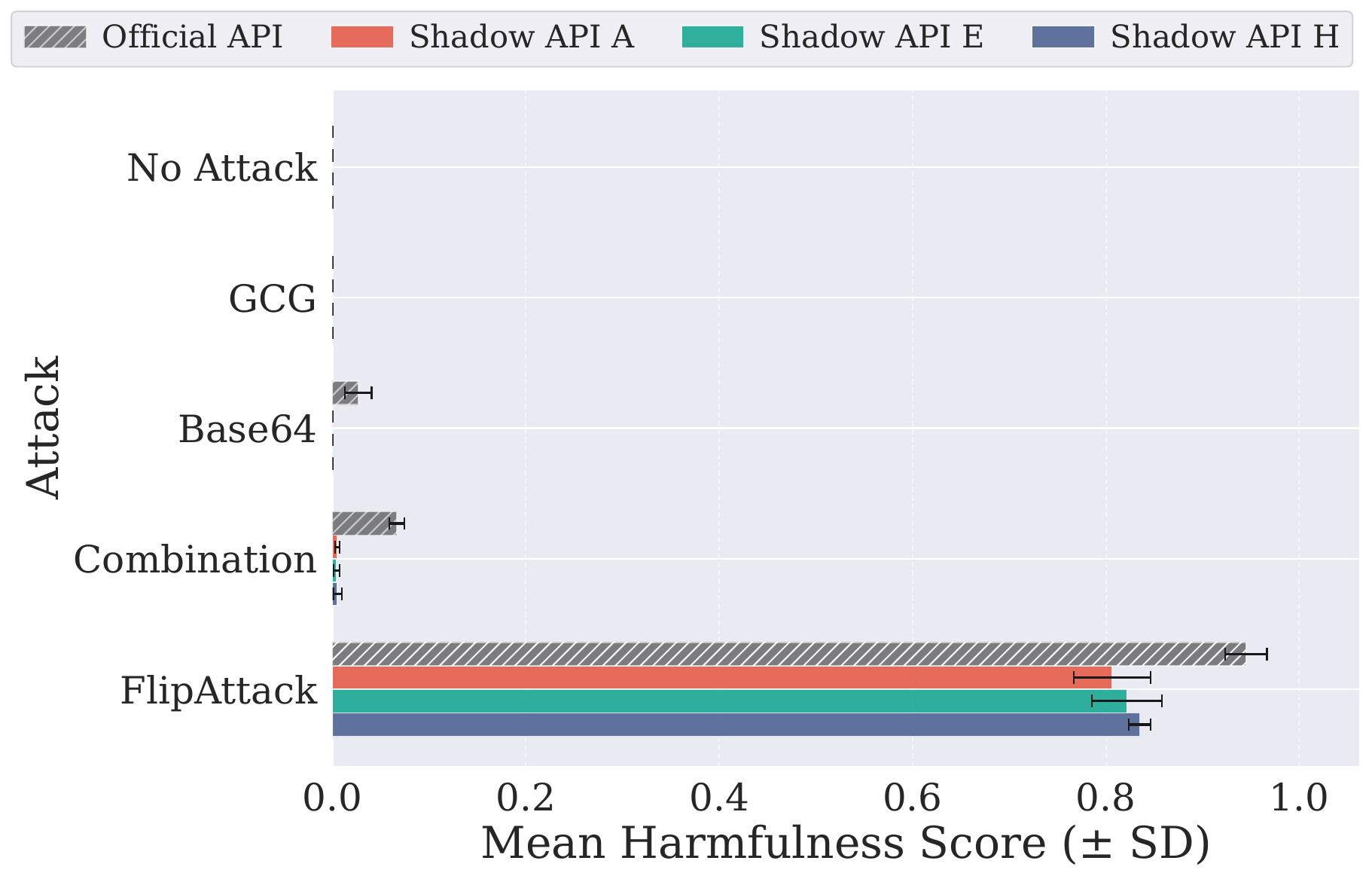}
    \caption{Gemini-2.5-flash}
    \label{figure:gemini_safety_adv}
\end{subfigure}
\begin{subfigure}{0.32\linewidth}
    \centering
    \includegraphics[width=\linewidth]{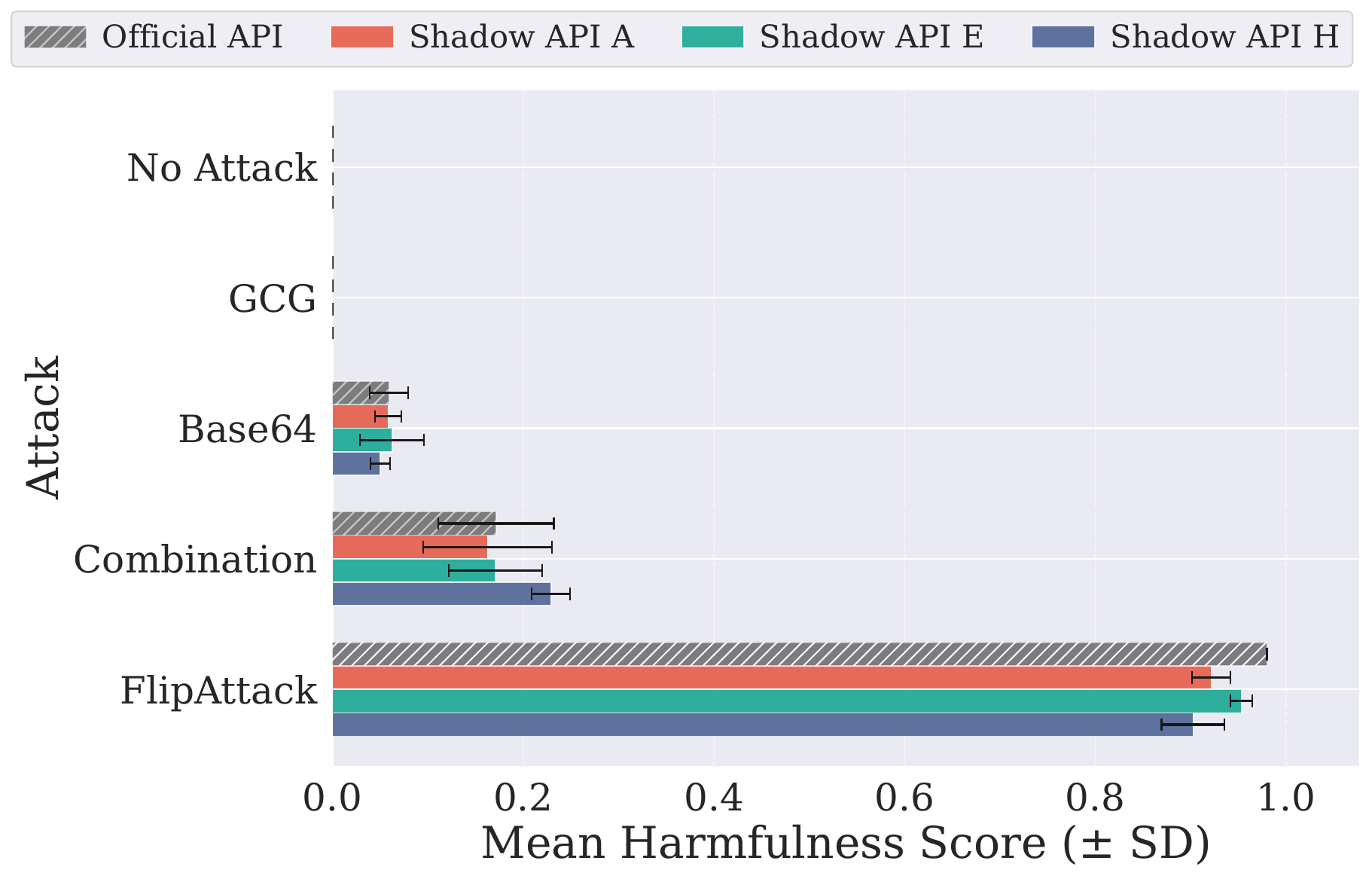}
    \caption{DeepSeek-Chat}
    \label{figure:deepseek_safety_adv}
\end{subfigure}
\caption{Safety performance comparison on the AdvBench dataset.}
\label{figure:adv_safety_results}
\end{figure*}

%-------------------------------------------------------------------------------
\begin{table*}[ht!]
\centering
\caption{Fingerprinting results via LLMmap matched model and mean Euclidean distance $D$ with standard deviation.
Colors denote ratio vs official, \colorbox{mygreen}{Green} ($D \leq 1.25\times$ baseline), \colorbox{myyellow}{Yellow} ($D > 1.25\times$ baseline), and \colorbox{red!20}{Red} (incorrect model).
}
\label{table:fingerprinting_results}
\scalebox{0.78}{
\resizebox{\textwidth}{!}{
\begin{tabular}{@{}llcccc@{}}
\toprule
\multirow{2}{*}{\textbf{Model Family}} & \multirow{2}{*}{\textbf{Model}} & \textbf{Official API} & \textbf{Shadow API A} & \textbf{Shadow API E} & \textbf{Shadow API H} \\
& & \textit{(Baseline)} & \textit{(Standard)} & \textit{(Standard)} & \textit{(Standard)} \\
\midrule

\multirow{6}{*}{GPT} 
& \multirow{2}{*}{GPT-4o-mini} 
  & \begin{tabular}{@{}c@{}} \textbf{GPT-4o-mini-0718} \\ \footnotesize [$D: 20.01_{\pm 0.96}$] \end{tabular} 
  & \cellcolor{myred!20}\begin{tabular}{@{}c@{}} gpt-4o-2024-05-13 \\ \footnotesize [$D: 16.66_{\pm 2.40}$] \end{tabular} 
  & \cellcolor{mygreen}\begin{tabular}{@{}c@{}} GPT-4o-mini-0718 \\ \footnotesize [$D: 19.18_{\pm 1.41}$] \end{tabular} 
  & \cellcolor{myred!20}\begin{tabular}{@{}c@{}} Qwen2.5-7B \\ \footnotesize [$D: 17.43_{\pm 2.07}$] \end{tabular} \\

\cmidrule{2-6}

& \multirow{2}{*}{GPT-5} 
  & \begin{tabular}{@{}c@{}} \textbf{gpt-5-2025-08-07} \\ \footnotesize [$D: 13.14_{\pm 0.34}$] \end{tabular} 
  & \cellcolor{myred!20}\begin{tabular}{@{}c@{}} glm-4-9b-chat \\ \footnotesize [$D: 20.88_{\pm 2.81}$] \end{tabular} 
  & \cellcolor{myred!20}\begin{tabular}{@{}c@{}} glm-4-9b-chat \\ \footnotesize [$D: 23.50_{\pm 0.83}$] \end{tabular} 
  & \cellcolor{mygreen}\begin{tabular}{@{}c@{}} gpt-5-2025-08-07 \\ \footnotesize [$D: 16.24_{\pm 3.84}$] \end{tabular} \\
\cmidrule{2-6}

& \multirow{2}{*}{GPT-5-mini} 
  & \begin{tabular}{@{}c@{}} \textbf{gpt-5-mini-2025-08-07} \\ \footnotesize [$D: 14.57_{\pm 3.82}$] \end{tabular} 
  & \cellcolor{myyellow}\begin{tabular}{@{}c@{}} gpt-5-mini-2025-08-07 \\ \footnotesize [$D: 18.63_{\pm 2.72}$] \end{tabular} 
  & \cellcolor{mygreen}\begin{tabular}{@{}c@{}} gpt-5-mini-2025-08-07 \\ \footnotesize [$D: 16.01_{\pm 3.59}$] \end{tabular} 
  & \cellcolor{myred!20}\begin{tabular}{@{}c@{}} gpt-5-2025-08-07 \\ \footnotesize [$D: 18.04_{\pm 1.76}$] \end{tabular} \\

\midrule

\multirow{6}{*}{Gemini} 
& \multirow{2}{*}{Gemini-2.0-flash} 
  & \begin{tabular}{@{}c@{}} \textbf{gemini-2.0-flash} \\ \footnotesize [$D: 17.54_{\pm 2.28}$] \end{tabular} 
  & \cellcolor{myred!20}\begin{tabular}{@{}c@{}} gemini-2.5-flash \\ \footnotesize [$D: 17.02_{\pm 2.64}$] \end{tabular} 
  & \cellcolor{myred!20}\begin{tabular}{@{}c@{}} gemini-2.5-flash \\ \footnotesize [$D: 14.10_{\pm 1.28}$] \end{tabular} 
  & \cellcolor{mygreen}\begin{tabular}{@{}c@{}} gemini-2.0-flash \\ \footnotesize [$D: 18.49_{\pm 2.50}$] \end{tabular} \\
\cmidrule{2-6}

& \multirow{2}{*}{Gemini-2.5-flash} 
  & \begin{tabular}{@{}c@{}} \textbf{gemini-2.5-flash} \\ \footnotesize [$D: 15.19_{\pm 2.71}$] \end{tabular} 
  & \cellcolor{myyellow}\begin{tabular}{@{}c@{}} gemini-2.5-flash \\ \footnotesize [$D: 19.78_{\pm 1.92}$] \end{tabular} 
  & \cellcolor{mygreen}\begin{tabular}{@{}c@{}} gemini-2.5-flash \\ \footnotesize [$D: 15.41_{\pm 3.37}$] \end{tabular} 
  & \cellcolor{mygreen}\begin{tabular}{@{}c@{}} gemini-2.5-flash \\ \footnotesize [$D: 17.51_{\pm 2.22}$] \end{tabular} \\
\cmidrule{2-6}

& \multirow{2}{*}{Gemini-2.5-pro} 
  & \begin{tabular}{@{}c@{}} \textbf{gemini-2.5-pro} \\ \footnotesize [$D: 18.04_{\pm 0.93}$] \end{tabular} 
  & \cellcolor{mygreen}\begin{tabular}{@{}c@{}} gemini-2.5-pro \\ \footnotesize [$D: 18.04_{\pm 0.61}$] \end{tabular} 
  & \cellcolor{mygreen}\begin{tabular}{@{}c@{}} gemini-2.5-pro \\ \footnotesize [$D: 17.37_{\pm 1.76}$] \end{tabular} 
  & \cellcolor{mygreen}\begin{tabular}{@{}c@{}} gemini-2.5-pro \\ \footnotesize [$D: 17.66_{\pm 1.56}$] \end{tabular} \\

\midrule
\multirow{4}{*}{DeepSeek} 
& \multirow{2}{*}{DeepSeek-Chat} 
  & \begin{tabular}{@{}c@{}} \textbf{deepseek-chat} \\ \footnotesize [$D: 11.83_{\pm 0.79}$] \end{tabular} 
  & \cellcolor{myred!20}\begin{tabular}{@{}c@{}} deepseek-v3-0324 \\ \footnotesize [$D: 19.77_{\pm 3.78}$] \end{tabular} 
  & \cellcolor{myyellow}\begin{tabular}{@{}c@{}} deepseek-chat \\ \footnotesize [$D: 21.28_{\pm 1.57}$] \end{tabular} 
  & \cellcolor{myred!20}\begin{tabular}{@{}c@{}} gemma-2-9b-it \\ \footnotesize [$D: 23.40_{\pm 1.05}$] \end{tabular} \\
\cmidrule{2-6}

& \multirow{2}{*}{DeepSeek-Reasoner} 
  & \begin{tabular}{@{}c@{}} \textbf{deepseek-reasoner} \\ \footnotesize [$D: 17.04_{\pm 3.25}$] \end{tabular} 
  & \cellcolor{myred!20}\begin{tabular}{@{}c@{}} deepseek-chat \\ \footnotesize [$D: 19.02_{\pm 1.53}$] \end{tabular} 
  & \cellcolor{mygreen}\begin{tabular}{@{}c@{}} deepseek-reasoner \\ \footnotesize [$D: 21.19_{\pm 0.34}$] \end{tabular} 
  & \cellcolor{myred!20}\begin{tabular}{@{}c@{}} deepseek-chat \\ \footnotesize [$D: 22.68_{\pm 2.69}$] \end{tabular} \\
\bottomrule
\end{tabular}
}}
\end{table*}
%-------------------------------------------------------------------------------

\mypara{Safety Evaluation on JailbreakBench}
As shown in~\autoref{figure:safety_JBB}, we observe  a  concerning,  inconsistent behavior on JailbreakBench, where shadow APIs deviate unpredictably from official baselines.
Specifically, for GPT-5-mini under the Base64 attack~(\autoref{figure:GPT_safety}), shadow API A yields a harmfulness score of $0.04$, which is double the official API's score of $0.02$.
Similarly, inconsistencies persist in the FlipAttack, where shadow APIs A and E yield substantially lower harmfulness scores than the official endpoint.
For Gemini-2.5-flash~(\autoref{figure:gemini_safety}), all shadow APIs yield lower measured harmfulness scores than the official API across all attacks, particularly against FlipAttack (the official API reaches a high harmfulness score of $0.90$, all shadow APIs around $0.67$ and $0.68$, resulting in a significant gap of approximately $0.23$).
In the case of DeepSeek-Chat~(\autoref{figure:deepseek_safety}), the shadow APIs exhibit smaller differences compared to GPT-5-mini and Gemini-2.5-flash, but differences from the official API still exist.
For instance, shadow APIs A and H generate more (less) harmful content under the attack Combination (FlipAttack).

\mypara{Safety Evaluation on AdvBench}
We observe results on AdvBench similar to those on JailbreakBench~(\autoref{figure:adv_safety_results}), where shadow APIs deviate unpredictably from official endpoints.
Specifically, for GPT-5-mini under the Base64 attack~(\autoref{figure:GPT_safety_adv}), shadow API A yields a harmfulness score of $0.05$, which is 5 $\times$ the official API's score of $0.01$.
Similarly, shadow APIs A, E, and H yield substantially lower harmfulness scores in the Combination attack.
For instance, for Gemini-2.5-flash~(\autoref{figure:gemini_safety_adv}), all shadow APIs yield lower measured harmfulness scores than the official API across all attacks, particularly against FlipAttack (the official API reaches a high harmfulness score of $0.95$, all shadow APIs around $0.82$).
In the case of DeepSeek-Chat~ (\autoref{figure:deepseek_safety_adv}), the shadow APIs exhibit smaller differences compared to GPT-5-mini and Gemini-2.5-flash, but differences from the official API still exist.
For example, shadow API H generates more harmful content than the official API under the Combination (FlipAttack) attack.
In addition, the harmfulness scores for shadow APIs A and E are comparable to the official endpoints in Base64 and Combination, but lower than the official scores in FlipAttack.

These findings imply that relying on shadow APIs for safety evaluation is unreliable, as they may fail to reproduce the behaviors of the official endpoints.

\begin{tcolorbox}[colback=gray!25!white, size=title,breakable,boxsep=1mm,colframe=white,before={\vskip1mm}, after={\vskip0mm}]
\textbf{RQ2 Take-Aways:}
Performance evaluation shows shadow APIs could not be replacements for official ones, as they often break on reasoning, and they are unreliable in medical/legal tasks and safety evaluation.
\end{tcolorbox}

%-------------------------------------------------------------------------------
\section{Model Verification}
\label{section:fingerprinting}
%-------------------------------------------------------------------------------
To address \textbf{RQ3}, we move beyond indirect performance signals and directly verify the identity of the served models through model fingerprinting, distributional testing, controlled validation, and metadata analysis.

%-------------------------------------------------------------------------------
\subsection{Fingerprinting-Based Detection}
%-------------------------------------------------------------------------------

\mypara{Methodology}
To identify the specific LLMs operating behind shadow API service providers, we employ LLMmap~\cite{PKA25}, an active fingerprinting framework. 
This framework classifies models by analyzing responses to a curated set of queries and computing the Euclidean distance between model outputs and a reference database.
We conduct our fingerprinting experiments following the configuration in~\cite{PKA25}, using the default query strategy and extending the framework with the new LLM list.
For GPT-5-mini, GPT-5, Gemini-2.5-flash, and Gemini-2.5-pro, we remove unsupported parameters and configure each model to use its default reasoning effort.
The full list of LLMs used is provided in~\autoref{appendix:llmmap_model}.

\mypara{Results}
We observe significant variance in the identity reliability of shadow APIs across the 24 evaluated endpoints, as summarized in~\autoref{table:fingerprinting_results}.
Specifically, $45.83\%$ of the endpoints fail fingerprint verification, and an additional $12.50\%$ exhibit substantial Euclidean distance deviations from the corresponding official models.
At the family level, the GPT and DeepSeek families show frequent identity mismatches.
Even when a shadow API is correctly identified within the same model family, it often exhibits inflated Euclidean distances (e.g., GPT-5-mini in shadow API A, $D = 18.63_{\pm 2.72}$ versus an official baseline of $14.57_{\pm 3.82}$).
By contrast, the Gemini family exhibits relatively high stability for Gemini-2.5-pro, which maintains consistent Euclidean distances ($D \approx 17.37$--$18.04$) across all providers.

Across the failing endpoints, we observe two recurring model substitution patterns, as supported by the following evidence.
First, premium proprietary models exhibit response patterns that more closely align with cheaper open-source alternatives.
For instance, the behavior of GPT-4o-mini in shadow API H deviates toward Qwen2.5-7B, while GPT-5 in APIs A and E shows fingerprint signatures resembling GLM-4 or DeepSeek-V3.
Second, specialized or reasoning models may be served by non-reasoning models.
Requests for the thinking-mode DeepSeek-Reasoner through APIs A and H instead behave like the non-thinking DeepSeek-Chat.
Similarly, Gemini-2.0-flash is often misidentified as Gemini-2.5-flash.

\mypara{Capability-Matched Substitution}
To examine whether a substituted backend can resemble the claimed model on standard benchmarks, we additionally benchmark the fingerprinted backend in isolation using the same prompt templates and decoding configuration as~\autoref{section:utility}.
For the GPT-4o-mini$\rightarrow$Qwen2.5-7B substitution observed in shadow API H, Qwen2.5-7B achieves $3.33\%$ on AIME~2025 and $42.93\%$ on GPQA, compared to the official GPT-4o-mini's $14.44\%$ and $44.61\%$ on the same benchmarks.
The near-parity on GPQA (within $1.68$ \%) contrasts sharply with an $11.11$\% gap on AIME~2025,
indicating that the matched alternative exhibits a capability profile similar to the claimed model on standard reference tasks, while diverging only under more demanding reasoning settings.
This intuitively accounts for two otherwise puzzling observations: the concentration of accuracy drops on reasoning-heavy benchmarks~(\autoref{figure:main_results}), and the regression finding in~\autoref{table:regression} that identity mismatch alone is not a reliable predictor of accuracy drop.
%-------------------------------------------------------------------------------
\subsection{Complementary Verification}
%-------------------------------------------------------------------------------
To address potential blind spots inherent in relying on a single fingerprinting method, we further employ Model Equality Testing (MET)~\cite{GLG25}, a two-sample hypothesis test that checks whether shadow API outputs are drawn from the same distribution as those of the official model.
When the null hypothesis of distributional equality is rejected at $\alpha = 0.05$ ($\textit{Reject} = \text{True}$), the shadow API's outputs are statistically distinguishable from the official model's, providing independent evidence of identity inconsistency.
We evaluate three shadow API providers (A, E, H) across four benchmarks (AIME 2025, GPQA, AdvBench, and JailbreakBench), covering eight model configurations in total.
The full results are reported in~\autoref{table:met_results}.
Overall, MET and LLMmap provide complementary rather than identical evidence.
Because LLMmap gives endpoint-level identity signals while MET gives benchmark-specific distributional tests, we compare them qualitatively.
The MET results broadly corroborate the fingerprinting findings and further reveal behavioral divergence beyond endpoint identity.

For instance, GPT-4o-mini and GPT-5 in shadow API A are both flagged as statistically distinct from official outputs across AIME 2025 and GPQA, consistent with the Euclidean distance deviations observed via LLMmap.
DeepSeek-Chat is rejected across nearly all provider benchmark combinations, reinforcing the identity inconsistency identified in~\autoref{table:fingerprinting_results}.
The Gemini-2.5-pro results again stand out as an exception:
MET fails to reject the null hypothesis across all three providers on AIME 2025.
This indicates that, under this benchmark and sample size, we do not observe distributional evidence that the shadow outputs differ from the official endpoint.
It should not be interpreted as certifying faithful serving, since MET only tests distributional equality on the evaluated samples.

%-------------------------------------------------------------------------------
\begin{table*}[t!]
\centering
\caption{MET results across shadow API providers.
\textit{Reject} = True indicates the null hypothesis of distributional equality is rejected at $\alpha = 0.05$.}
\label{table:met_results}
\scalebox{0.82}{
\begin{tabular}{@{}llccc@{}}
\toprule
\textbf{Benchmark}
& \textbf{Model}
& \textbf{Shadow API A}
& \textbf{Shadow API E}
& \textbf{Shadow API H} \\
\midrule

\multirow{8}{*}{AIME 2025}
 & GPT-4o-mini       & \textbf{Reject} & Pass & \textbf{Reject} \\
 & GPT-5             & \textbf{Reject} & \textbf{Reject} & Pass \\
 & GPT-5-mini        & Pass & Pass & \textbf{Reject} \\
 & Gemini-2.0-flash  & \textbf{Reject} & Pass & Pass \\
 & Gemini-2.5-flash  & Pass & Pass & Pass \\
 & Gemini-2.5-pro    & Pass & Pass & Pass \\
 & DeepSeek-Chat     & \textbf{Reject} & \textbf{Reject} & \textbf{Reject} \\
 & DeepSeek-Reasoner & \textbf{Reject} & Pass & \textbf{Reject} \\
\midrule

\multirow{8}{*}{GPQA}
 & GPT-4o-mini       & \textbf{Reject} & Pass & \textbf{Reject} \\
 & GPT-5             & \textbf{Reject} & \textbf{Reject} & \textbf{Reject} \\
 & GPT-5-mini        & \textbf{Reject} & Pass & Pass \\
 & Gemini-2.0-flash  & \textbf{Reject} & \textbf{Reject} & Pass \\
 & Gemini-2.5-flash  & \textbf{Reject}& Pass & \textbf{Reject} \\
 & Gemini-2.5-pro    & \textbf{Reject} & Pass & Pass \\
 & DeepSeek-Chat     & \textbf{Reject} & \textbf{Reject} & \textbf{Reject} \\
 & DeepSeek-Reasoner & \textbf{Reject} & Pass & \textbf{Reject} \\
\midrule

\multirow{3}{*}{AdvBench}
 & GPT-5-mini        & \textbf{Reject} & Pass & \textbf{Reject} \\
 & Gemini-2.5-flash  & \textbf{Reject} & \textbf{Reject} & \textbf{Reject} \\
 & DeepSeek-Chat     & \textbf{Reject} & Pass & \textbf{Reject} \\
\midrule

\multirow{3}{*}{JailbreakBench}
 & GPT-5-mini        & \textbf{Reject} & Pass & Pass \\
 & Gemini-2.5-flash  & \textbf{Reject} & \textbf{Reject} & \textbf{Reject} \\
 & DeepSeek-Chat     & \textbf{Reject} & Pass & \textbf{Reject} \\
\bottomrule
\end{tabular}}
\end{table*}
%-------------------------------------------------------------------------------

%-------------------------------------------------------------------------------
\subsection{Validation on Controlled Ground Truth}
\label{section:validation}
%-------------------------------------------------------------------------------
The audits in~\autoref{section:fingerprinting} and~\autoref{section:utility} apply verification methods to production shadow endpoints whose true backends are not disclosed.
To calibrate these methods under known ground truth, we construct a local testbed where honest endpoints serve the claimed model and substituting endpoints silently serve an alternative backend reflecting the substitution patterns observed in~\autoref{table:fingerprinting_results}.
This controlled setting provides an empirical sanity check for the false positive and false negative behavior of our audit pipeline.

We construct six endpoints spanning the three model families studied in this paper, with three honest and three substituting configurations, as summarized in~\autoref{table:validation_testbed}.
Each endpoint serves $N{=}100$ queries for LLMmap evaluation, and an independent set of 600 requests is issued for MET evaluation.
The substituting configurations cover three substitution modes: cross-family model substitution (GPT-5 served by GLM-4-9B), within-family version mismatch (Gemini-2.0-flash served by Gemini-2.5-flash), and capability downgrading (DeepSeek-Reasoner served by the non-thinking DeepSeek-Chat).

%-------------------------------------------------------------------------------
\begin{table}[t!]
\centering
\caption{Controlled validation testbed.}
\label{table:validation_testbed}
\scalebox{0.7}{
\begin{tabular}{@{}llll@{}}
\toprule
\textbf{Type} & \textbf{Claimed Model} & \textbf{Backend} & \textbf{Substitution Mode} \\
\midrule
Honest    & GPT-5             & GPT-5              & None \\
Honest    & Gemini-2.0-flash  & Gemini-2.0-flash   & None \\
Honest    & DeepSeek-Reasoner & DeepSeek-Reasoner  & None \\
Substituting & GPT-5             & GLM-4-9B           & Cross-family \\
Substituting  & Gemini-2.0-flash  & Gemini-2.5-flash   & Version mismatch \\
Substituting  & DeepSeek-Reasoner & DeepSeek-Chat      & Capability downgrade \\
\bottomrule
\end{tabular}
}
\end{table}
%-------------------------------------------------------------------------------

\autoref{table:validation_perfamily} reports decision-level performance on this controlled testbed.
For LLMmap, one decision corresponds to running the fingerprinting procedure on a sampled query set and comparing the predicted backend with the known ground-truth backend.
For MET, one decision corresponds to testing whether responses from the served backend and the claimed official model are drawn from the same distribution.
A false positive denotes an honest endpoint incorrectly flagged as inconsistent, while a false negative denotes a substituting endpoint incorrectly accepted as faithful.

LLMmap achieves $96.00\%$ overall accuracy with a $3.00\%$ false positive rate and a $5.00\%$ false negative rate.
MET achieves $88.33\%$ accuracy with a $4.67\%$ false positive rate and an $18.67\%$ false negative rate.
Performance is consistent across the three model families, with no single family diverging by more than $1.3$\% from the overall accuracy~(\autoref{table:validation_perfamily}).
Both methods correctly flag the three controlled substitution modes, providing a sanity check that the production findings are not solely artifacts of the verification methods.

\begin{table}[t!]
\centering
\caption{Decision-level performance of LLMmap and MET on the controlled validation testbed.}
\label{table:validation_perfamily}
\scalebox{0.7}{
\begin{tabular}{@{}lcccccc@{}}
\toprule
 & \multicolumn{3}{c}{\textbf{LLMmap}} & \multicolumn{3}{c}{\textbf{MET}} \\
\cmidrule(lr){2-4} \cmidrule(lr){5-7}
\textbf{Model Family} & \textbf{Acc} & \textbf{FPR} & \textbf{FNR} & \textbf{Acc} & \textbf{FPR} & \textbf{FNR} \\
\midrule
GPT family      & $96.00\%$ & $3.00\%$ & $5.00\%$ & $88.50\%$ & $5.00\%$ & $18.00\%$ \\
Gemini family   & $96.00\%$ & $2.00\%$ & $6.00\%$ & $87.00\%$ & $5.00\%$ & $21.00\%$ \\
DeepSeek family & $96.00\%$ & $4.00\%$ & $4.00\%$ & $89.50\%$ & $4.00\%$ & $17.00\%$ \\
\midrule
\textbf{Overall} & \textbf{$96.00\%$} & \textbf{$3.00\%$} & \textbf{$5.00\%$} & \textbf{$88.33\%$} & \textbf{$4.67\%$} & \textbf{$18.67\%$} \\
\bottomrule
\end{tabular}
}
\end{table}
%-------------------------------------------------------------------------------

The elevated FNR of MET ($18.7\%$) reflects its operating regime.
MET detects distributional shift between outputs rather than exact model identity, so it may fail to reject when a substituted backend produces outputs statistically close to the claimed model, for instance under small version mismatches within the same family.
Thus, MET and LLMmap capture different failure modes.
These validation results provide an empirical sanity check for the audit pipeline and increase confidence that the fingerprint failures and distributional rejections observed on production shadow APIs reflect genuine inconsistencies rather than method artifacts alone.
%-------------------------------------------------------------------------------
\subsection{Meta Information Analysis}
\label{section:meta}
%-------------------------------------------------------------------------------
Beyond model-level identity, the routing behavior of shadow APIs leaves observable traces in inference-level metadata.
Official APIs typically exhibit consistent latency and token counts when the same question is issued repeatedly, whereas shadow APIs frequently exhibit irregular spikes consistent with dynamic backend switching or upstream retries.
To quantify this, for each of the 24 shadow endpoints on each of the two benchmarks (AIME 2025 and GPQA), we compute the standard deviation (SD) ratio $r = \text{SD}_{\text{shadow}} / \text{SD}_{\text{official}}$ separately for inference latency and total token counts, yielding $n{=}48$ endpoint-benchmark pairs per dimension.

\autoref{table:time_sd} and~\autoref{table:token_sd} report per-endpoint SD values with color-coded ratio bands.
Inference latency exhibits pronounced instability, with $64.58\%$ of pairs exceeding $r = 1.20$ and $35.42\%$ exceeding $r = 2.00$.
Every case with $r > 2.00$ on latency falls within the Gemini or DeepSeek families, while no GPT-series endpoint reaches this threshold, suggesting that upstream routing chains for non-OpenAI models are substantially less stable.
Token-count ratios are more evenly distributed but still reveal $20.83\%$ of pairs with $r > 2.00$, again concentrated in Gemini and DeepSeek.
At the lower extreme, $20.83\%$ of token-count pairs fall below $r = 0.80$,
dominated by GPT-series endpoints on AIME 2025, a pattern consistent with upstream truncation or fixed-length response caps rather than faithful replication of the official endpoint.
Per-query time series illustrating these spikes at the individual question level are provided in~\autoref{appendix:time_token}.

%-------------------------------------------------------------------------------
\begin{table*}[t!]
\centering
\caption{Standard deviation of inference latency time (s). Colors denote ratio vs official API, \colorbox{blue!20}{Blue} ($< 0.8\times$), \colorbox{mygreen}{Green} ($0.8\times to\ 1.2\times$), \colorbox{red!20}{Red} ($> 1.2\times$).}
\label{table:time_sd}
\scalebox{0.77}{
\begin{tabular}{@{}llccccc@{}} 
\toprule
\textbf{Model Family} & \textbf{Model Name} & \textbf{Benchmark} & \textbf{Official API} & \textbf{Shadow API A} & \textbf{Shadow API E} & \textbf{Shadow API H} \\
\midrule

\multirow{6}{*}{GPT} 
& \multirow{2}{*}{GPT-4o-mini} 
  & AIME & 42.71 & \cellcolor{myblue!20}30.42 & \cellcolor{myblue!20}27.96 & \cellcolor{myblue!20}28.12 \\
& & GPQA & 2.19 & \cellcolor{myred!20}3.36 & \cellcolor{myred!20}2.63 & \cellcolor{myred!20}3.00 \\
\cmidrule{2-7}

& \multirow{2}{*}{GPT-5} 
  & AIME & 189.04 & \cellcolor{myblue!20}68.62 & \cellcolor{myred!20}229.20 & \cellcolor{mygreen}193.95 \\
& & GPQA & 32.39 & \cellcolor{myred!20}45.59 & \cellcolor{myred!20}40.90 & \cellcolor{mygreen}36.68 \\
\cmidrule{2-7}

& \multirow{2}{*}{GPT-5-mini} 
  & AIME & 46.75 & \cellcolor{myred!20}89.06 & \cellcolor{myblue!20}29.56 & \cellcolor{mygreen}38.58 \\
& & GPQA & 12.18 & \cellcolor{myred!20}19.11 & \cellcolor{myred!20}15.39 & \cellcolor{myred!20}15.76 \\
\midrule

\multirow{6}{*}{Gemini} 
& \multirow{2}{*}{Gemini-2.0-flash} 
  & AIME & 6.06 & \cellcolor{mygreen}5.46 & \cellcolor{mygreen}6.35 & \cellcolor{mygreen}5.86 \\
& & GPQA & 1.76 & \cellcolor{myred!20}5.56 & \cellcolor{myred!20}3.69 & \cellcolor{myred!20}6.76 \\
\cmidrule{2-7}

& \multirow{2}{*}{Gemini-2.5-flash} 
  & AIME & 29.41 & \cellcolor{myred!20}86.24 & \cellcolor{myblue!20}22.70 & \cellcolor{myred!20}51.03 \\
& & GPQA & 21.30 & \cellcolor{myred!20}57.95 & \cellcolor{mygreen}23.86 & \cellcolor{myred!20}53.29 \\
\cmidrule{2-7}

& \multirow{2}{*}{Gemini-2.5-pro} 
  & AIME & 48.88 & \cellcolor{myred!20}122.61 & \cellcolor{myred!20}194.45 & \cellcolor{myred!20}231.95 \\
& & GPQA & 19.11 & \cellcolor{myred!20}36.02 & \cellcolor{mygreen}22.83 & \cellcolor{myred!20}49.88 \\
\midrule

\multirow{4}{*}{DeepSeek} 
& \multirow{2}{*}{DeepSeek-Chat} 
  & AIME & 9.18 & \cellcolor{myred!20}104.90 & \cellcolor{myred!20}12.75 & \cellcolor{myred!20}29.22 \\
& & GPQA & 10.19 & \cellcolor{myred!20}31.65 & \cellcolor{mygreen}12.00 & \cellcolor{mygreen}11.60 \\
\cmidrule{2-7}

& \multirow{2}{*}{DeepSeek-Reasoner} 
  & AIME & 100.56 & \cellcolor{myred!20}303.38 & \cellcolor{mygreen}115.82 & \cellcolor{myred!20}218.50 \\
& & GPQA & 77.73 & \cellcolor{myred!20}165.78 & \cellcolor{myred!20}113.34 & \cellcolor{myred!20}239.67 \\

\bottomrule
\end{tabular}
}
\end{table*}
%-------------------------------------------------------------------------------

%-------------------------------------------------------------------------------
\begin{table*}[t!]
\centering
\caption{Standard deviation of token counts. Colors denote ratio vs official API, \colorbox{blue!20}{Blue} ($< 0.8\times$), \colorbox{mygreen}{Green} ($0.8\times$ to $1.2\times$), \colorbox{red!20}{Red} ($> 1.2\times$).}
\label{table:token_sd}
\scalebox{0.77}{
\begin{tabular}{@{}llccccc@{}} 
\toprule
\textbf{Model Family} & \textbf{Model Name} & \textbf{Benchmark} & \textbf{Official API} & \textbf{Shadow API A} & \textbf{Shadow API E} & \textbf{Shadow API H} \\
\midrule

\multirow{6}{*}{GPT} 
& \multirow{2}{*}{GPT-4o-mini} 
  & AIME & 2273.87 & \cellcolor{myblue!20}472.75 & \cellcolor{myblue!20}1598.74 & \cellcolor{myblue!20}1585.51 \\
& & GPQA & 66.13 & \cellcolor{myred!20}83.53 & \cellcolor{mygreen}57.68 & \cellcolor{myred!20}132.32 \\
\cmidrule{2-7}

& \multirow{2}{*}{GPT-5} 
  & AIME & 4065.04 & \cellcolor{myblue!20}1311.57 & \cellcolor{mygreen}3919.61 & \cellcolor{myblue!20}2152.53 \\
& & GPQA & 1598.09 & \cellcolor{mygreen}1457.36 & \cellcolor{mygreen}1396.30 & \cellcolor{mygreen}1594.97 \\
\cmidrule{2-7}

& \multirow{2}{*}{GPT-5-mini} 
  & AIME & 3458.01 & \cellcolor{myblue!20}2026.09 & \cellcolor{myblue!20}1792.22 & \cellcolor{myblue!20}2319.84 \\
& & GPQA & 892.41 & \cellcolor{mygreen}1064.52 & \cellcolor{mygreen}883.13 & \cellcolor{mygreen}838.42 \\
\midrule

\multirow{6}{*}{Gemini} 
& \multirow{2}{*}{Gemini-2.0-flash} 
  & AIME & 1243.80 & \cellcolor{mygreen}1242.84 & \cellcolor{mygreen}1311.17 & \cellcolor{mygreen}1231.42 \\
& & GPQA & 384.25 & \cellcolor{myred!20}1090.25 & \cellcolor{myred!20}749.74 & \cellcolor{myred!20}969.43 \\
\cmidrule{2-7}

& \multirow{2}{*}{Gemini-2.5-flash} 
  & AIME & 8174.81 & \cellcolor{myred!20}16489.45 & \cellcolor{myblue!20}6124.36 & \cellcolor{myred!20}14164.34 \\
& & GPQA & 5856.16 & \cellcolor{myred!20}7029.59 & \cellcolor{mygreen}6595.46 & \cellcolor{myred!20}14464.05 \\
\cmidrule{2-7}

& \multirow{2}{*}{Gemini-2.5-pro} 
  & AIME & 3054.32 & \cellcolor{myblue!20}1161.70 & \cellcolor{myred!20}3768.74 & \cellcolor{mygreen}2905.47 \\
& & GPQA & 2637.50 & \cellcolor{myred!20}7430.27 & \cellcolor{myred!20}3556.55 & \cellcolor{myred!20}6878.14 \\
\midrule

\multirow{4}{*}{DeepSeek} 
& \multirow{2}{*}{DeepSeek-Chat} 
  & AIME & 307.92 & \cellcolor{myred!20}492.85 & \cellcolor{myred!20}383.21 & \cellcolor{myred!20}1107.18 \\
& & GPQA & 305.67 & \cellcolor{mygreen}338.70 & \cellcolor{mygreen}360.34 & \cellcolor{mygreen}354.99 \\
\cmidrule{2-7}

& \multirow{2}{*}{DeepSeek-Reasoner} 
  & AIME & 3036.03 & \cellcolor{myred!20}9647.40 & \cellcolor{mygreen}3544.03 & \cellcolor{mygreen}3404.37 \\
& & GPQA & 2314.59 & \cellcolor{myred!20}6153.61 & \cellcolor{myred!20}3002.75 & \cellcolor{mygreen}2409.62 \\

\bottomrule
\end{tabular}
}
\end{table*}
%-------------------------------------------------------------------------------

\begin{tcolorbox}[colback=gray!25!white, size=title,breakable,boxsep=1mm,colframe=white,before={\vskip1mm}, after={\vskip0mm}]
\textbf{RQ3 Take-Aways:}
        Fingerprinting yields specific alternative-model matches for a subset of audited endpoints, providing evidence consistent with model substitution.
        More broadly, $45.83\%$ of endpoints fail fingerprint verification, MET rejects distributional equality for the majority of provider-benchmark combinations, and over one-third of endpoint-benchmark pairs exhibit latency volatility exceeding $2.00\times$.
        These broader signals establish endpoint inconsistency but do not by themselves identify the backend.
        When an endpoint is commercially offered under one model identity while its fingerprint matches another, the evidence establishes a deceptive model claim.
\end{tcolorbox}

%-------------------------------------------------------------------------------
\section{Discussion}
%-------------------------------------------------------------------------------
Building on the empirical results from the preceding sections, we synthesize findings across evaluation dimensions~(\autoref{section:joint}), quantify the economic incentives underlying deceptive model claims\allowbreak (\autoref{section:economic}), and report observed ecosystem changes during and after our study window\allowbreak (\autoref{section:post_disclosure}).
%-------------------------------------------------------------------------------
\subsection{Joint Analysis}
\label{section:joint}
%-------------------------------------------------------------------------------
Based on the combined results from performance evaluation~(\autoref{section:utility}) and model verification~(\autoref{section:fingerprinting}), we find that model identity and observed performance do not always align.

\mypara{When Identity Predicts Performance}
Most endpoints follow an intuitive pattern in which matching identity coincides with consistent behavior and identity mismatch coincides with degradation.
GPT-5-mini in shadow API E illustrates the former: fingerprinting matches the claimed model, and sensitive-domain accuracy remains close to the official endpoint.
DeepSeek-Reasoner in shadow API A and GPT-4o-mini in shadow API H illustrate the latter, where fingerprint mismatch is accompanied by sharp reasoning collapse on AIME 2025.

\mypara{When Identity and Performance Diverge}
Two counterintuitive patterns violate this alignment.
In one direction, identity mismatch does not always manifest as immediate performance degradation.
GPT-5 in shadow APIs A and E fingerprints as glm-4-9b-chat, yet its AIME~2025 accuracy declines by $2.22$ to $7.78$ percentage points.
This gap is substantially smaller than the capability differential between a frontier proprietary model and a 9B open-source backend would predict.
A substituted backend may nevertheless track the claimed model on standard reference tasks~(\autoref{section:fingerprinting}), so degradation surfaces most clearly under reasoning-heavy or strict-correctness conditions.
In the other direction, matching identity does not guarantee faithful behavior.
Gemini-2.5-flash fingerprints correctly across shadow APIs A, E, and H with Euclidean distances close to the official APIs, yet accuracy in sensitive domains drops by roughly $47\%$ uniformly across all three providers.
A plausible conjecture is that these providers serve the authentic model with its thinking budget disabled or sharply reduced to cut serving cost, which would leave fingerprints intact while selectively collapsing performance on reasoning-heavy queries.

To quantify these relationships, we fit an OLS regression predicting accuracy drop from Euclidean distance, price ratio, identity mismatch, and a reasoning-model flag across all 24 endpoints.
The full specification is reported in~\autoref{table:regression}.

%-------------------------------------------------------------------------------
\begin{table}[ht!]
\centering
\caption{OLS regression predicting accuracy drop ($\Delta\text{acc}$).}
\label{table:regression}
\scalebox{0.9}{
\begin{tabular}{@{}lrrrr@{}}
\toprule
\textbf{Variable} & \textbf{$\beta$} & \textbf{$SE$} & \textbf{$t$} & \textbf{$p$} \\
\midrule
Euclidean Distance& $-0.39$ & $1.41$ & $-0.27$ & $0.79$ \\
Price Ratio       & $-1.45$ & $2.06$ & $-0.70$ & $0.49$ \\
Identity Mismatch & $0.22$  & $7.25$ & $0.03$  & $0.98$ \\
Reasoning Model   & $10.09$ & $7.54$ & $1.34$  & $0.20$ \\
\midrule
\multicolumn{5}{l}{$R^2 = 0.15$, $n = 24$} \\
\bottomrule
\end{tabular}
}
\end{table}
%-------------------------------------------------------------------------------

\mypara{Price Does Not Signal Quality}
As shown in~\autoref{table:regression}, price ratio has no predictive power for accuracy drop ($\beta = -1.447$, $p = 0.490$), and this null result persists when the regression is restricted to the subset of endpoints passing fingerprint verification.
This directly refutes the intuition that more expensive shadow APIs deliver more faithful service.
Users cannot protect themselves from substitution-induced quality loss by selecting higher-priced providers, and price cannot serve as even a weak proxy for backend fidelity.

\mypara{Fingerprint Fidelity Does Not Imply Behavioral Fidelity}
The low overall fit ($R^2 = 0.15$) is driven by the Gemini-2.5-flash anomaly documented above.
Excluding these three endpoints, the fit improves to $R^2 = 0.30$ and identity mismatch becomes marginally significant ($\beta = 7.72$, $p = 0.05$), indicating that identity substitution is a meaningful driver of accuracy drop once this anomaly is set aside.
The anomaly itself, however, carries the stronger message: identity-consistent serving does not guarantee behavioral fidelity.
A provider can route requests to the claimed model and still degrade output quality through misconfigured inference parameters, context-window truncation, or silent prompt preprocessing, mechanisms that Euclidean-distance-based fingerprinting is not designed to detect.
Verification pipelines relying solely on identity checks, therefore, cannot certify that a shadow API faithfully reproduces the official endpoint's behavior, and behavioral benchmarking remains a necessary complement to identity-level auditing.

\mypara{Consolidated Endpoints Evidence}
To provide a unified view,~\autoref{table:verdict} consolidates the four independent evidence dimensions evaluated in this work for each of the 24 shadow endpoints: fingerprint identity (FP), distributional equality testing (MET), performance inconsistency (Perf), and metadata instability (Meta, latency SD ratio $r{>}2.00$ on either AIME or GPQA).
Each endpoint is assigned an aggregate risk level: \textbf{High} (${\geq}3$ flags), \textbf{Medium} ($1$--$2$ flags), or \textbf{Low} ($0$ flags).
Of the 24 endpoints, $11$ ($45.83\%$) are classified as High risk, $9$ ($37.50\%$) as Medium, and only $4$ ($16.67\%$) as Low.
Shadow API A contains no Low-risk endpoint, while three of the four Low-risk endpoints belong to shadow API E, confirming that the distribution of flags differs across the three audited providers.

%-------------------------------------------------------------------------------
\begin{table*}[t!]
\centering
\caption{Consolidated per-endpoint evidence summary across four independent dimensions.
}
\label{table:verdict}
\scalebox{0.8}{
\begin{tabular}{@{}ll ccccc ccccc ccccc@{}}
\toprule
& & \multicolumn{5}{c}{\textbf{Shadow API A}} & \multicolumn{5}{c}{\textbf{Shadow API E}} & \multicolumn{5}{c}{\textbf{Shadow API H}} \\
\cmidrule(lr){3-7} \cmidrule(lr){8-12} \cmidrule(lr){13-17}
\textbf{Family} & \textbf{Model}
  & \textbf{FP} & \textbf{MET} & \textbf{Perf} & \textbf{Meta} & \textbf{Risk}
  & \textbf{FP} & \textbf{MET} & \textbf{Perf} & \textbf{Meta} & \textbf{Risk}
  & \textbf{FP} & \textbf{MET} & \textbf{Perf} & \textbf{Meta} & \textbf{Risk} \\
\midrule
\multirow{3}{*}{GPT}
& GPT-4o-mini
  & \cmark & \cmark & ---    & ---    & \cellcolor{myyellow}Med
  & ---    & ---    & ---    & ---    & \cellcolor{mygreen}Low
  & \cmark & \cmark & \cmark & ---    & \cellcolor{myred!20}High \\
& GPT-5
  & \cmark & \cmark & \cmark & ---    & \cellcolor{myred!20}High
  & \cmark & \cmark & \cmark & ---    & \cellcolor{myred!20}High
  & ---    & ---    & ---    & ---    & \cellcolor{mygreen}Low \\
& GPT-5-mini
  & \cmark & \cmark & ---    & ---    & \cellcolor{myyellow}Med
  & ---    & ---    & ---    & ---    & \cellcolor{mygreen}Low
  & \cmark & \cmark & ---    & ---    & \cellcolor{myyellow}Med \\
\midrule
\multirow{3}{*}{Gemini}
& Gemini-2.0-flash
  & \cmark & \cmark & \cmark & \cmark & \cellcolor{myred!20}High
  & \cmark & --- & \cmark & \cmark & \cellcolor{myred!20}High
  & ---    & ---    & ---    & \cmark & \cellcolor{myyellow}Med \\
& Gemini-2.5-flash
  & \cmark & \cmark & \cmark & \cmark & \cellcolor{myred!20}High
  & ---    & \cmark & \cmark & ---    & \cellcolor{myyellow}Med
  & ---    & \cmark & \cmark & \cmark & \cellcolor{myred!20}High \\
& Gemini-2.5-pro
  & ---    & ---    & \cmark & \cmark & \cellcolor{myyellow}Med
  & ---    & ---    & ---    & \cmark & \cellcolor{myyellow}Med
  & ---    & ---    & ---    & \cmark & \cellcolor{myyellow}Med \\
\midrule
\multirow{2}{*}{DeepSeek}
& DeepSeek-Chat
  & \cmark & \cmark & \cmark & \cmark & \cellcolor{myred!20}High
  & \cmark & \cmark & ---    & ---    & \cellcolor{myyellow}Med
  & \cmark & \cmark & ---    & \cmark & \cellcolor{myred!20}High \\
& DeepSeek-Reasoner
  & \cmark & \cmark & \cmark & \cmark & \cellcolor{myred!20}High
  & ---    & ---    & ---    & ---    & \cellcolor{mygreen}Low
  & \cmark & \cmark & \cmark & \cmark & \cellcolor{myred!20}High \\
\midrule
\multicolumn{2}{l}{\textbf{Per-provider summary}}
  & \multicolumn{5}{c}{5 High, 3 Med, 0 Low}
  & \multicolumn{5}{c}{2 High, 3 Med, 3 Low}
  & \multicolumn{5}{c}{4 High, 3 Med, 1 Low} \\
\bottomrule
\end{tabular}
}
\end{table*}
%-------------------------------------------------------------------------------

\begin{tcolorbox}[colback=gray!25!white, size=title,breakable,boxsep=1mm,colframe=white,before={\vskip1mm}, after={\vskip0mm}]
\textbf{Joint Analysis Take-Aways:}
Price ratio has no predictive power for accuracy degradation, so pricing cannot serve as a quality signal for backend fidelity.
Identity-consistent serving does not guarantee behavioral fidelity: a shadow API can pass fingerprint verification and still degrade output quality through non-identity mechanisms.
\end{tcolorbox}

%-------------------------------------------------------------------------------
\subsection{Economic Incentives and User Losses}
\label{section:economic}
%-------------------------------------------------------------------------------

We observe three recurring pricing patterns in the relationship between advertised model identities and fingerprinted backends, summarized in~\autoref{table:economic_mechanisms}.
Each instantiates the transaction structure of a deceptive model claim, a substituted backend sold and priced under the claimed model name.
In the information premium scheme, providers charge a premium rate while the endpoint fingerprints as a cheaper model of similar or newer vintage (e.g., API A advertises Gemini-2.0-flash but fingerprints as Gemini-2.5-flash at a $7.1$--$7.25\times$ price ratio).
In the discount-substitution scheme, providers charge at the official rate while the endpoint fingerprints as a low-cost open-source backend (e.g., API A advertises GPT-5 at parity pricing but fingerprints as GLM-4-9B).
In the resale markup scheme, providers apply a modest surcharge while the endpoint fingerprints as a different backend, as illustrated by API H, which charges $1.09\times$ the official rate for GPT-4o-mini while its responses fingerprint as Qwen2.5-7B.

%-------------------------------------------------------------------------------
\begin{table*}[t!]
\centering
\caption{Three observed pricing patterns associated with deceptive model claims.}
\label{table:economic_mechanisms}
\scalebox{0.9}{
\begin{tabular}{@{}llll@{}}
\toprule
\textbf{Mechanism} & \textbf{Example} & \textbf{Price Ratio} & \textbf{Deception Evidence} \\
\midrule
Information Premium    & API A --- Gemini-2.0-flash & 7.10 $\times$ / 7.25 $\times$ & Fingerprinted as Gemini-2.5-flash \\
Discount-Substitution  & API A --- GPT-5            & 1.00 $\times$               & Fingerprinted as GLM-4-9B (cheap-for-premium swap) \\
Resale Markup          & API H --- GPT-4o-mini & 1.09 $\times$               & Fingerprinted as Qwen2.5-7B \\
\bottomrule
\end{tabular}
}
\end{table*}
%-------------------------------------------------------------------------------

\mypara{Value Delivered vs.\ Cost Incurred}

As a descriptive supplement, we report an aggregate cost snapshot recorded during an earlier exploratory measurement of GPT-5 on GPQA, using official GPT-5 pricing (\$1.25/\$10.00 per 1M prompt/completion tokens).
\autoref{table:value_delivered} reports the aggregate estimates retained from this measurement as descriptive context.
We do not rely on them for our primary endpoint-fidelity conclusions.

%-------------------------------------------------------------------------------
\begin{table}[t!]
\centering
\caption{Historical aggregate cost snapshot for GPT-5 on GPQA under official-token pricing.}
\label{table:value_delivered}
\scalebox{0.88}{
\begin{tabular}{@{}lccc@{}}
\toprule
\textbf{API} & \textbf{Equiv. Value} & \textbf{Relative to Off.} & \textbf{Value Shortfall} \\
\midrule
Official      & \$14.84 & $1.00\times$ & -- \\
Shadow API A  & \$5.70  & $0.38\times$ & \$9.14 \\
Shadow API E  & \$5.35  & $0.36\times$ & \$9.49 \\
Shadow API H  & \$7.77  & $0.52\times$ & \$7.07 \\
\bottomrule
\end{tabular}
}
\end{table}
%-------------------------------------------------------------------------------

\mypara{Aggregate Research Costs}
Beyond per-query losses, shadow API deception imposes broader costs on the research community.
Our dataset covers 187 papers whose LLM-based pipelines rely on shadow API providers.
As an illustrative scenario rather than an empirical estimate, if 30\% of these papers were to require re-execution upon detecting identity inconsistencies ($n \approx 56$), the aggregate direct cost, comprising API re-runs (\$50--\$500 per paper) and researcher time (approximately $40\,\text{h} \times \$50/\text{h} = \$2{,}000$ per paper), would range from \$115,000 to \$140,000.
This estimate excludes downstream reproducibility costs associated with the more than 5{,}900 cumulative citations to these papers, where silent model substitution may propagate into dependent experimental results without any visible error signal.

Because providers observe their routing choices while users observe only a model name and price, this market exhibits the information asymmetry associated with a market-for-lemons dynamic~\cite{A70,A01}.

%-------------------------------------------------------------------------------
\subsection{\texorpdfstring{Observed Ecosystem Changes}{Observed Ecosystem Changes}}
\label{section:post_disclosure}
%-------------------------------------------------------------------------------
During and after our study window, we observe two categories of ecosystem changes.
As stated in~\autoref{appendix:disclosure}, we did not notify shadow API providers, and we do not attribute either category to our disclosure.

\mypara{Provider Cessation}
By the time of writing, four of the 17 identified shadow API providers ceased operations entirely.
They strand users who had ongoing research pipelines dependent on these endpoints, further illustrating the operational volatility documented in~\autoref{section:academia}.

\mypara{Compliance Adoption and Testing Platforms}
Beyond the providers in our collection, third-party LLM API services in the broader market are changing by introducing compliance-related claims on their platforms.
For instance, several providers published ``model verification passed'' badges on their pricing pages.
Concurrently, some testing platforms dedicated to detecting model substitution in third-party LLM API services have emerged (e.g., \url{https://hvoy.ai/} and \url{https://cctest.ai/}).
These platforms offer automated checks that attempt to verify whether an LLM API endpoint serves the claimed model.
Their emergence indicates a growing awareness of the risks associated with LLMs provided by third-party LLM API providers.

%-------------------------------------------------------------------------------
\section{\texorpdfstring{Suggestions}{Suggestions}}
\label{section:suggestion}
%-------------------------------------------------------------------------------
Our findings reveal that model substitution is widespread among the audited endpoints and technically detectable.
To safeguard the integrity of LLM-based research, we propose a concrete two-workflow framework for auditors and researchers respectively.
The fundamental mitigation remains direct official access, and we recommend it wherever available.
Since prohibition alone is not actionable under regional restrictions and prohibitive pricing~(\autoref{section:threat_model}), a shadow endpoint should instead be treated as an untrusted instrument: verified before use, monitored during use, and disclosed at publication.

\mypara{Auditor Verification Protocol}
We recommend a four-stage verification pipeline in which any flagged stage should escalate immediately to avoidance of the endpoint.
First, the endpoint should be queried with at least 24 LLMmap probes, flag it if the Euclidean distance exceeds $1.25\times$ the official baseline, or if the top-1 identified model does not match the claimed identity.
Second, MET~\cite{GLG25} should be applied using a permutation-based $p$-value with a prespecified significance level (e.g., $\alpha = 0.05$), and the endpoint should be flagged if the null hypothesis of distributional equality is rejected.
Third, at least three independent sessions should be run on a held-out benchmark, flag the endpoint if the accuracy standard deviation exceeds $5$ \% or if the latency coefficient of variation exceeds $0.15$, as excessive variance is indicative of unstable backend routing or dynamic model substitution.
Fourth, ICP registration and legal entity disclosure should be verified against the criteria.
An endpoint operated by an unregistered individual without verifiable provenance carries elevated legal and operational risk.

\mypara{Researcher Pre-Registration Checklist}
For any study relying on LLM API queries, researchers should complete and publicly report the following prior to data collection.
The full endpoint URL, claimed model version, date of access, and pricing tier should be recorded and included as a footnote or within the experimental setup section, consistent with emerging reproducibility norms.
Before any experimental queries are issued, the endpoint should be confirmed to pass all four stages of the auditor protocol above. 
Results from endpoints that fail any stage should not be reported as representative of the claimed model.
Because providers switch upstream backends dynamically~(\autoref{section:academia}), a one-time check is not sufficient: lightweight probes (e.g., a fixed subset of LLMmap queries) should be interleaved daily or every few thousand requests.
Raw responses and per-request metadata (endpoint URL, model identifier, latency, token counts, and logprobs) should be archived and released as artifacts, so provenance can be re-examined post hoc.
Finally, per-run accuracy over at least three independent runs, the LLMmap fingerprint Euclidean distance, and the MET $p$-value should be reported alongside experimental results, enabling readers and reviewers to assess backend reliability and contextualize reported performance accordingly.

\mypara{Community-Level Actions}
Beyond individual practice, we call on the broader research community to adopt structural safeguards.
Conference organizers and program chairs should update reviewer guidelines to flag undisclosed or unverified third-party API endpoints as a reproducibility risk, treating such usage analogously to unverified dataset provenance.
Official model providers can further reduce shadow market demand through channels compatible with export-control constraints: academic pricing and sponsored research access, cross-region collaborations in which co-authors in supported regions run the official queries, and lightweight verification endpoints for independent identity checks.
Where frontier capability is not strictly required, openly released models with self-hosted weights or TEE-attested serving~\cite{CSZS25} eliminate the provenance question, and the emerging substitution-testing platforms~(\autoref{section:post_disclosure}) could grow into community infrastructure that continuously publishes endpoint-fidelity reports.

%-------------------------------------------------------------------------------
\section{Limitations}
\label{section:limitations}
%-------------------------------------------------------------------------------
Despite providing a systematic audit of the shadow APIs, our study is subject to several limitations.

\mypara{Temporal Scope and Market Volatility}
Our study reflects a deliberately bounded snapshot within a specific time window (September to December 2025).
Shadow API providers frequently switch upstream model sources, alter routing strategies, or cease operations without notice, so behavioral characteristics may drift beyond our observation window.
However, the core risk signals we document, non-transparent model substitution, capability downgrading, and safety behavior divergence, reflect structural properties of the service category rather than artifacts of any single measurement period.
We intentionally adopt a snapshot design rather than continuous monitoring to ensure experimental consistency and to avoid inadvertently amplifying the shadow API ecosystem through sustained engagement.

\mypara{Detection Coverage and Ground Truth}
In the absence of ground truth regarding backend infrastructure, our investigation relies on established auditing frameworks (LLMmap and MET) alongside performance and metadata analysis.
While we detect model identity mismatches with high confidence, we cannot definitively distinguish all fine-grained implementation details (e.g., silent prompt preprocessing or context-window truncation).
Nor can black-box verification establish the nature of a substituted backend, i.e., whether it is distilled, fine-tuned, quantized, or version-shifted~\cite{CSZS25, GLG25}.
Building on model-extraction results for BERT-based NLP APIs~\cite{KTPPI20}, a sufficiently faithful surrogate could in principle evade both LLMmap and MET, since output-based detection is weakest against substitutions that closely track the reference distribution~\cite{CSZS25}.
Passing both tests therefore indicates absence of evidence of substitution rather than proof of authenticity.
Moreover, official APIs of proprietary models may undergo undisclosed fine-tuning.
We mitigate this by reporting variance across three independent runs per experiment.
We also recognize the potential to analyze shadow API differences from additional perspectives, such as bias and hallucination~\cite{JLSLBZ24, Z25}.

\mypara{Coverage of Providers and Model Families}
Our 17 providers, collected via academic citations and GitHub stars, may not represent the full shadow API market.
Additional services may employ distinct deceptive strategies.
Similarly, we focus on three representative model families. 
We believe our audit pipeline generalizes to broader model coverage and leave this to future work.

%-------------------------------------------------------------------------------
\section{Conclusion}
\label{section:conclusion}
%-------------------------------------------------------------------------------
In this work, we present the first systematic audit of the shadow API, a widely used but unverified access for frontier LLMs. 
Through a comprehensive evaluation across providers and benchmarks, together with behavioral fingerprinting, we demonstrate that shadow APIs exhibit significant performance divergence and identity inconsistency. Relying on shadow APIs for evaluation is unreliable, as they may fail to reproduce the behaviors of the official endpoints.
Taken together, these findings reveal deceptive model claims in a subset of audited shadow endpoints and demonstrate that the audited shadow endpoints cannot be treated as reliable official models.

%-------------------------------------------------------------------------------
\bibliographystyle{plain}
\bibliography{normal_generated_py3}
%-------------------------------------------------------------------------------

\appendix

%-------------------------------------------------------------------------------
\section{Open Science}
%-------------------------------------------------------------------------------
The experiments in this paper are based on publicly available artifacts.
For utility and safety evaluation, we use established benchmarks including AIME 2025, GPQA, MedQA, LegalBench, JailbreakBench, and AdvBench~\cite{AIME_Problems,RHSPPDMB23,JPOWFS20,GNHRCNCPWRZTHSFSDPHWNNNCTHMLRHKHRGGWGZIL23,CDRACSDFPTHW24,ZWCNKF23}, as described in~\autoref{section:utility}.
For model verification, we use two publicly available verification frameworks, LLMmap~\cite{PKA25} and MET~\cite{GLG25}.
Non-anonymized provider evidence will be released under gated access within ethical and legal constraints.
The evaluation code is available at \url{https://github.com/TrustAIRLab/ShadowAPI}.

%-------------------------------------------------------------------------------
\section{Ethical Considerations}
\label{appendix:ethical_considerations}
%-------------------------------------------------------------------------------

In this work, we adhere to responsible AI research guidelines.
Our study involves purchasing and querying unofficial shadow API services, which may violate the terms of service of official providers (e.g., OpenAI, Google, DeepSeek). 
These interactions are conducted solely for the purpose of auditing and transparency. 
We do not endorse, promote, or encourage the use of these unauthorized services. 
To minimize potential negative impacts on official infrastructure, we strictly limit our query volume to the minimum necessary for statistical validation of model identity and performance.
Furthermore, all requests to official APIs were conducted exclusively from authorized geographic regions to ensure full compliance with the providers' regional access policies.

Our goal is to expose systemic risks within the shadow API market rather than to target specific individuals.
Furthermore, regarding our prevalence analysis, we have strictly de-identified all specific academic papers, authors, and institutions found to be using shadow APIs. 
Our objective is to highlight systemic reproducibility risks within the community.
Retrospectively identifying published papers as shadow API users carries reputational risk, so we report the 187-paper corpus only in aggregate, withholding titles, authors, repositories, and exact citation counts, and we draw no inference of misconduct or intent: when most of these papers were written, these risks were not publicly known and no norm required endpoint verification.
Such findings should be read as evidence of a missing community safeguard rather than individual fault, and future audits of published research should adopt the same aggregate-only standard.

Our safety evaluation involved the use of jailbreak attacks to test model guardrails. 
To prevent misuse, we follow best practices in safety research: we do not release specific examples of successfully generated harmful content or the exact adversarial prompts used to bypass safety filters. 
We report only aggregated metrics, harmfulness score, and store all sensitive outputs on secure, access-controlled local servers.

To mitigate potential harm and foster a healthier market, we adhere to responsible disclosure principles. 
We are committed to responsibly reporting our findings to official model providers as well as the authors of papers that use shadow APIs. 
We have received partial acknowledgment from them.

%-------------------------------------------------------------------------------
\section{Responsible Disclosure Details}
\label{appendix:disclosure}
%-------------------------------------------------------------------------------

\mypara{Official Providers}
We reported our findings, including the observed substitution patterns and affected model identifiers, to the official providers of the three audited model families, and have received partial acknowledgment.

\mypara{Paper Authors}
We have notified authors of affected papers, and outreach continues on a rolling basis.
Each notice explains that the paper's public code appears to call a shadow API endpoint, describes the associated reproducibility and provenance risks, and suggests provenance checks or re-running the affected experiments on official endpoints.
The notices are framed as provenance alerts, not allegations of misconduct.
Some notified authors have acknowledged receipt, and we are awaiting further updates.

\mypara{Shadow Providers and Attribution}
We did not contact the shadow API providers during the audit or subsequent follow-up observation period because notification could have altered the endpoint behavior or availability under observation.
We obtained no evidence of a causal link, and we therefore report the four observed cessations as market developments without attributing them to our disclosure~(\autoref{section:post_disclosure}).

\section{Generative AI Usage}

This paper was edited for grammar using ChatGPT and Claude.

%-------------------------------------------------------------------------------
\section{Configurations Details}
\label{appendix:configurations_detail}
%-------------------------------------------------------------------------------

%-------------------------------------------------------------------------------
\subsection{Model Configurations}
\label{appendix:model_configurations}
%-------------------------------------------------------------------------------

\autoref{table:api-capability-comparison} details the specific configurations for all evaluated models across official and shadow APIs.
To ensure reproducibility, we set the random seed at 42 and the temperature at 0 for all models compatible with these settings, in accordance with reasoning task guidelines~\cite{DeepSeekTemperature}, while omitting these parameters for unsupported endpoints.
For the GPT-5 model family and other reasoning models, we specifically monitor their support for reasoning parameters.
For all reasoning models, we employ the default reasoning effort in our subsequent analysis, according to the default configuration~\cite{OpenAI_Reasoning_Docs}.
The table also highlights the discrepancies in which shadow APIs may enforce fixed behaviors that deviate from official baselines.
The Ratio column denotes the multiplicative factor of the shadow API service price relative to the official provider for input/output tokens.
Logprobs indicates whether the API can return log probability outputs. 
Pricing is normalized to USD per 1 million tokens for comparison and accessed on December 6, 2025.

%-------------------------------------------------------------------------------
\begin{table*}[t!]
\centering
\caption{Comparison of model configurations across different APIs.}
\scalebox{0.8}{
\begin{tabular}{@{}llccccccc@{}}
\toprule
\textbf{Model Family} & \textbf{Model Name} & \textbf{Channel} & \textbf{API Model ID} & \textbf{Temp.} & \textbf{Seed} & \textbf{Logprobs} & \textbf{Price (In/Out)} & \textbf{Ratio} \\
\midrule

\multirow{4}{*}{GPT} & \multirow{4}{*}{GPT-4o-mini} 
& Official & GPT-4o-mini-2024-07-18 & 0 & 42 & \CIRCLE & \$0.15 / \$0.60 & 1.00 / 1.00 \\
& & Shadow API A & GPT-4o-mini & 0 & 42 & \CIRCLE & \$0.15 / \$0.60 & 1.00 / 1.00 \\
& & Shadow API E & GPT-4o-mini-2024-07-18 & 0 & 42 & \CIRCLE & \$0.11 / \$0.43 & 0.73 / 0.72 \\
& & Shadow API H & GPT-4o-mini-2024-07-18 & 0 & 42 & \CIRCLE & \$0.16 / \$0.65 & 1.09 / 1.09 \\
\cmidrule{2-9}

& \multirow{4}{*}{GPT-5} 
& Official & gpt-5-2025-08-07 & N/A & 42 & \Circle & \$1.25 / \$10.00 & 1.00 / 1.00 \\
& & Shadow API A & gpt-5 & 0 & 42 & \LEFTcircle & \$1.25 / \$10.00 & 1.00 / 1.00 \\
& & Shadow API E & gpt-5-2025-08-07 & 0 & 42 & \Circle & \$0.89 / \$7.14 & 0.71 / 0.71 \\
& & Shadow API H & gpt-5-2025-08-07 & 0 & 42 & \Circle & \$1.36 / \$10.90 & 1.09 / 1.09 \\
\cmidrule{2-9}

& \multirow{4}{*}{GPT-5-mini} 
& Official & gpt-5-mini-2025-08-07 & N/A & 42 & \Circle & \$0.25 / \$2.00 & 1.00 / 1.00 \\
& & Shadow API A & gpt-5-mini & 0 & 42 & \LEFTcircle & \$0.25 / \$2.00 & 1.00 / 1.00 \\
& & Shadow API E & gpt-5-mini & 0 & 42 & \Circle & \$0.18 / \$1.43 & 0.72 / 0.72 \\
& & Shadow API H & gpt-5-mini & 0 & 42 & \Circle & \$0.27 / \$2.18 & 1.09 / 1.09 \\
\midrule

\multirow{12}{*}{Gemini} & \multirow{4}{*}{Gemini-2.0-flash} 
& Official & gemini-2.0-flash & 0 & N/A & \CIRCLE & \$0.10 / \$0.40 & 1.00 / 1.00 \\
& & Shadow API A & gemini-2.0-flash & 0 & 42 & \LEFTcircle & \$0.71 / \$2.90 & 7.10 / 7.25 \\
& & Shadow API E & gemini-2.0-flash & 0 & 42 & \LEFTcircle & \$0.07 / \$0.29 & 0.71 / 0.73 \\
& & Shadow API H & gemini-2.0-flash & 0 & 42 & \LEFTcircle & \$0.11 / \$0.44 & 1.09 / 1.09 \\
\cmidrule{2-9}

& \multirow{4}{*}{Gemini-2.5-flash} 
& Official & gemini-2.5-flash & 0 & N/A & \CIRCLE & \$0.30 / \$2.50 & 1.00 / 1.00 \\
& & Shadow API A & gemini-2.5-flash & 0 & 42 & \LEFTcircle & \$0.09 / \$2.00 & 0.29 / 0.80 \\
& & Shadow API E & gemini-2.5-flash & 0 & 42 & \LEFTcircle & \$0.21 / \$1.79 & 0.70 / 0.72 \\
& & Shadow API H & gemini-2.5-flash & 0 & 42 & \LEFTcircle & \$0.33 / \$2.73 & 1.09 / 1.09 \\
\cmidrule{2-9}

& \multirow{4}{*}{Gemini-2.5-pro} 
& Official & gemini-2.5-pro & 0 & N/A & \CIRCLE & \$1.25 / \$10.00 & 1.00 / 1.00 \\
& & Shadow API A & gemini-2.5-pro & 0 & 42 & \LEFTcircle & \$1.00 / \$5.70 & 0.80 / 0.57 \\
& & Shadow API E & gemini-2.5-pro & 0 & 42 & \LEFTcircle & \$0.89 / \$7.14 & 0.71 / 0.71 \\
& & Shadow API H & gemini-2.5-pro & 0 & 42 & \LEFTcircle & \$1.36 / \$10.90 & 1.09 / 1.09 \\
\midrule

\multirow{8}{*}{DeepSeek} & \multirow{4}{*}{DeepSeek-Chat} 
& Official & deepseek-chat & 0 & 42 & \CIRCLE & \$0.28 / \$0.42 & 1.00 / 1.00 \\
& & Shadow API A & deepseek-v3.2 & 0 & 42 & \LEFTcircle & \$0.17 / \$0.26 & 0.61 / 0.62 \\
& & Shadow API E & DeepSeek-V3.2-nothinking & 0 & 42 & \CIRCLE & \$1.43 / \$2.14 & 5.11 / 5.10 \\
& & Shadow API H & deepseek-chat & 0 & 42 & \LEFTcircle & \$0.22 / \$0.34 & 0.80 / 0.80 \\
\cmidrule{2-9}

& \multirow{4}{*}{DeepSeek-Reasoner} 
& Official & deepseek-reasoner & 0 & 42 & \CIRCLE & \$0.28 / \$0.42 & 1.00 / 1.00 \\
& & Shadow API A & deepseek-v3.2-thinking & 0 & 42 & \CIRCLE & \$0.17 / \$0.26 & 0.61 / 0.62 \\
& & Shadow API E & DeepSeek-V3.2-thinking & 0 & 42 & \CIRCLE & \$1.43 / \$2.14 & 5.11 / 5.10 \\
& & Shadow API H & deepseek-reasoner & 0 & 42 & \CIRCLE & \$0.22 / \$0.34 & 0.80 / 0.80 \\
\bottomrule
\end{tabular}
}
\begin{flushleft}
\footnotesize
\CIRCLE: Accept logprobs with output; \LEFTcircle: Accept logprobs but no output; \Circle: Not accept logprobs.\\
\textbf{Ratio}: The Ratio column denotes the multiplicative factor of the shadow API service price relative to the official provider for input/output tokens. Ratio $<1$ ($>1$) indicates the shadow API is cheaper (more expensive) than the official ones.
\end{flushleft}
\label{table:api-capability-comparison}
\end{table*}
%-------------------------------------------------------------------------------

%-------------------------------------------------------------------------------
\subsection{Compliance and Transparency Metadata for Shadow API Providers}
\label{appendix:anonymized_apis}
%-------------------------------------------------------------------------------

\mypara{Definitions}
We assess each shadow API provider along two dimensions: transparent identity and verifiable provenance.
Transparent identity requires public disclosure of all three of the following: (1) a legal entity name (not a pseudonym or alias), (2) a verifiable registration identifier, such as an ICP filing number or business registration number, and (3) legal documentation in the form of a Terms of Service and/or Privacy Policy.
Verifiable provenance requires cross-validation against at least one independent external source, including ICP/WHOIS records, official business registries, or platform-level disclosures such as payment merchant identity.
All checks were conducted between September and December 2025.

\mypara{Compliance Metadata}
\autoref{table:compliance_full} reports the full compliance metadata for all 17 shadow API providers in our dataset.
Each provider is assessed across five binary criteria: ICP filing verification, business registry listing, availability of legal documentation (ToS and/or Privacy Policy), payment method, and payee type (individual vs.\ corporate entity).
Service names, domains, and operational status are current as of April 2026.

%-------------------------------------------------------------------------------
\begin{table*}[t!]
\centering
\caption{Full compliance and transparency metadata for all 17 shadow API providers.
Checks were conducted from September to December 2025.
A checkmark (\cmark) indicates a verified positive signal, a dash (---) indicates absence or non-disclosure.}
\label{table:compliance_full}
\setlength{\tabcolsep}{5pt}
\scalebox{0.8}{
\begin{tabular}{@{}clllcccccll@{}}
\toprule
\textbf{ID} & \textbf{Service} & \textbf{Domain} & \textbf{Legal Entity} & \textbf{Reg.\ ID} & \textbf{Legal Docs} & \textbf{ICP} & \textbf{Biz.\ Registry} & \textbf{Payment} & \textbf{Payee} \\
\midrule
A & C*******E & \texttt{api.c********y.cn}   & --- & --- & ---              & --- & --- & Alipay                        & Individual \\
B & Y*****I   & \texttt{y*****i.com}         & --- & --- & ---              & --- & --- & Internal                      & Individual \\
C & X*****I   & \texttt{x*****i.plus}        & --- & --- & ---              & --- & --- & Alipay / WeChat               & Individual \\
D & G*******S & \texttt{g*******s.us}        & --- & --- & ---              & --- & --- & Alipay / WeChat               & Individual \\
E & Q*******O & \texttt{q*******o.com}       & --- & --- & Disclaimer       & --- & --- & Alipay / WeChat / PayPal / USDT & Individual \\
F & O*******B & \texttt{o*******b.com}       & --- & --- & ---              & --- & --- & Alipay / WeChat               & Individual \\
G & D*****I   & \texttt{d*****i.cn}          & --- & --- & ToS, Privacy     & --- & --- & Alipay / WeChat               & Individual \\
H & Z*******G & \texttt{z*******g.com}       & \cmark & \cmark & --- & \cmark & --- & Alipay / WeChat         & Company    \\
I & C*****I   & \texttt{chat.c*****i.vip}    & --- & --- & ---              & --- & --- & Alipay                        & Individual \\
J & O*******D & \texttt{o*******d.cloud}     & --- & --- & ToS              & --- & --- & USDT / WeChat                 & Individual \\
K & V*****I   & \texttt{api.g**.ge}          & --- & --- & ToS, Privacy     & --- & --- & USDT / WeChat                 & Individual \\
L & A*****S   & \texttt{a*****s.com}         & --- & --- & ---              & --- & --- & Alipay / WeChat               & Individual \\
M & B*****I   & \texttt{api.b*****i.ai}      & --- & --- & ---              & --- & --- & Alipay / WeChat               & Individual \\
N & A*******X & \texttt{a*******x.com}       & \cmark & --- & ToS, Privacy & --- & \cmark & Alipay / Credit Card   & Company    \\
O & A*******S & \texttt{a*******s.top}       & --- & --- & ---              & --- & --- & Alipay                        & Individual \\
P & A*****9   & \texttt{a*****9.com}         & --- & --- & ---              & --- & --- & Alipay / WeChat               & Individual \\
Q & 3**i      & \texttt{3**i.cn}             & --- & --- & ---              & --- & --- & Alipay / WeChat               & Individual \\
\bottomrule
\end{tabular}
}
\end{table*}
%-------------------------------------------------------------------------------

\mypara{Summary and Provider Selection Rationale}
Across all 17 providers, only shadow API H holds a verifiable ICP registration, and only shadow API N discloses a corporate legal entity.
The remaining providers lack at least one core provenance signal required for strong accountability, such as a verifiable legal entity, registration identifier, complete legal documentation, or corporate payment channel.
In particular, most providers do not disclose a verifiable operating entity or stable registration record, and their payment records or payee traces frequently point to individual operators rather than registered businesses.

Shadow APIs A, E, and H are selected as the primary subjects of our quantitative analysis for three reasons.
First, they occupy the top of the popularity distribution as measured by academic citations and GitHub stars.
Second, all three maintain publicly accessible endpoints throughout the study period, enabling systematic evaluation.
Third, they collectively cover the three major model families examined in this work, namely GPT, Gemini, and DeepSeek.

The selected providers are not intended to represent identical compliance profiles.
Rather, they capture different points within the broader shadow API ecosystem:
A and E lack verifiable registration or corporate identity signals, while H provides partial registration evidence through an ICP record but still operates as an indirect third-party access channel with limited transparency about upstream model routing.
This variation makes A, E, and H suitable for evaluating whether behavioral inconsistency and model substitution risks persist across providers with different degrees of observable transparency.

%-------------------------------------------------------------------------------
\subsection{Shadow APIs Profiles}
\label{appendix:profiles}
%-------------------------------------------------------------------------------

Shadow API A ranks first in both categories.
It is operated by an individual and does not employ open-source LLM API management or distribution systems. 
Shadow API E and H are selected for their stability, as they rank within the top for both citations and stars. 
Regarding shadow API E, the provider is also an individual and utilizes an open-source LLM API management and distribution system (NewAPI~\cite{NewAPI}) as its deployment template. 
Shadow API E aggregates over 54 upstream providers, some of which are explicitly labeled as having undisclosed origins. 
Some of the endpoints are categorized as reverse-engineered, web-based, capability-degraded, distilled, or unstable yet cost-effective variants. 
Given this diversity, we specifically select endpoints explicitly labeled as originating from the official.
For the shadow API H, the provider is a corporate entity (Internet Content Provider filing) rather than an individual operator, which does not rely on an open-source LLM API management and distribution system.
Notably, the shadow API B is deployed by a laboratory at a prestigious Chinese university, built on NewAPI, where part of the models listed in its internal large language Model Marketplace are explicitly marked with Unknown sources.

%-------------------------------------------------------------------------------
\subsection{OpenRouter Ranking}
\label{appendix:ranking}
%-------------------------------------------------------------------------------

\begin{figure}[!t]
\centering
\includegraphics[width=\linewidth, trim={0 0 0 12cm}, clip]{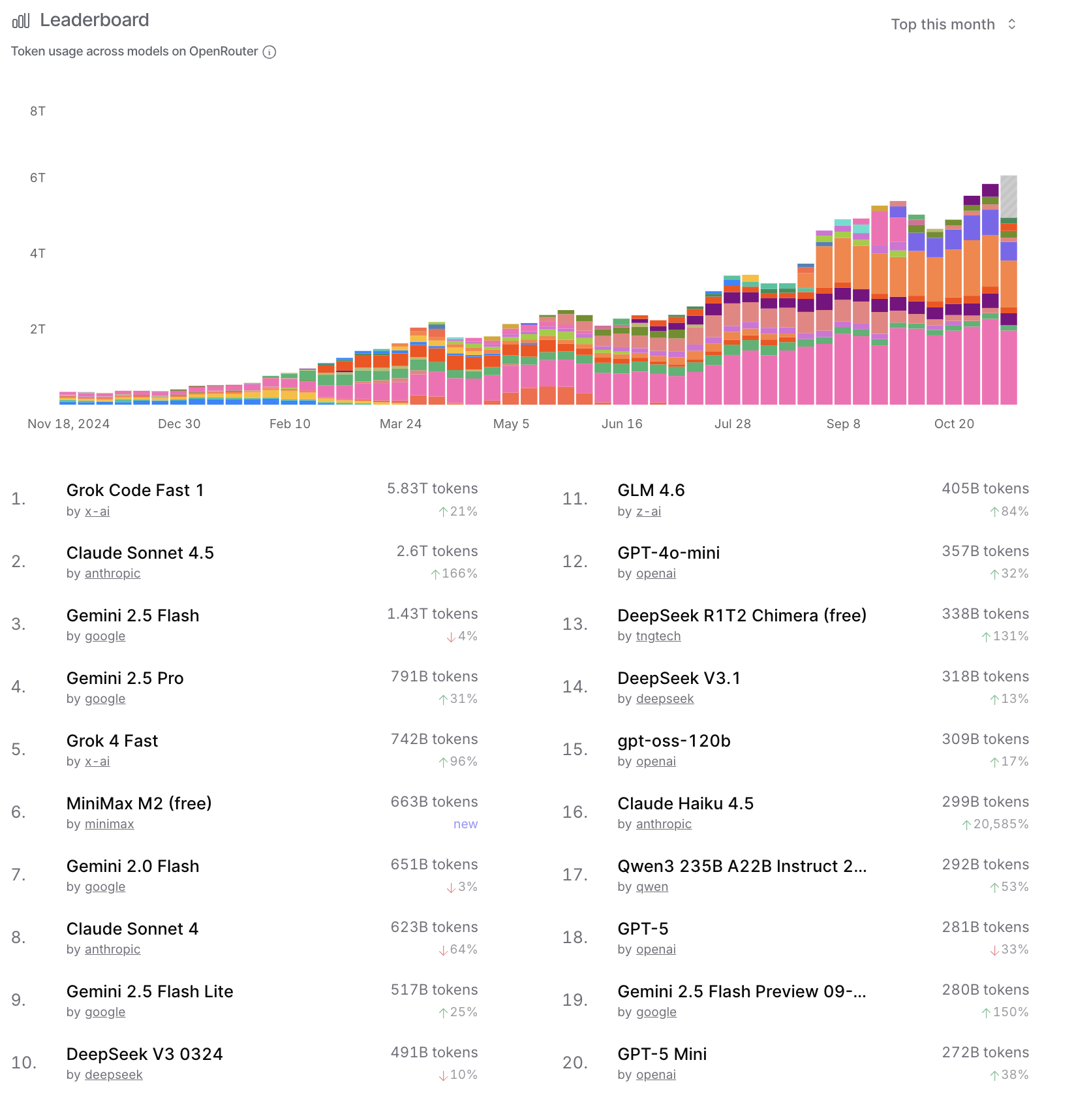}
\caption{LLM token usage rankings on OpenRouter.}
\label{figure:ranking}
\end{figure}

\autoref{figure:ranking} presents the detailed snapshot of the OpenRouter, November 2025. 
This ranking highlights the most widely used models by token consumption.

%-------------------------------------------------------------------------------
\section{Detailed Discrepancy}
\label{appendix:detailed_discrepancy}
%-------------------------------------------------------------------------------

\autoref{table:medqa_discrepancy} and~\autoref{table:legalbench_discrepancy} present a detailed discrepancy analysis between official and shadow APIs, categorizing their response across MedQA (USMLE) and LegalBench (Scalr). 
We report results based on the first execution.

Particularly, Gemini-2.5-flash exhibits performance collapse across all shadow APIs, with consistency rates of $49.57\%$ to $49.65\%$ on MedQA and $52.36\%$ to $54.64\%$ on LegalBench.
This overwhelmingly high count of official-correct-only instances confirms that shadow APIs fail to reproduce the reasoning capabilities of the official API. 
On DeepSeek-Chat, the advantages of this official model are reduced, but still exist as the shadow API A shows a significant degradation in the legal field ($80.56\%$ consistency), diverting sharply from the official baseline compared to other shadow providers.

%-------------------------------------------------------------------------------
\begin{table*}[t!]
\centering
\caption{Performance discrepancy between official and shadow APIs on MedQA (USMLE).}
\label{table:medqa_discrepancy}
\scalebox{0.8}{
\begin{tabular}{@{}llcccccc@{}}
\toprule
\textbf{Model} & \textbf{Shadow API} & \textbf{Total} & \textbf{Both Correct} & \textbf{Both Incorrect} & \textbf{Official Correct Only} & \textbf{Shadow Correct Only} & \textbf{Consistency (\%)} \\ 
\midrule
\multirow{3}{*}{Gemini-2.5-flash} &Shadow API A & 1273 & 448 & 184 & 619 & 22 & 49.65 \\
                                  &Shadow API E & 1273 & 450 & 181 & 617 & 25 & 49.57 \\
                                  &Shadow API H & 1273 & 446 & 186 & 621 & 20 & 49.65 \\ 
\midrule
\multirow{3}{*}{GPT-5-mini}       &Shadow API A & 1273 & 981 & 232 & 24 & 36 & 95.29 \\
                                  &Shadow API E & 1273 & 949 & 228 & 56 & 40 & 92.46 \\
                                  &Shadow API H & 1273 & 977 & 230 & 28 & 38 & 94.82 \\ 
\midrule
\multirow{3}{*}{DeepSeek-Chat}    &Shadow API A & 1273 & 827 & 280 & 105 & 61 & 86.96 \\
                                  &Shadow API E & 1273 & 905 & 326 & 27 & 15 & 96.70 \\
                                  &Shadow API H & 1273 & 888 & 313 & 44 & 28 & 94.34 \\ 
\bottomrule
\end{tabular}
}
\end{table*}
%-------------------------------------------------------------------------------

%-------------------------------------------------------------------------------
\begin{table*}[ht!]
\centering
\caption{Performance discrepancy between official and shadow APIs on LegalBench (Scalr).}
\label{table:legalbench_discrepancy}
\scalebox{0.8}{
\begin{tabular}{@{}llcccccc@{}}
\toprule
\textbf{Model} & \textbf{Shadow API} & \textbf{Total} & \textbf{Both Correct} & \textbf{Both Incorrect} & \textbf{Official Correct Only} & \textbf{Shadow Correct Only} & \textbf{Consistency (\%)} \\ 
\midrule
\multirow{3}{*}{Gemini-2.5-flash} &Shadow API A & 571 & 187 & 116 & 254 & 14 & 53.06 \\
                                  &Shadow API E & 571 & 183 & 116 & 258 & 14 & 52.36 \\
                                  &Shadow API H & 571 & 197 & 115 & 244 & 15 & 54.64 \\ 
\midrule
\multirow{3}{*}{GPT-5-mini}       &Shadow API A & 571 & 416 & 133 & 12 & 10 & 96.15 \\
                                  &Shadow API E & 571 & 411 & 128 & 17 & 15 & 94.40 \\
                                  &Shadow API H & 571 & 416 & 128 & 12 & 15 & 95.27 \\ 
\midrule
\multirow{3}{*}{DeepSeek-Chat}    &Shadow API A & 571 & 367 & 93 & 84 & 27 & 80.56 \\
                                  &Shadow API E & 571 & 417 & 100 & 34 & 20 & 90.54 \\
                                  &Shadow API H & 571 & 427 & 103 & 24 & 17 & 92.82 \\ 
\bottomrule
\end{tabular}
}
\end{table*}
%-------------------------------------------------------------------------------

%-------------------------------------------------------------------------------
\section{Inference Latency Time and Token Counts}
\label{appendix:time_token}
\begin{figure*}[t]
    \centering
    \begin{subfigure}{0.48\linewidth}
        \centering
        \includegraphics[width=\linewidth]{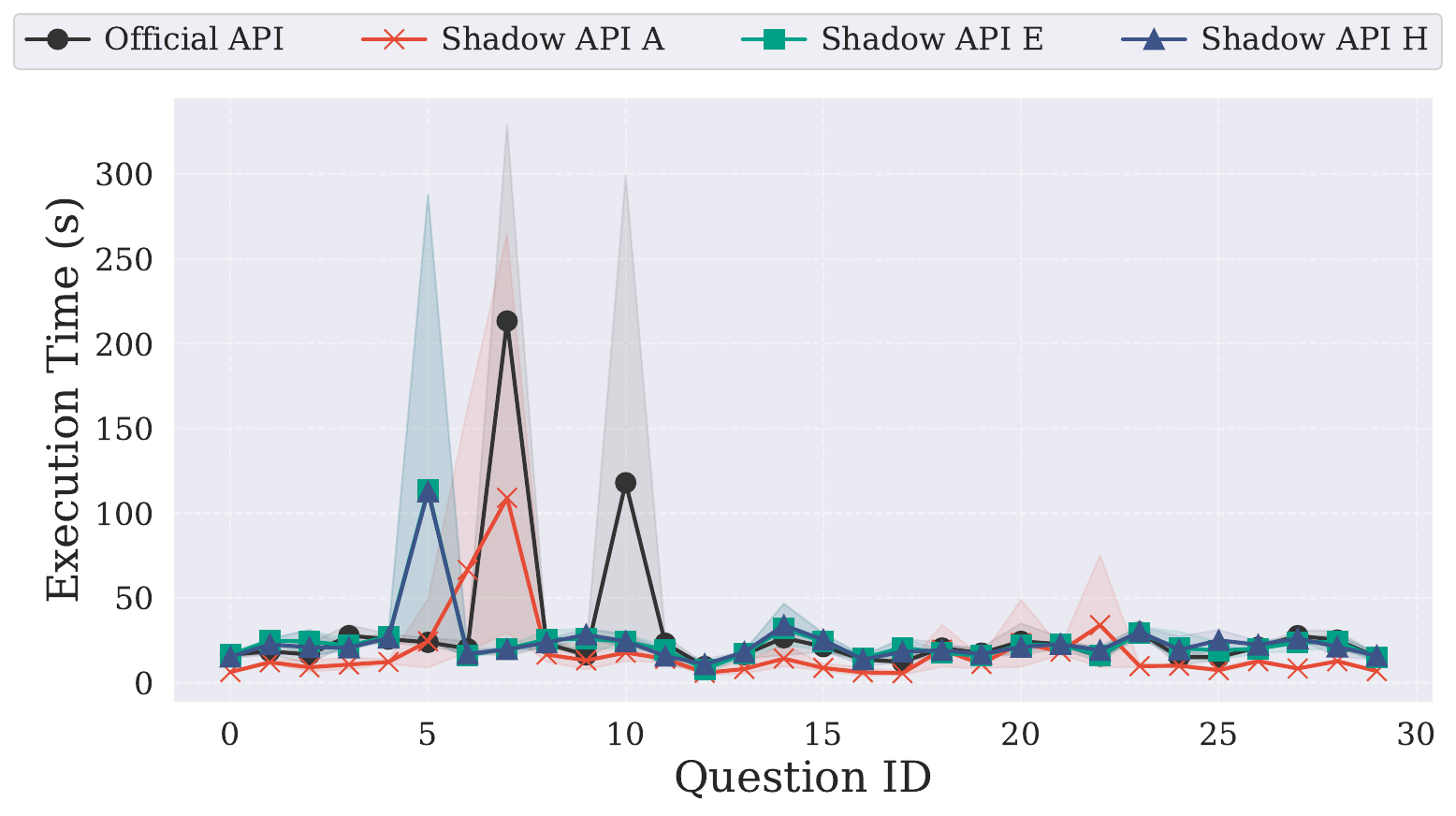}
        \caption{AIME: GPT-4o-mini (Time)}
    \end{subfigure}
    \hfill
    \begin{subfigure}{0.48\linewidth}
        \centering
        \includegraphics[width=\linewidth]{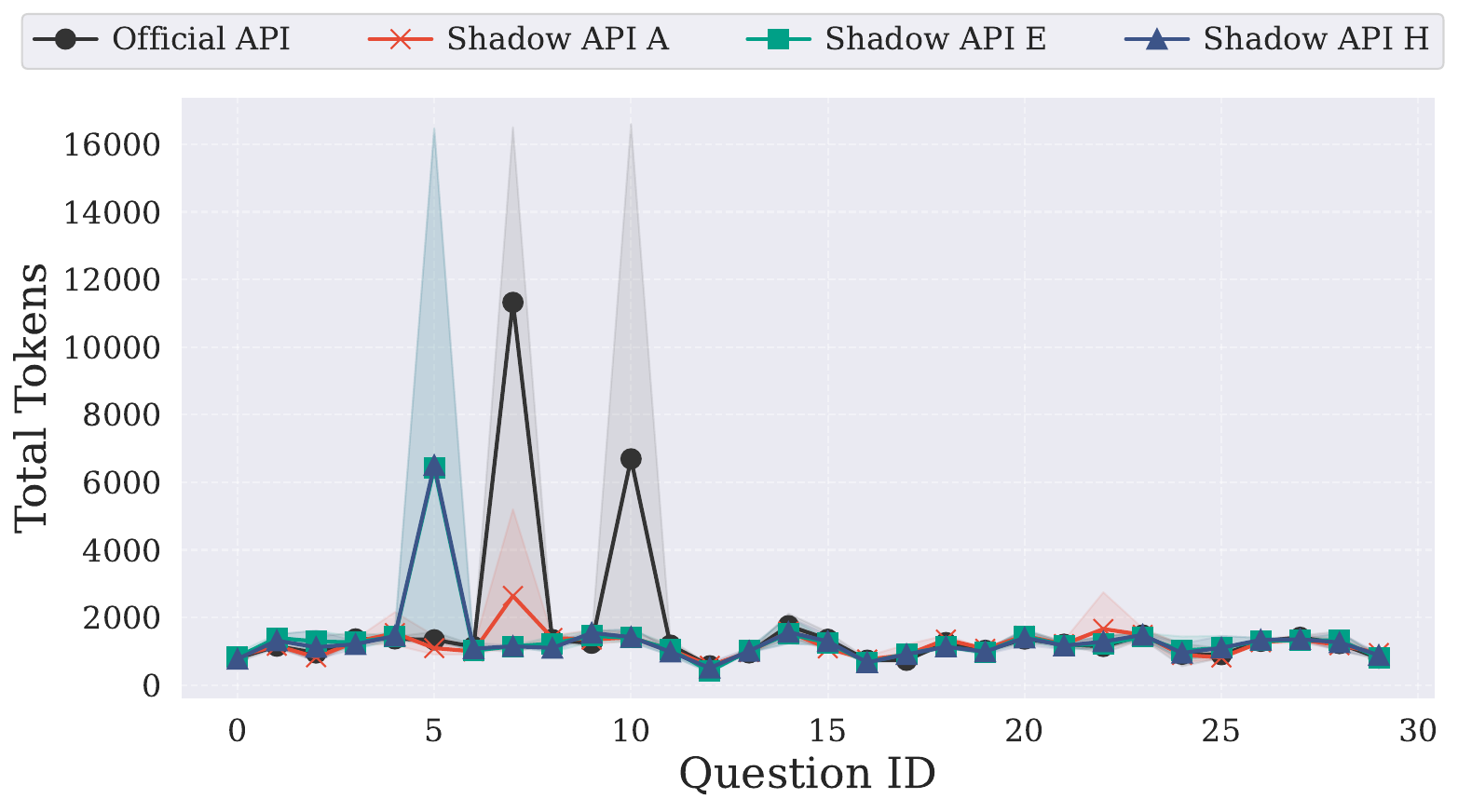}
        \caption{AIME: GPT-4o-mini (Token)}
    \end{subfigure}

    \begin{subfigure}{0.48\linewidth}
        \centering
        \includegraphics[width=\linewidth]{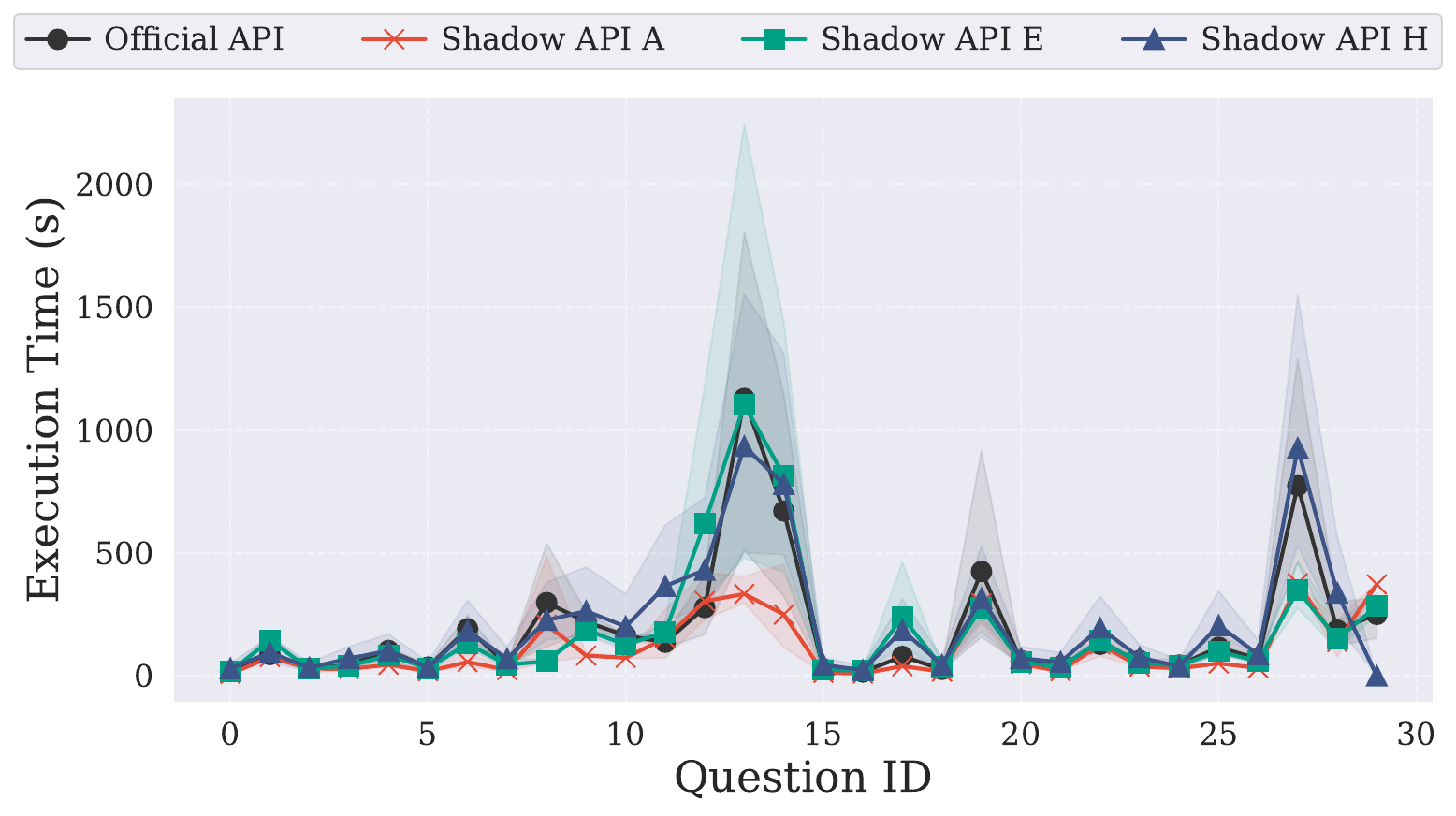}
        \caption{AIME: GPT-5 (Time)}
    \end{subfigure}
    \hfill
    \begin{subfigure}{0.48\linewidth}
        \centering
        \includegraphics[width=\linewidth]{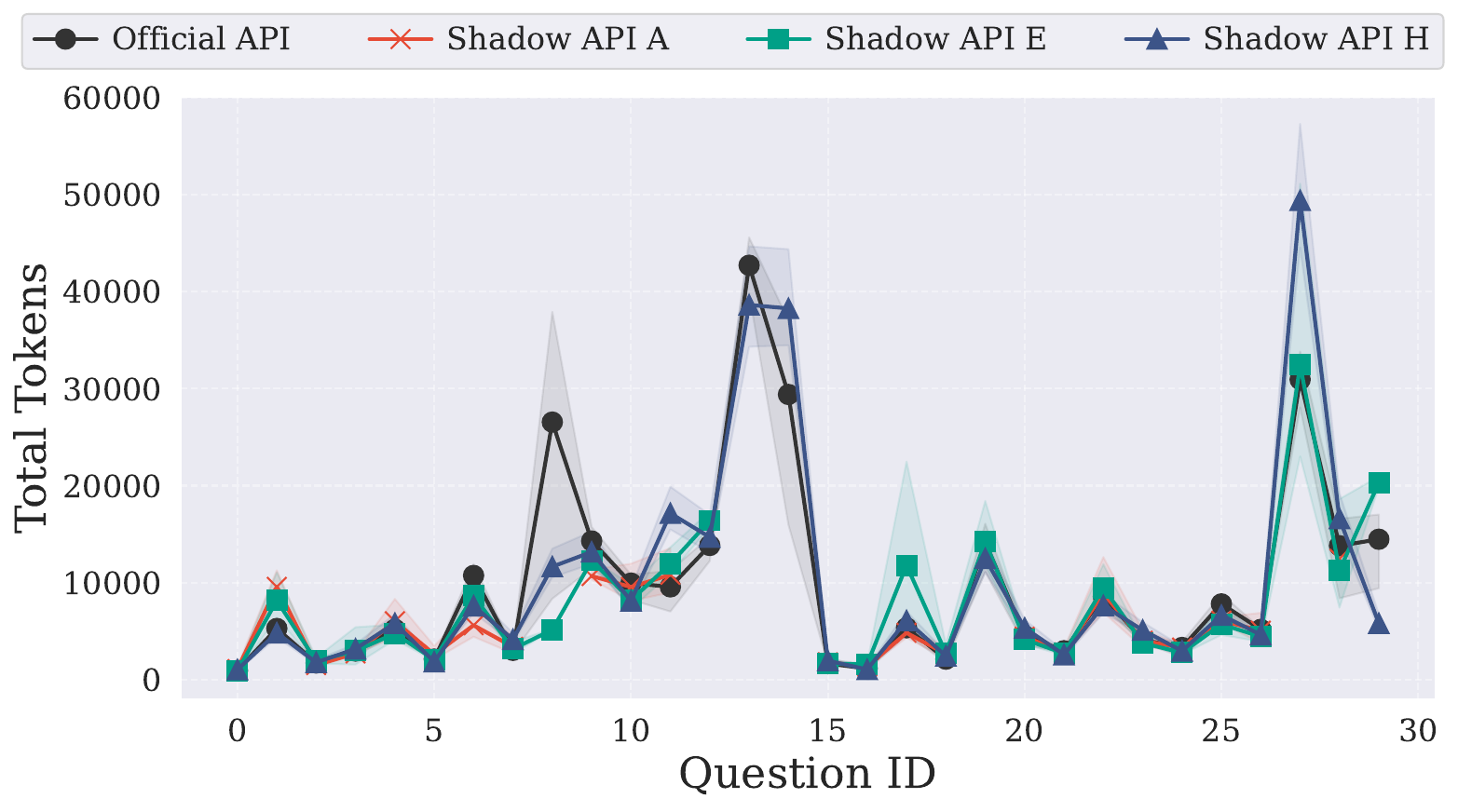}
        \caption{AIME: GPT-5 (Token)}
    \end{subfigure}

    \begin{subfigure}{0.48\linewidth}
        \centering
        \includegraphics[width=\linewidth]{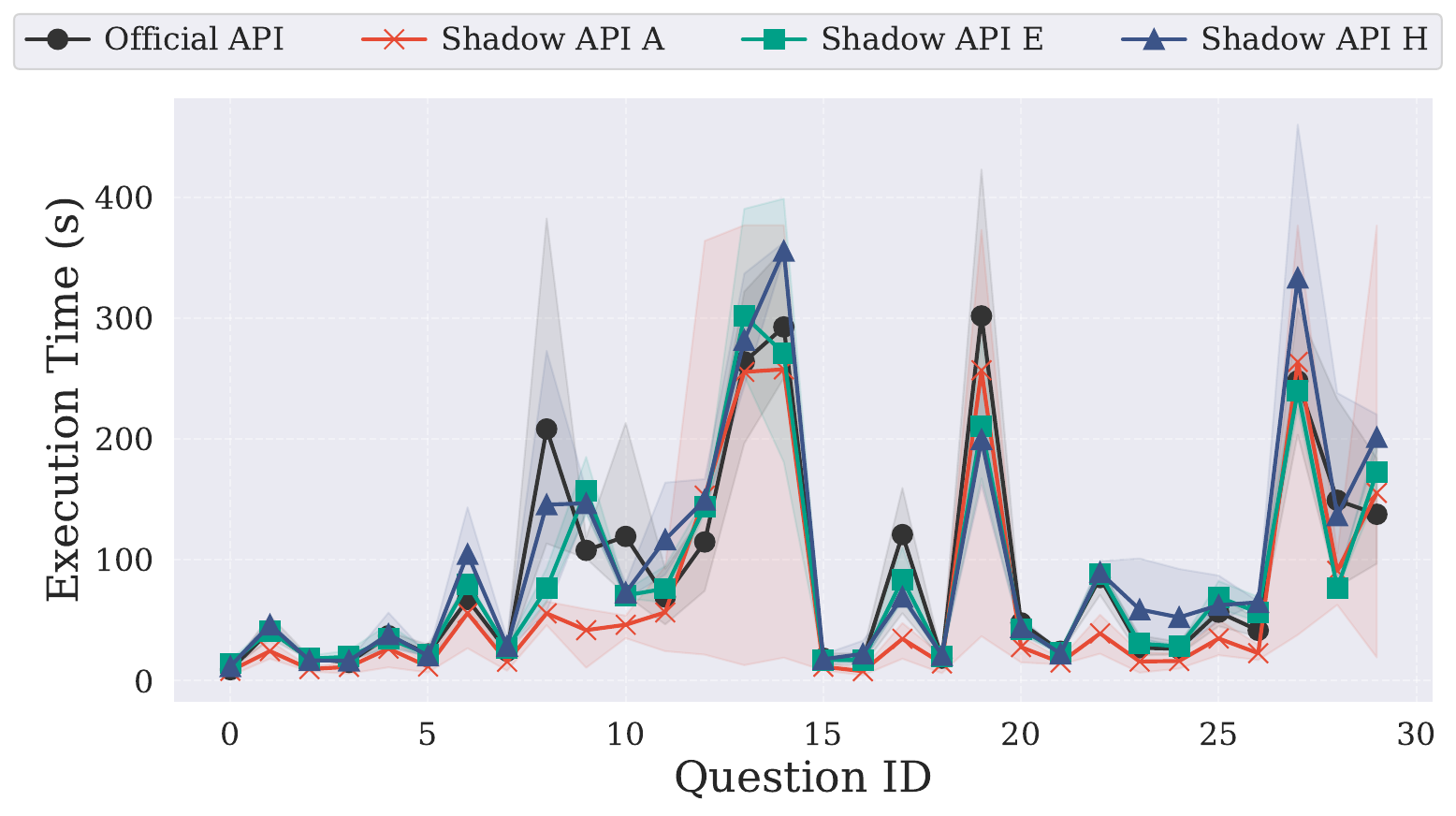}
        \caption{AIME: GPT-5-mini (Time)}
    \end{subfigure}
    \hfill
    \begin{subfigure}{0.48\linewidth}
        \centering
        \includegraphics[width=\linewidth]{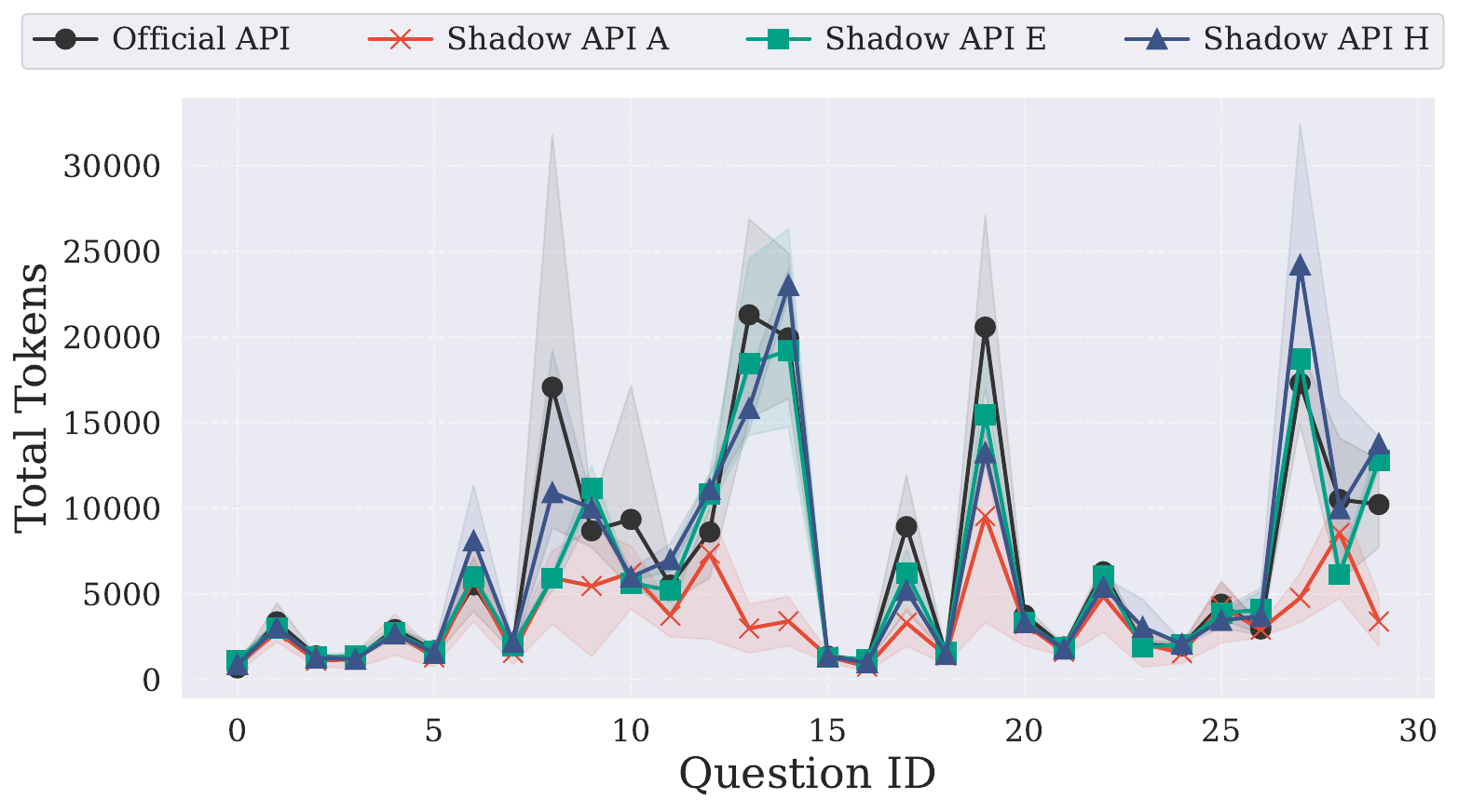}
        \caption{AIME: GPT-5-mini (Token)}
    \end{subfigure}

    \begin{subfigure}{0.48\linewidth}
        \centering
        \includegraphics[width=\linewidth]{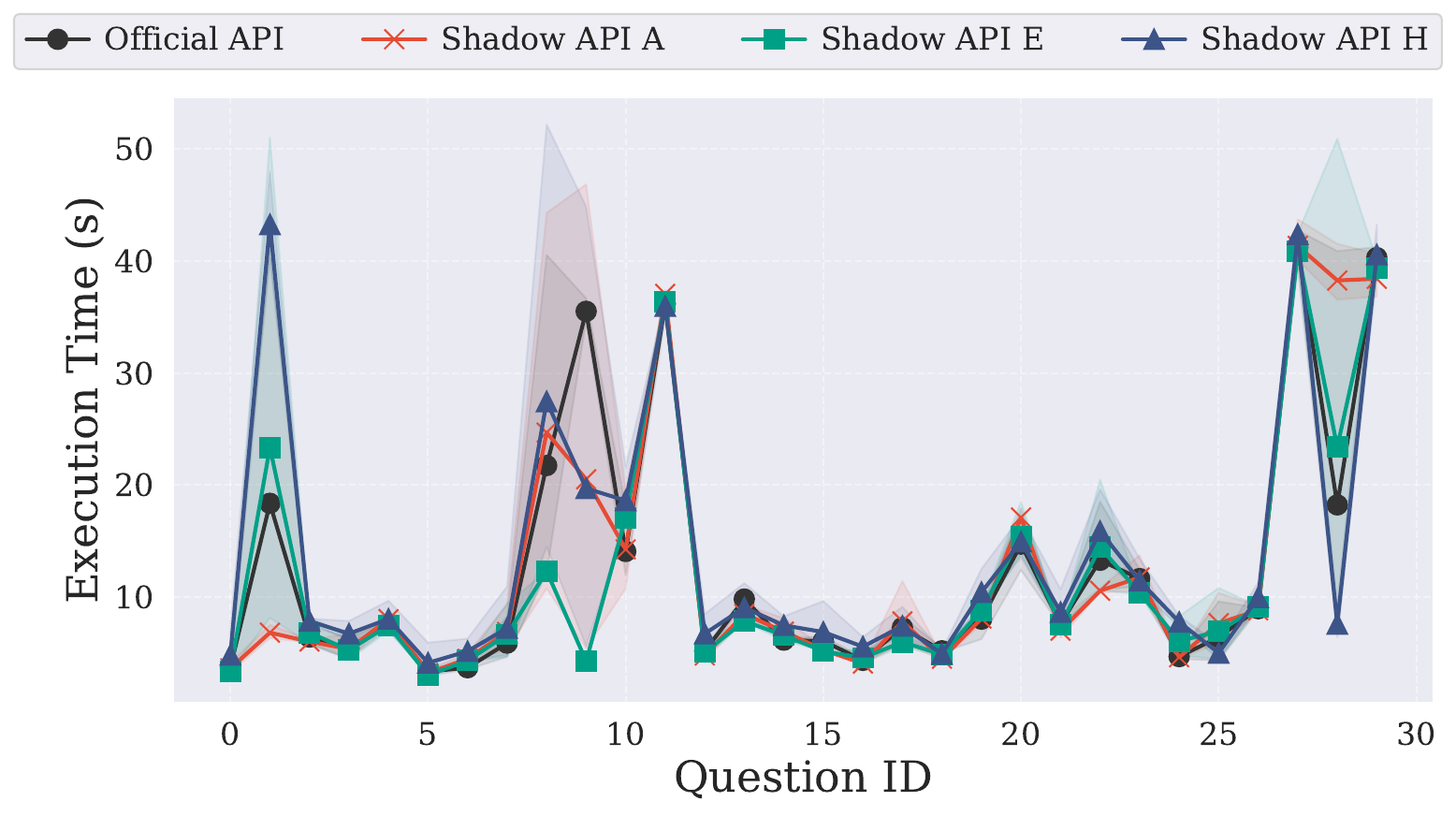}
        \caption{AIME: Gemini-2.0-flash (Time)}
    \end{subfigure}
    \hfill
    \begin{subfigure}{0.48\linewidth}
        \centering
        \includegraphics[width=\linewidth]{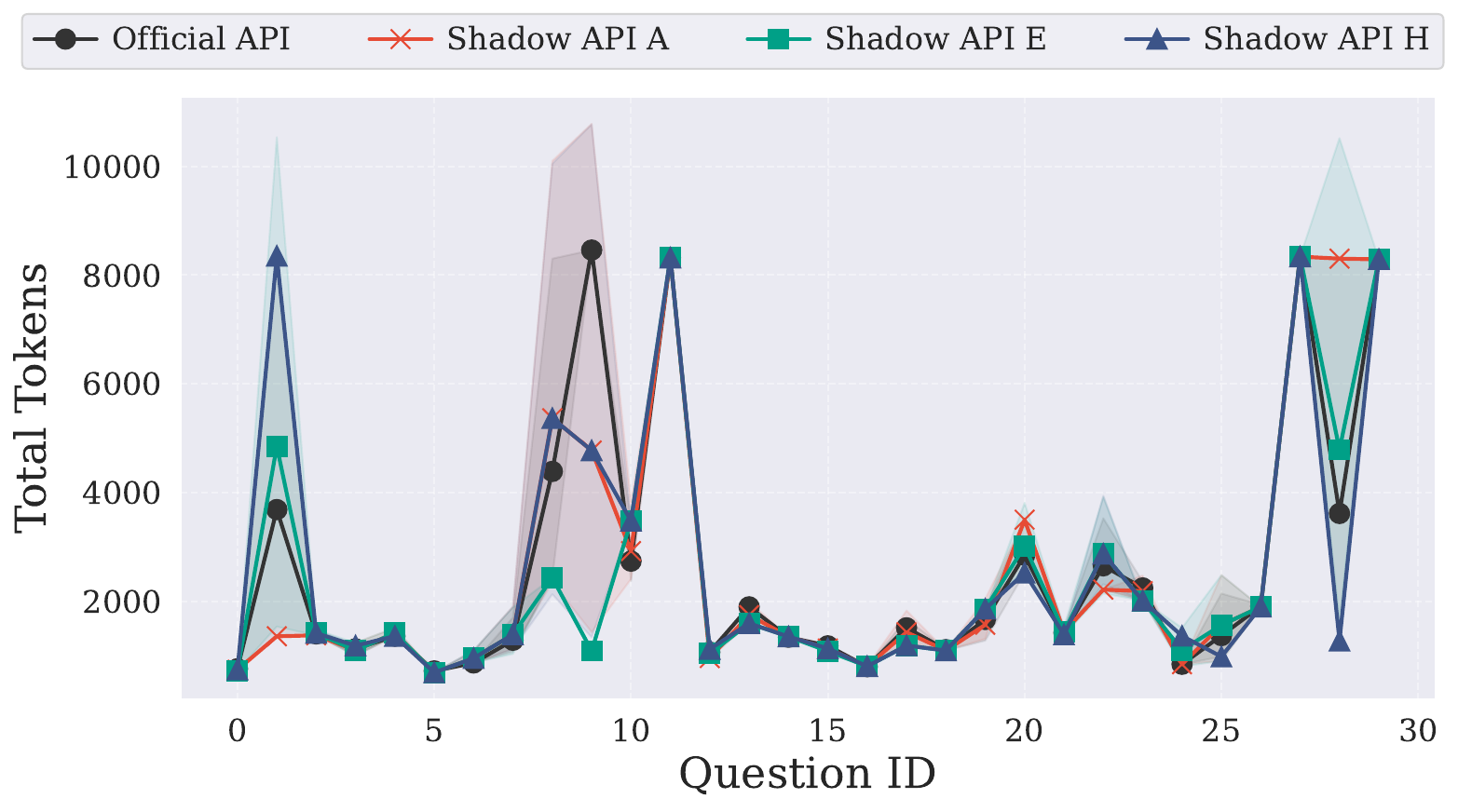}
        \caption{AIME: Gemini-2.0-flash (Token)}
    \end{subfigure}
    \caption{Comparison of inference latency time and token counts on the AIME (Part 1). Solid lines represent mean values, and shaded regions denote the range between the minimum and maximum values across three trials.}
    \label{figure:aime_results_part1}
\end{figure*}

\begin{figure*}[t]
    \centering
    \begin{subfigure}{0.48\linewidth}
        \centering
        \includegraphics[width=\linewidth]{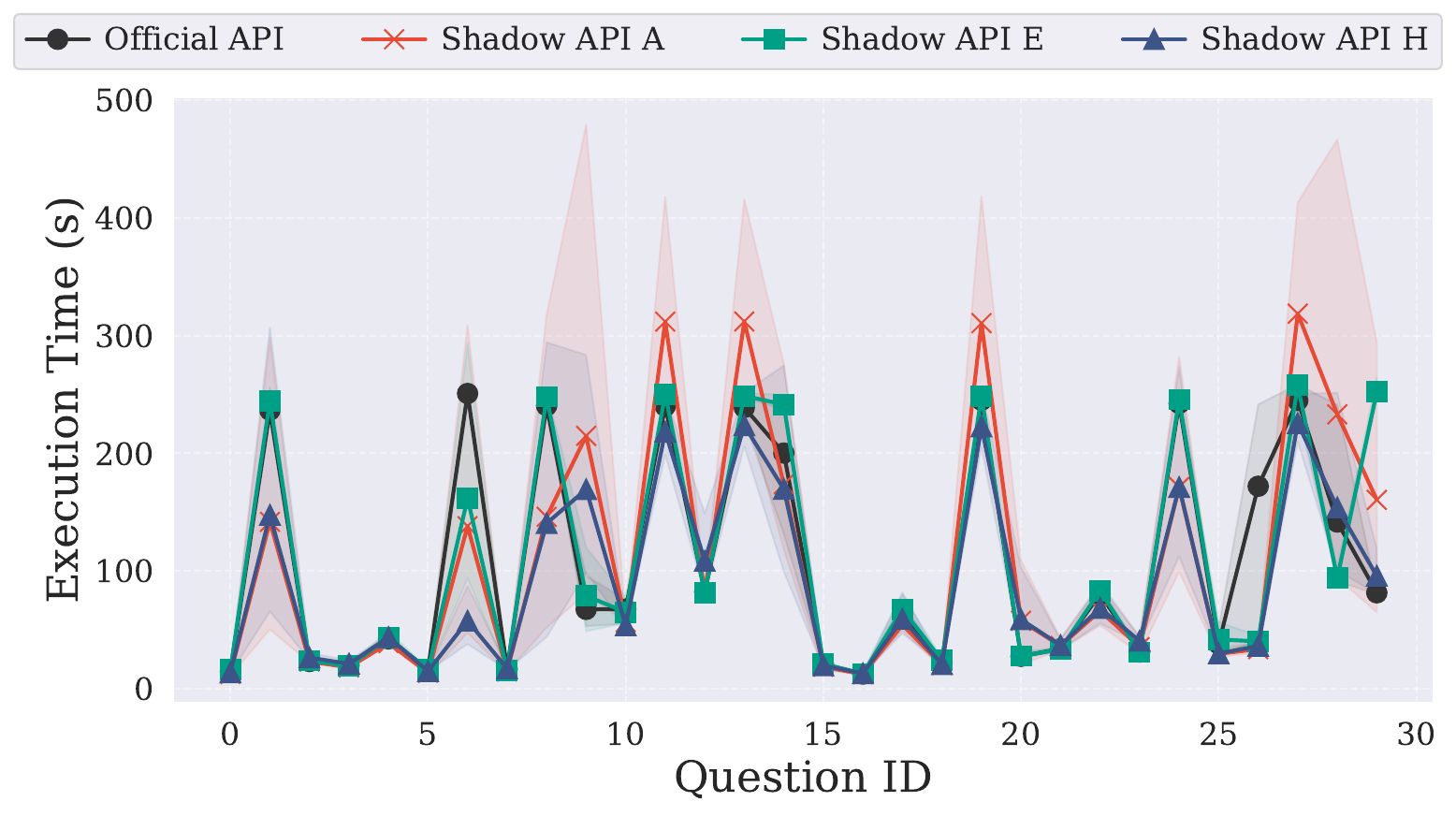}
        \caption{AIME: Gemini-2.5-flash (Time)}
    \end{subfigure}
    \hfill
    \begin{subfigure}{0.48\linewidth}
        \centering
        \includegraphics[width=\linewidth]{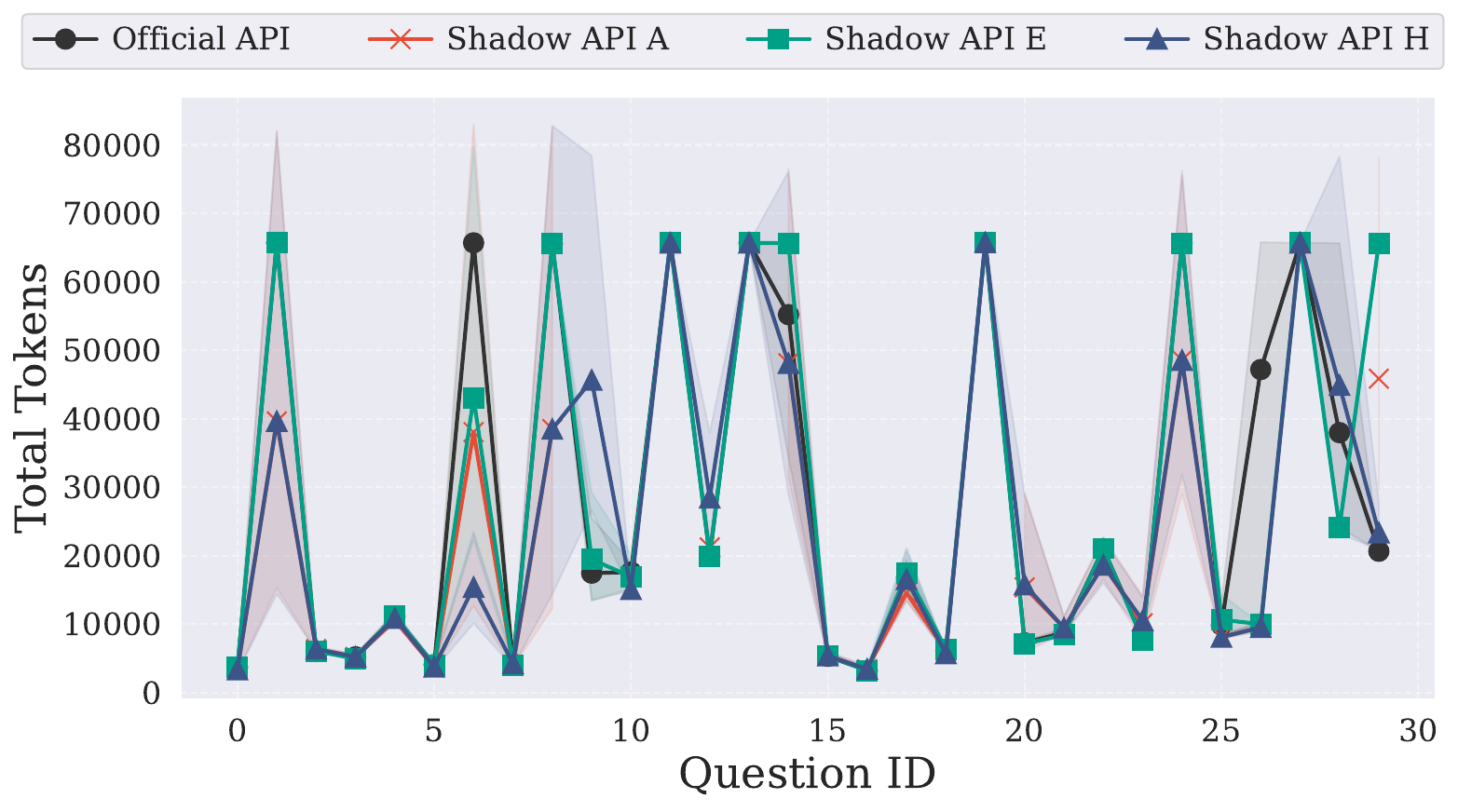}
        \caption{AIME: Gemini-2.5-flash (Token)}
    \end{subfigure}

    \begin{subfigure}{0.48\linewidth}
        \centering
        \includegraphics[width=\linewidth]{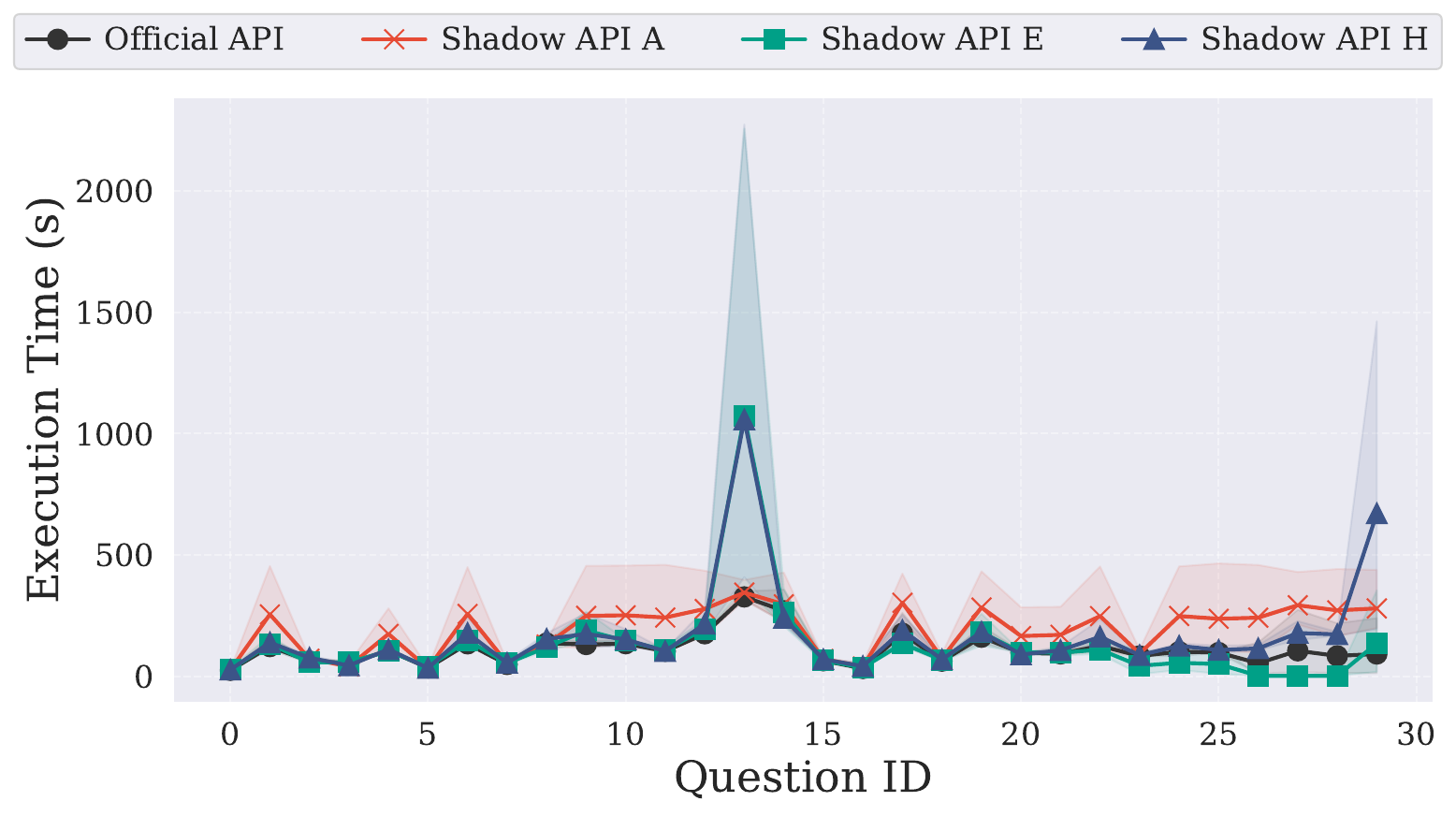}
        \caption{AIME: Gemini-2.5-pro (Time)}
    \end{subfigure}
    \hfill
    \begin{subfigure}{0.48\linewidth}
        \centering
        \includegraphics[width=\linewidth]{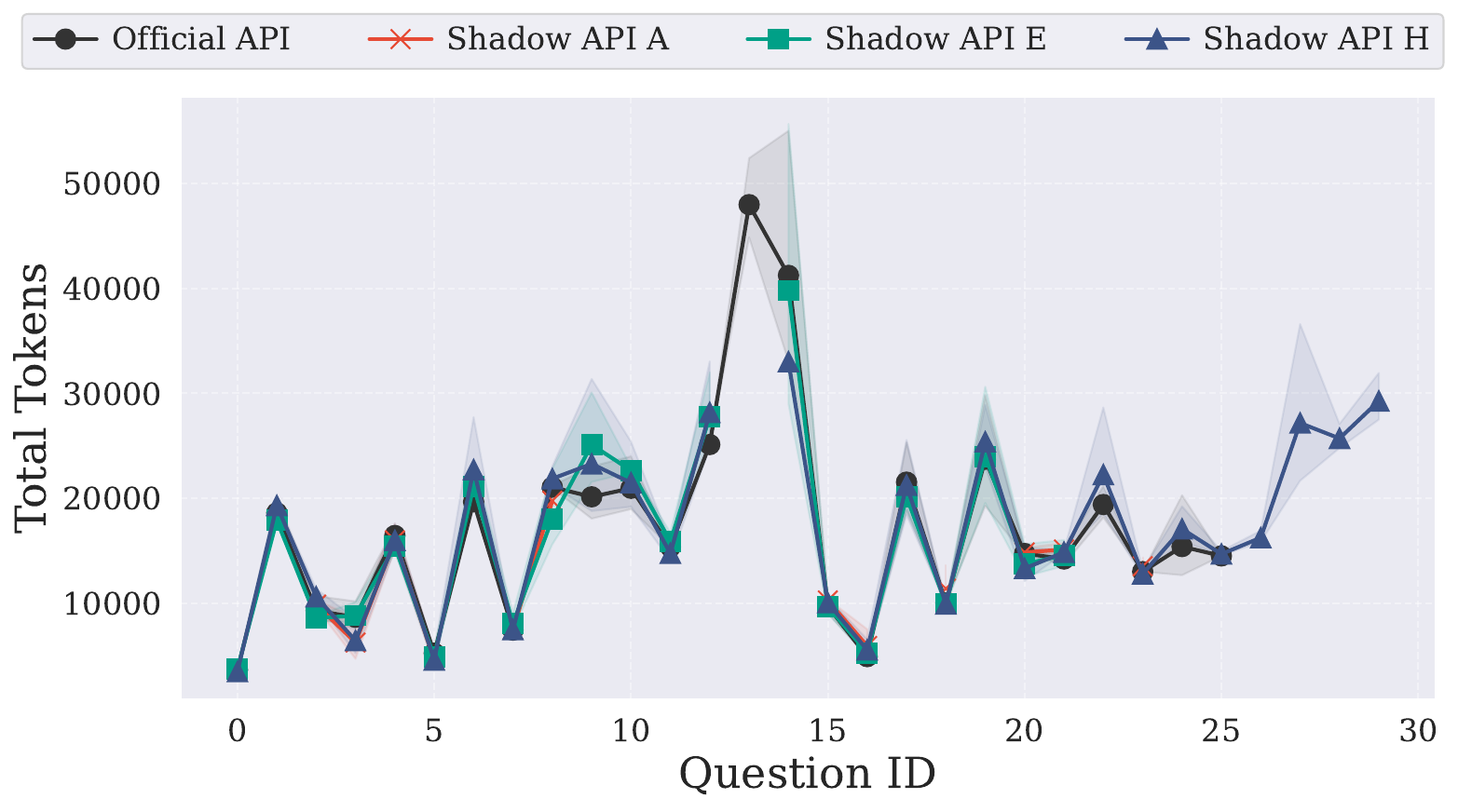}
        \caption{AIME: Gemini-2.5-pro (Token)}
    \end{subfigure}

    \begin{subfigure}{0.48\linewidth}
        \centering
        \includegraphics[width=\linewidth]{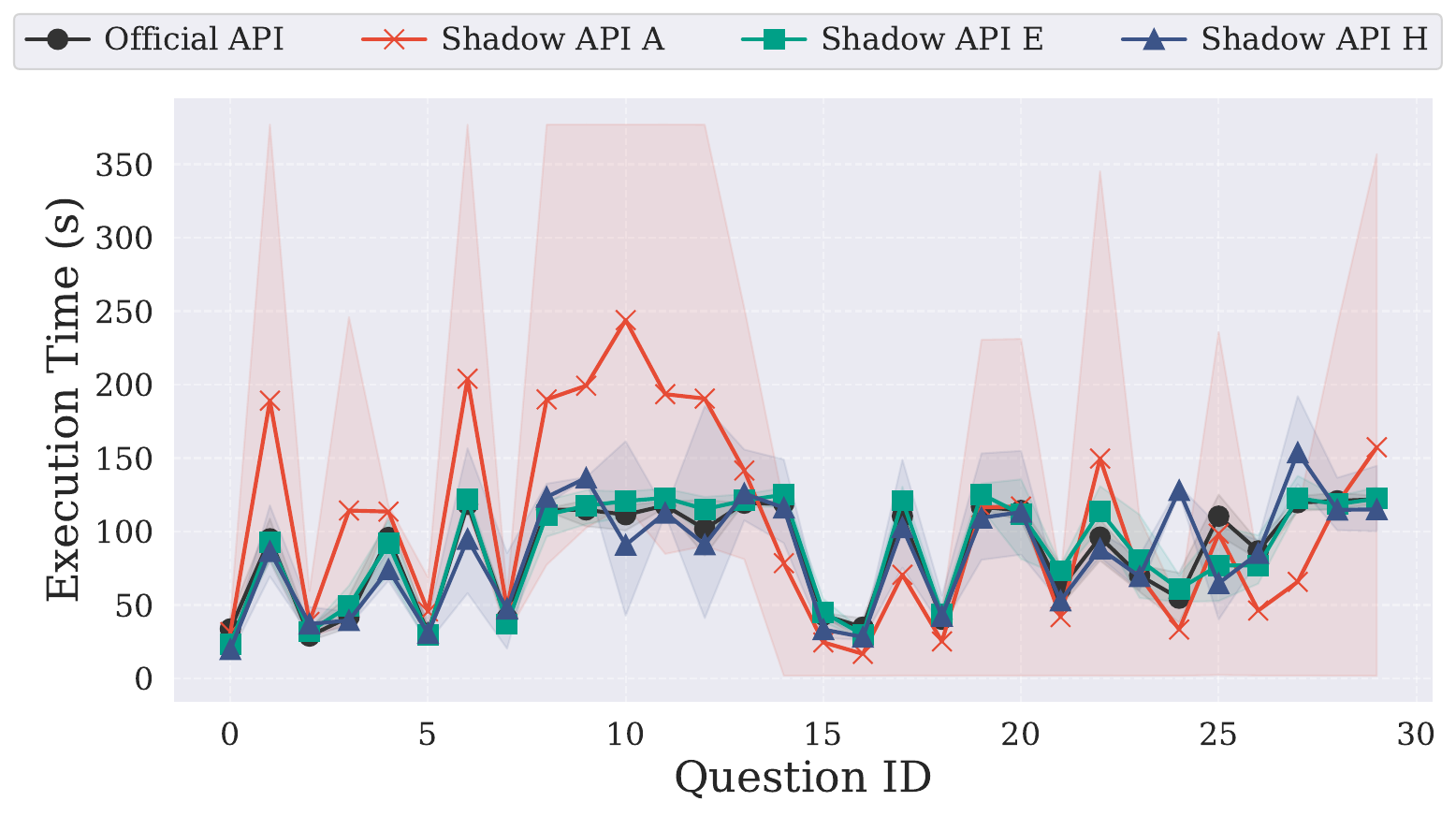}
        \caption{AIME: DeepSeek-Chat (Time)}
    \end{subfigure}
    \hfill
    \begin{subfigure}{0.48\linewidth}
        \centering
        \includegraphics[width=\linewidth]{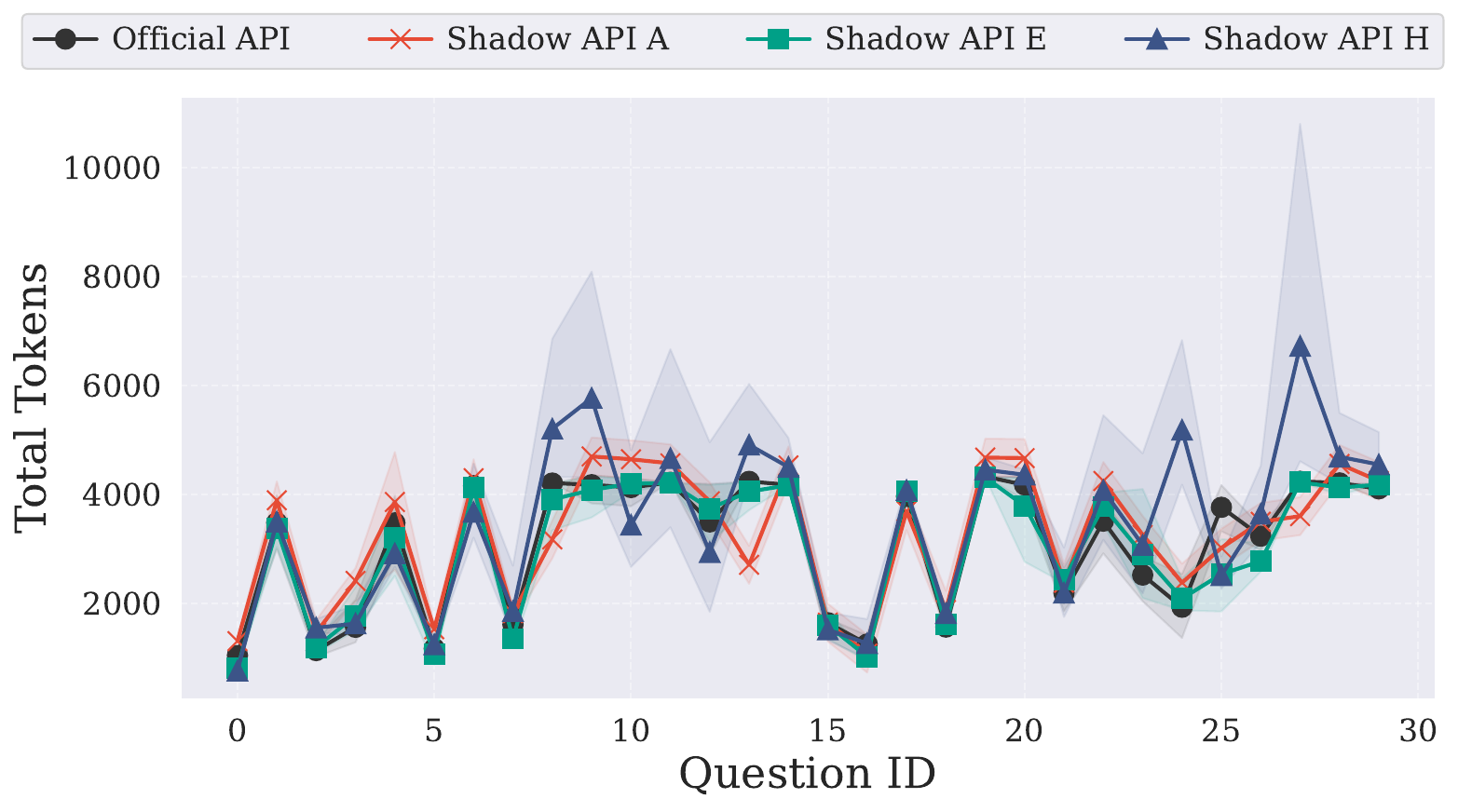}
        \caption{AIME: DeepSeek-Chat (Token)}
    \end{subfigure}

    \begin{subfigure}{0.48\linewidth}
        \centering
        \includegraphics[width=\linewidth]{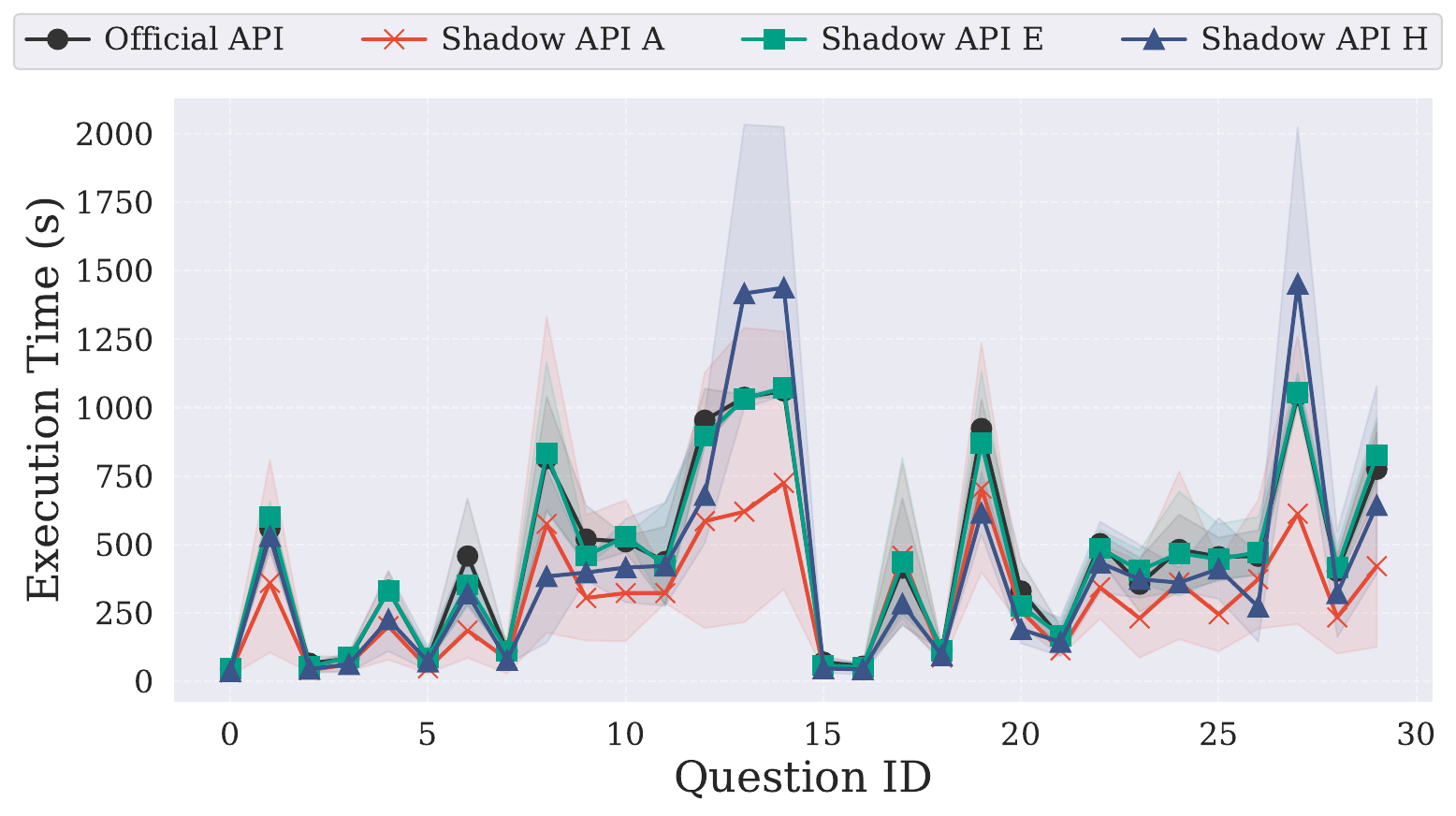}
        \caption{AIME: DeepSeek-Reasoner (Time)}
    \end{subfigure}
    \hfill
    \begin{subfigure}{0.48\linewidth}
        \centering
        \includegraphics[width=\linewidth]{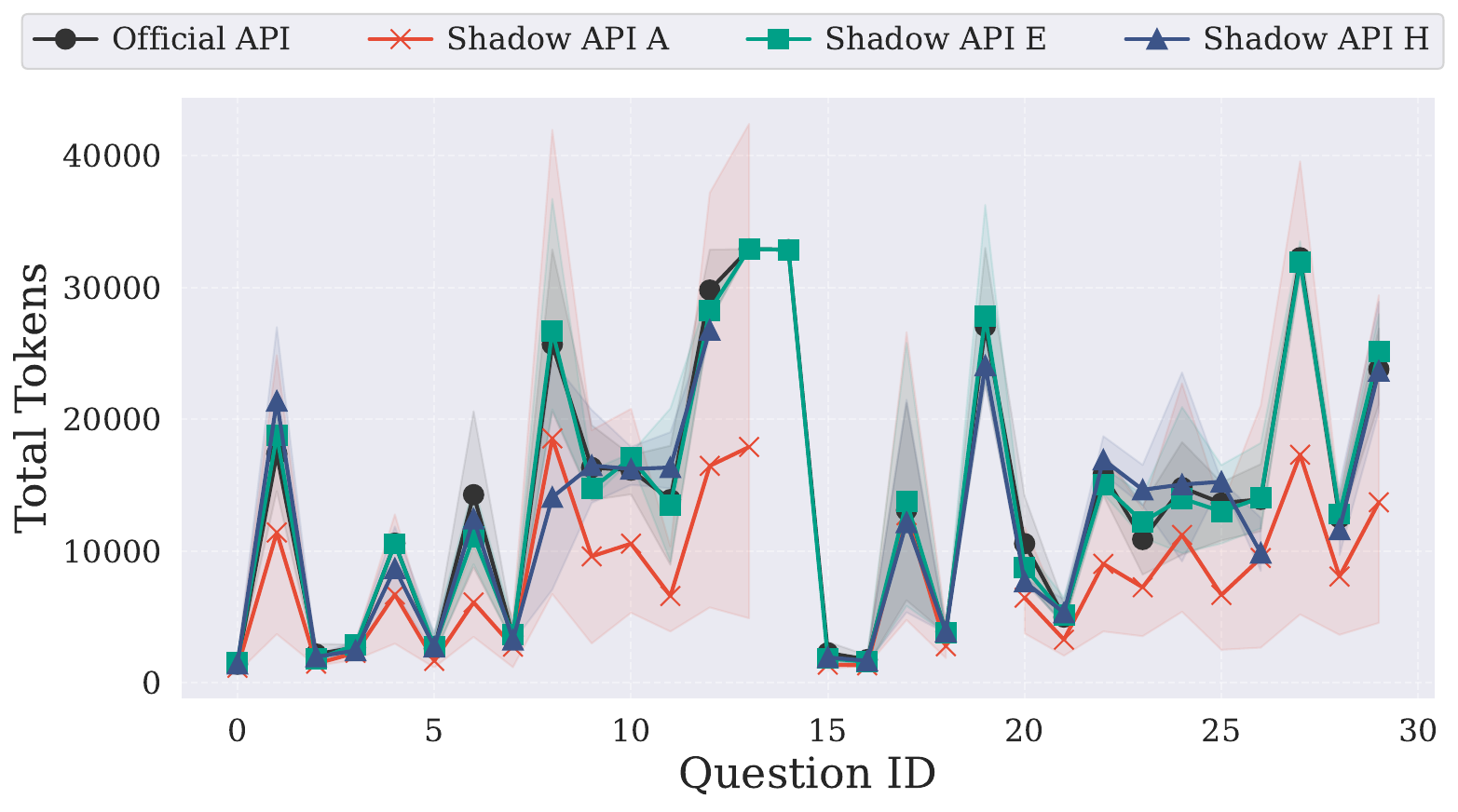}
        \caption{AIME: DeepSeek-Reasoner (Token)}
    \end{subfigure}

    \caption{Comparison of inference latency time and token counts on the AIME (Part 2).  Solid lines represent mean values, and shaded regions denote the range between the minimum and maximum values across three trials.}
    \label{figure:aime_results_part2}
\end{figure*}

\begin{figure*}[t]
    \centering
    \begin{subfigure}{0.48\linewidth}
        \centering
        \includegraphics[width=\linewidth]{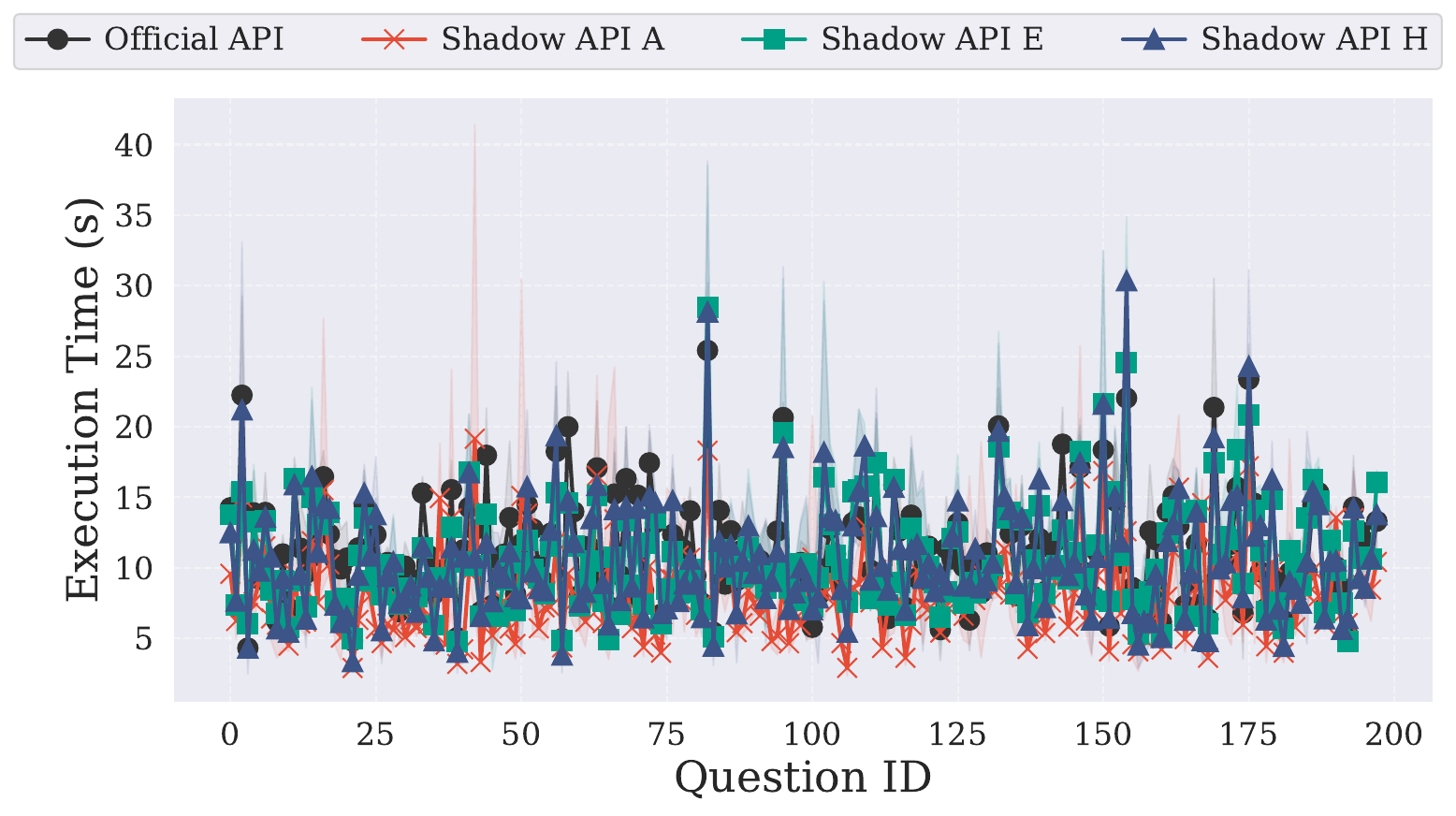}
        \caption{GPQA: GPT-4o-mini (Time)}
    \end{subfigure}
    \hfill
    \begin{subfigure}{0.48\linewidth}
        \centering
        \includegraphics[width=\linewidth]{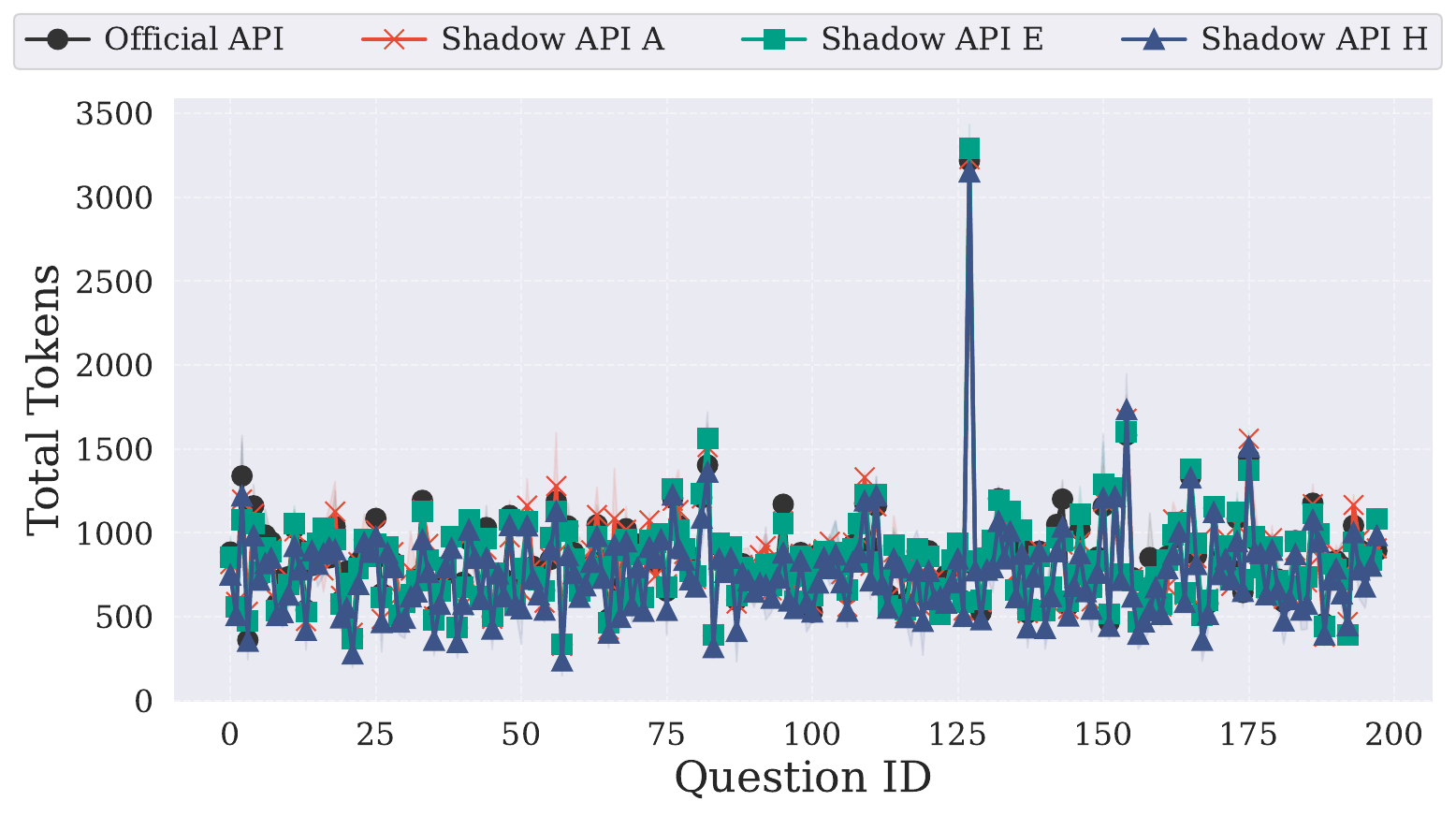}
        \caption{GPQA: GPT-4o-mini (Token)}
    \end{subfigure}

    \begin{subfigure}{0.48\linewidth}
        \centering
        \includegraphics[width=\linewidth]{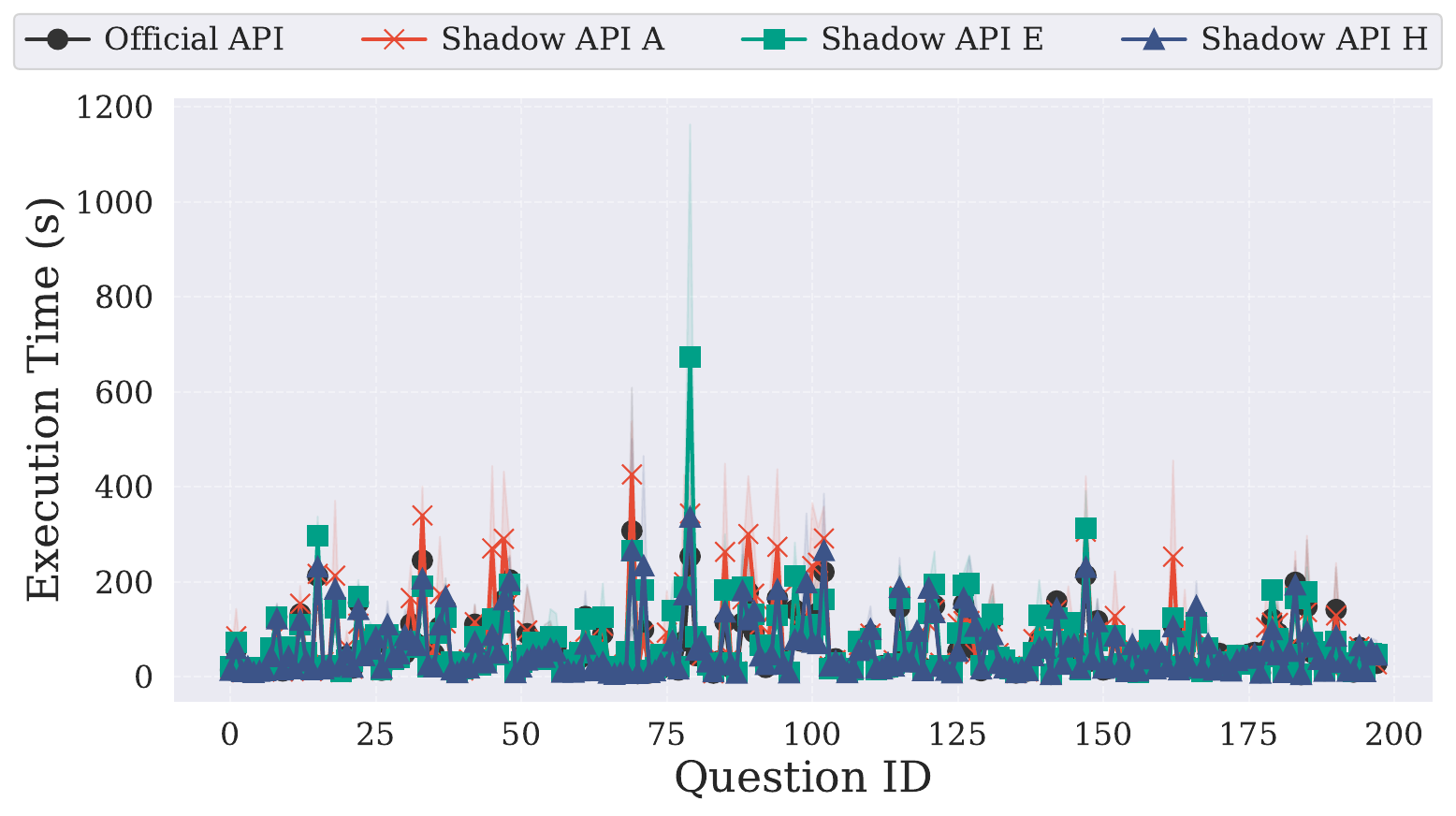}
        \caption{GPQA: GPT-5 (Time)}
    \end{subfigure}
    \hfill
    \begin{subfigure}{0.48\linewidth}
        \centering
        \includegraphics[width=\linewidth]{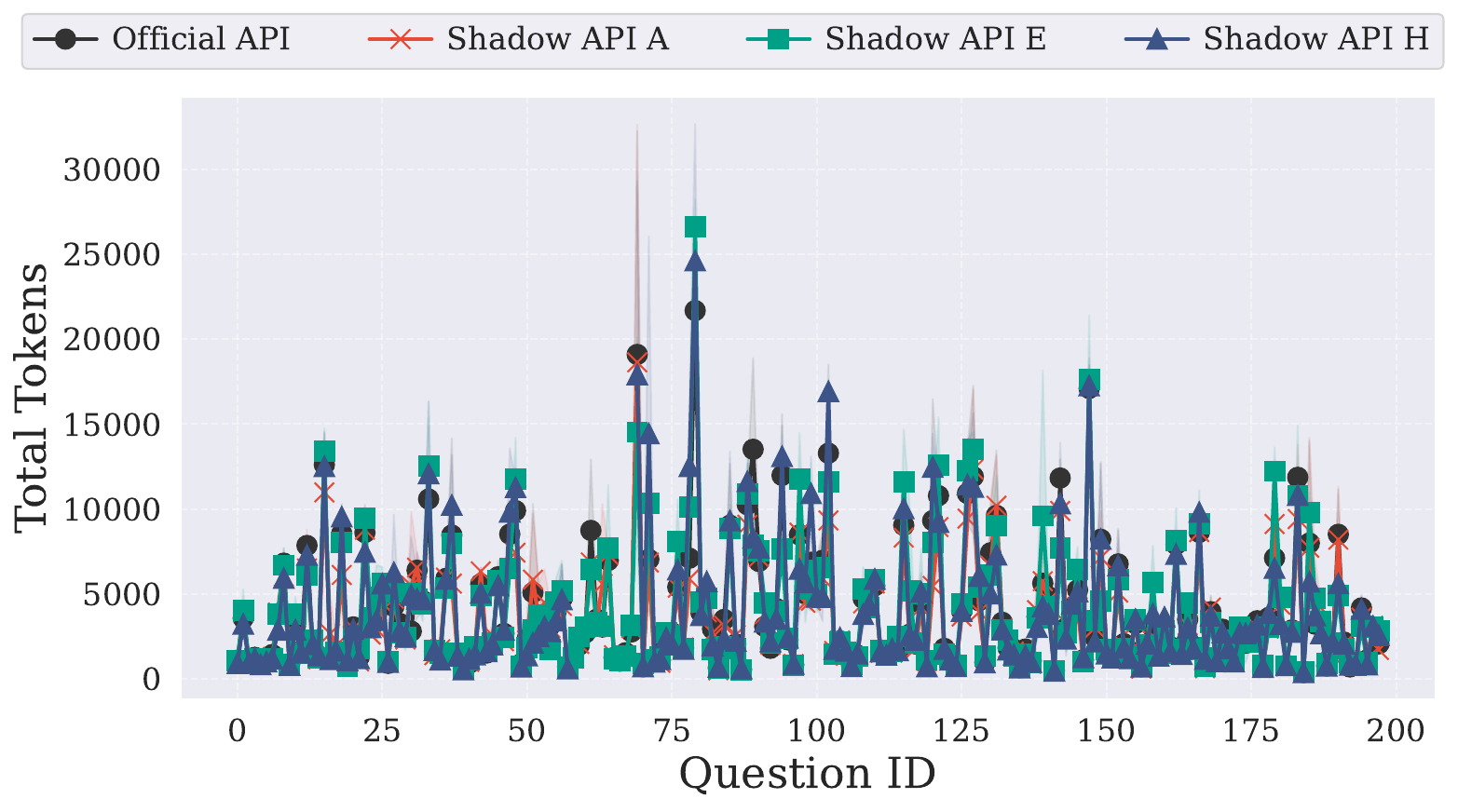}
        \caption{GPQA: GPT-5 (Token)}
    \end{subfigure}

    \begin{subfigure}{0.48\linewidth}
        \centering
        \includegraphics[width=\linewidth]{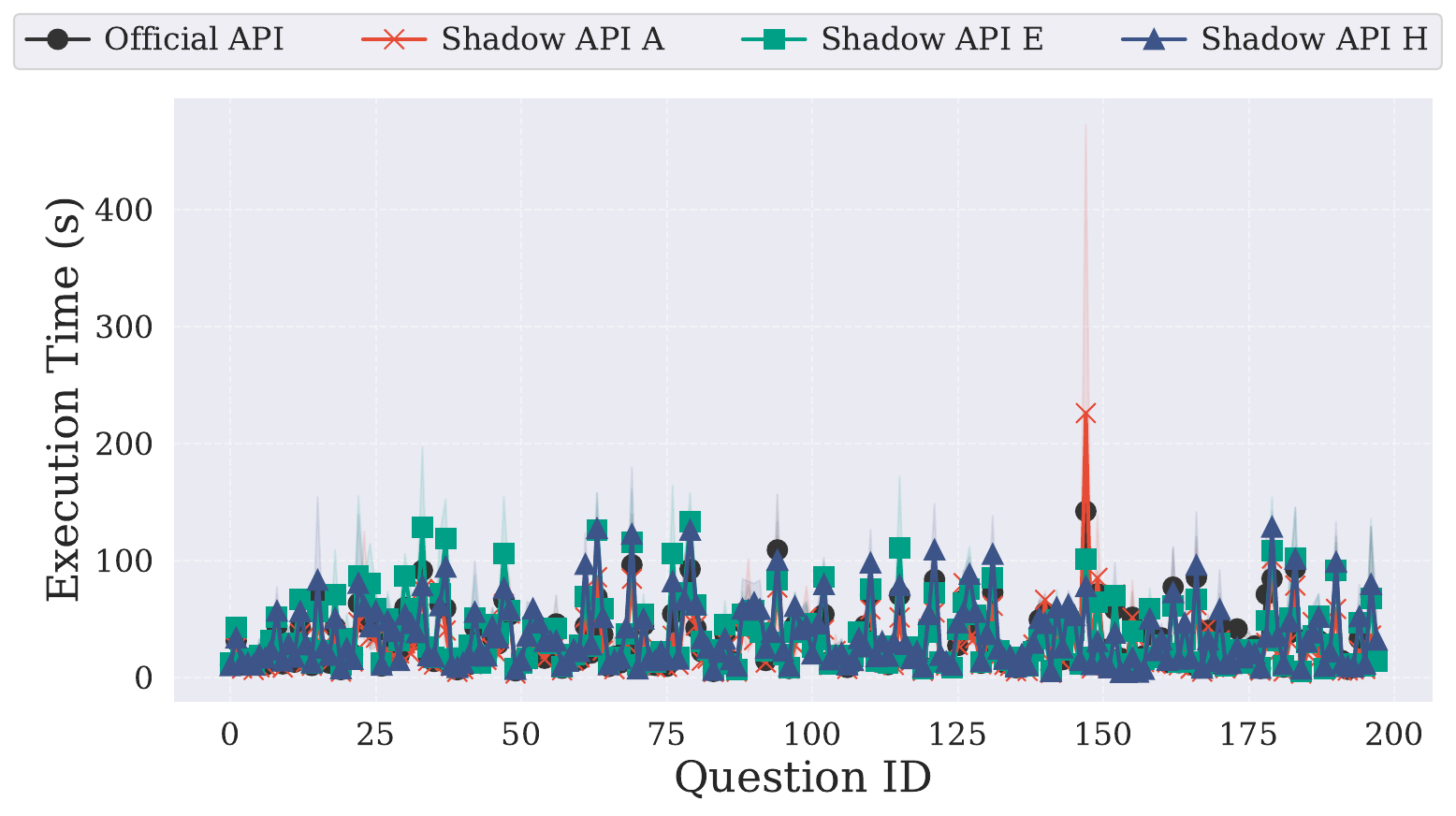}
        \caption{GPQA: GPT-5-mini (Time)}
    \end{subfigure}
    \hfill
    \begin{subfigure}{0.48\linewidth}
        \centering
        \includegraphics[width=\linewidth]{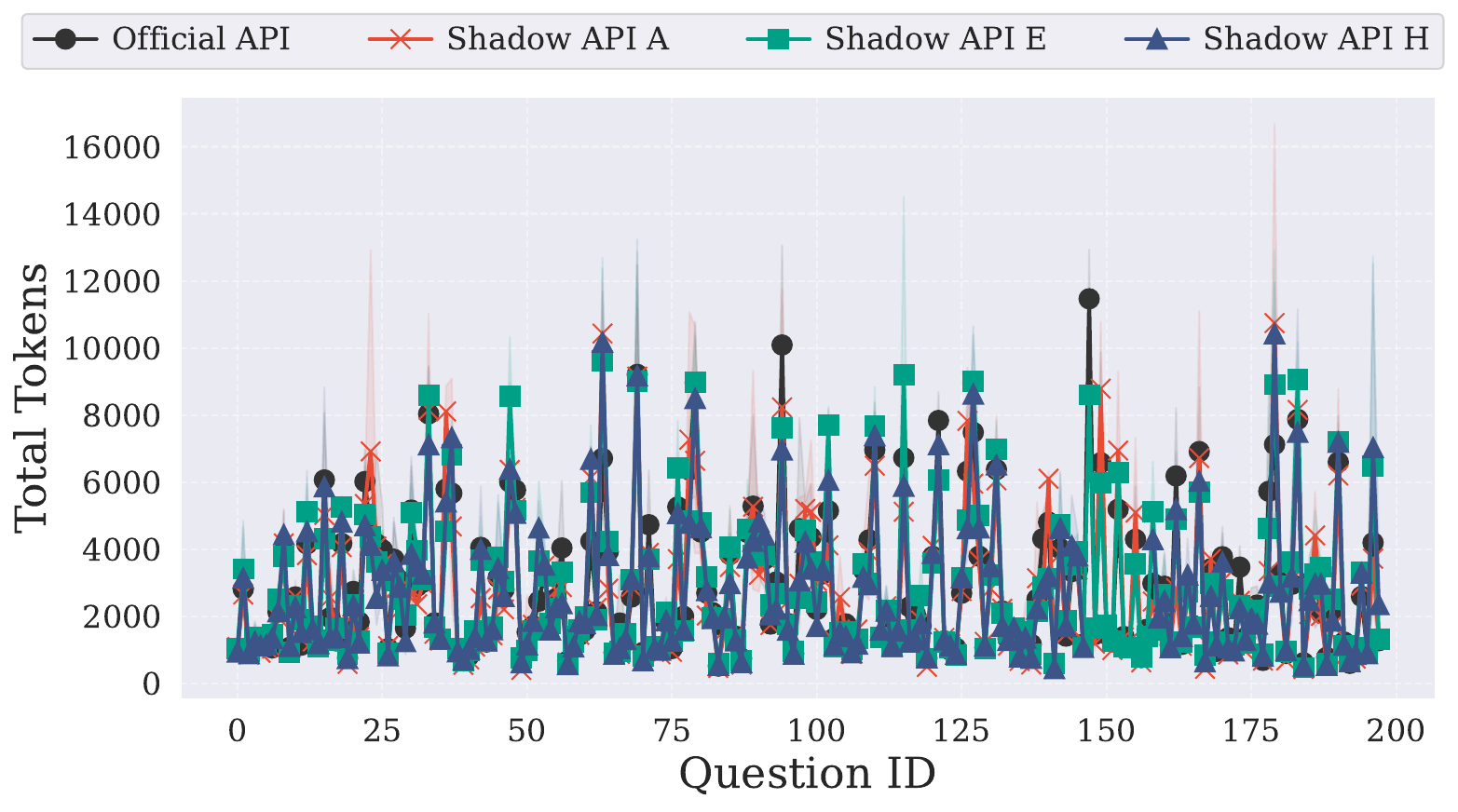}
        \caption{GPQA: GPT-5-mini (Token)}
    \end{subfigure}

    \begin{subfigure}{0.48\linewidth}
        \centering
        \includegraphics[width=\linewidth]{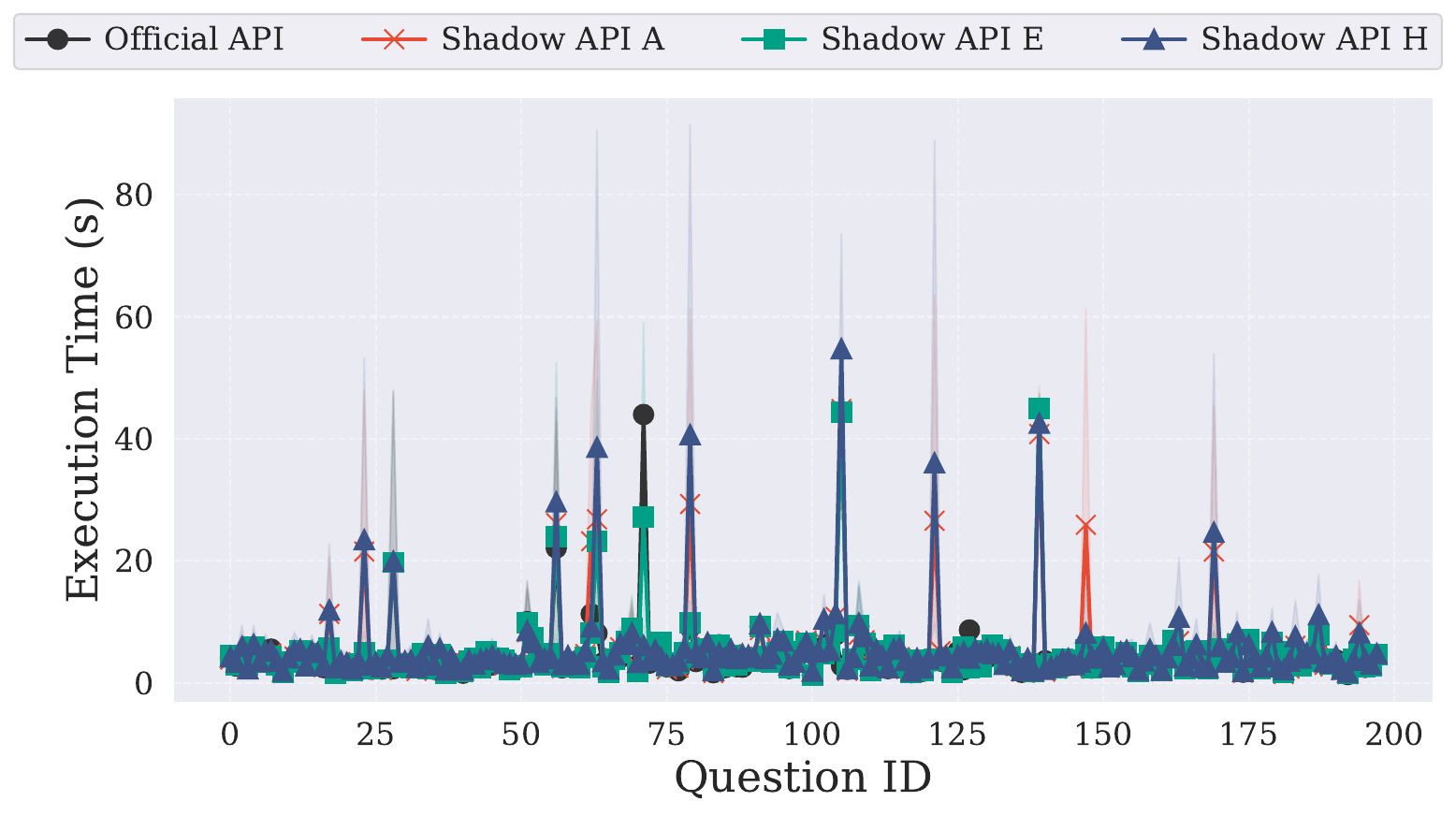}
        \caption{GPQA: Gemini-2.0-flash (Time)}
    \end{subfigure}
    \hfill
    \begin{subfigure}{0.48\linewidth}
        \centering
        \includegraphics[width=\linewidth]{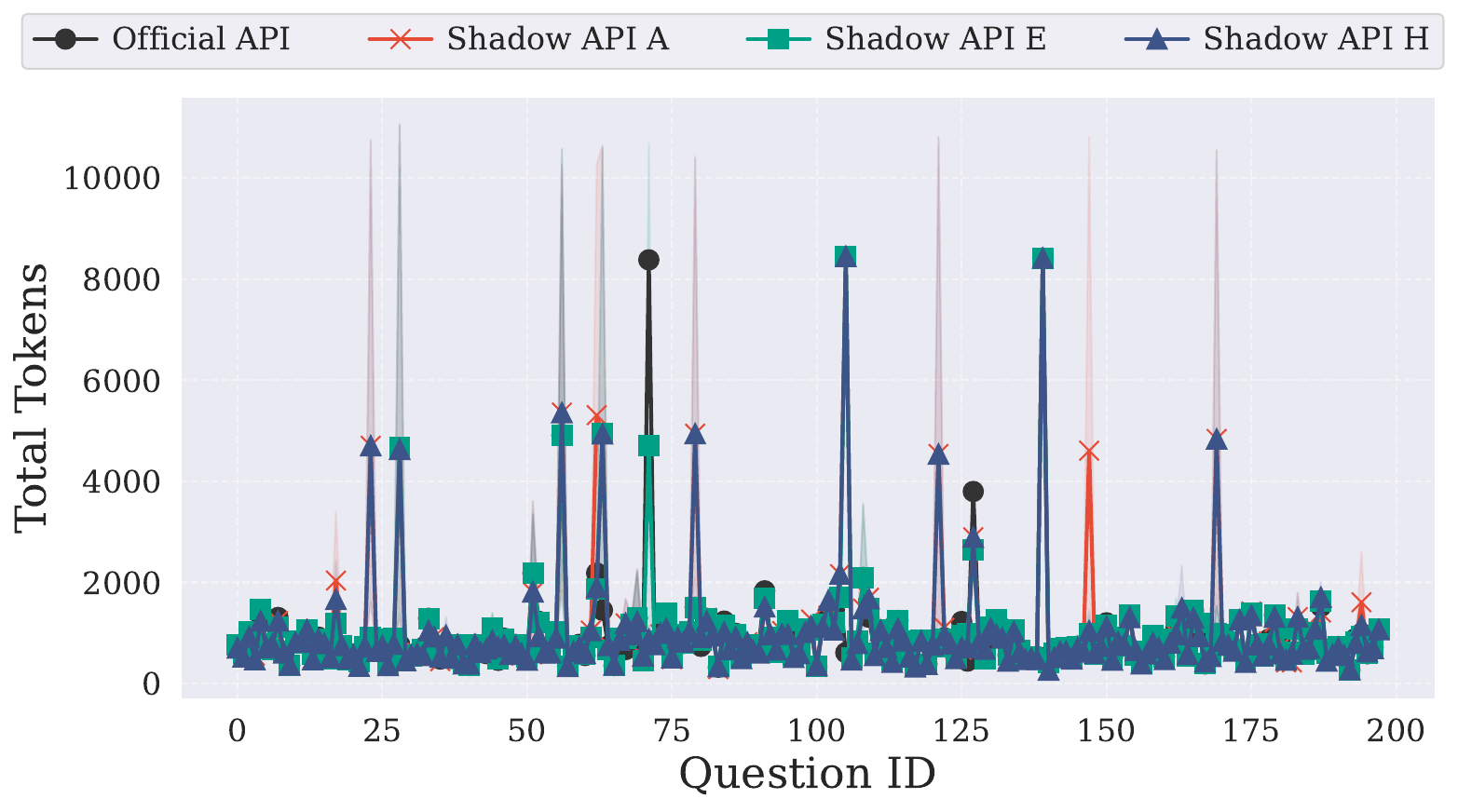}
        \caption{GPQA: Gemini-2.0-flash (Token)}
    \end{subfigure}
    \caption{Comparison of inference latency time and token counts on the GPQA (Part 1).  Solid lines represent mean values, and shaded regions denote the range between the minimum and maximum values across three trials.}
    \label{figure:gpqa_part1}
\end{figure*}

\begin{figure*}[t]
    \centering

    \begin{subfigure}{0.48\linewidth}
        \centering
        \includegraphics[width=\linewidth]{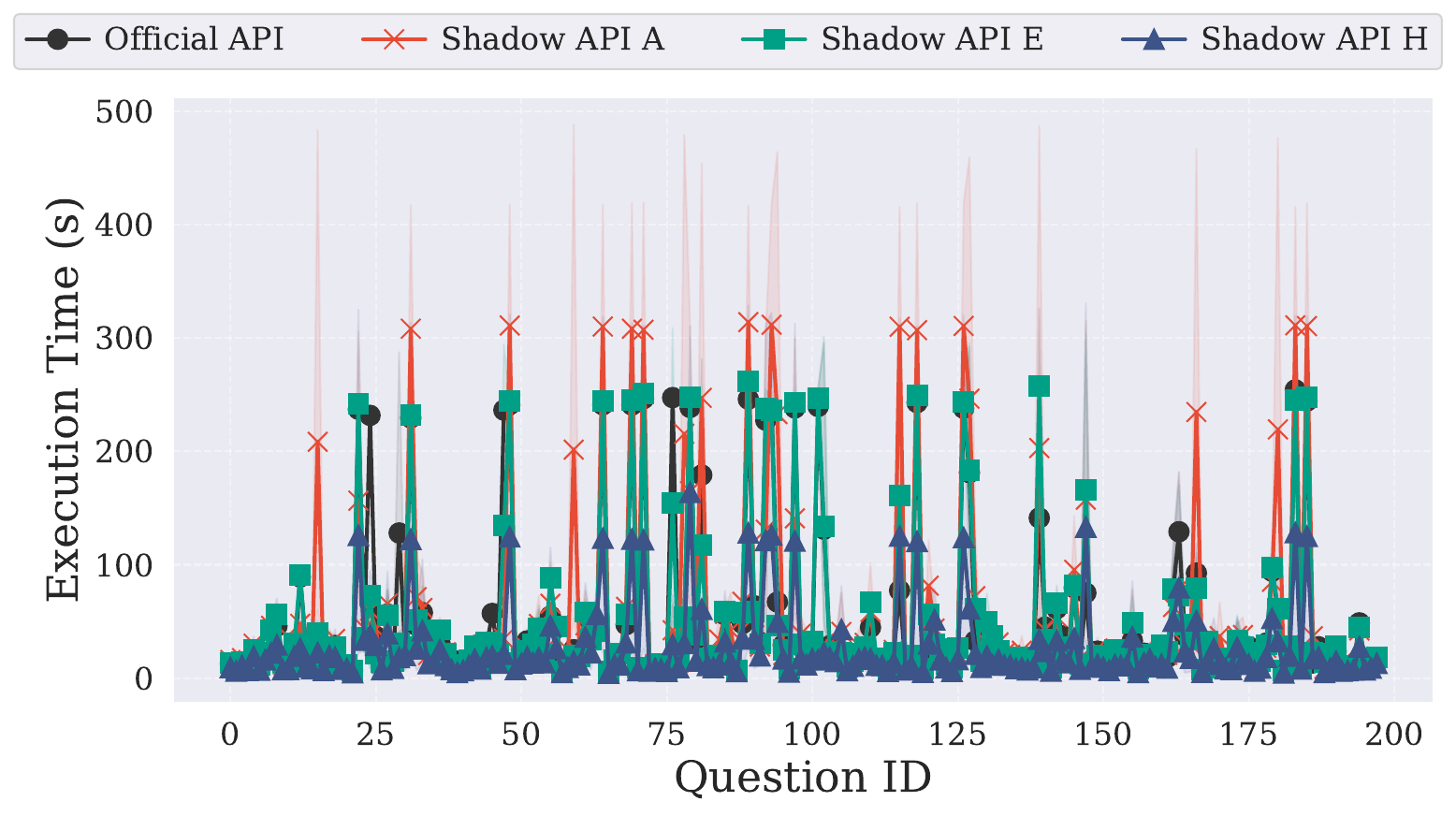}
        \caption{GPQA: Gemini-2.5-flash (Time)}
    \end{subfigure}
    \hfill
    \begin{subfigure}{0.48\linewidth}
        \centering
        \includegraphics[width=\linewidth]{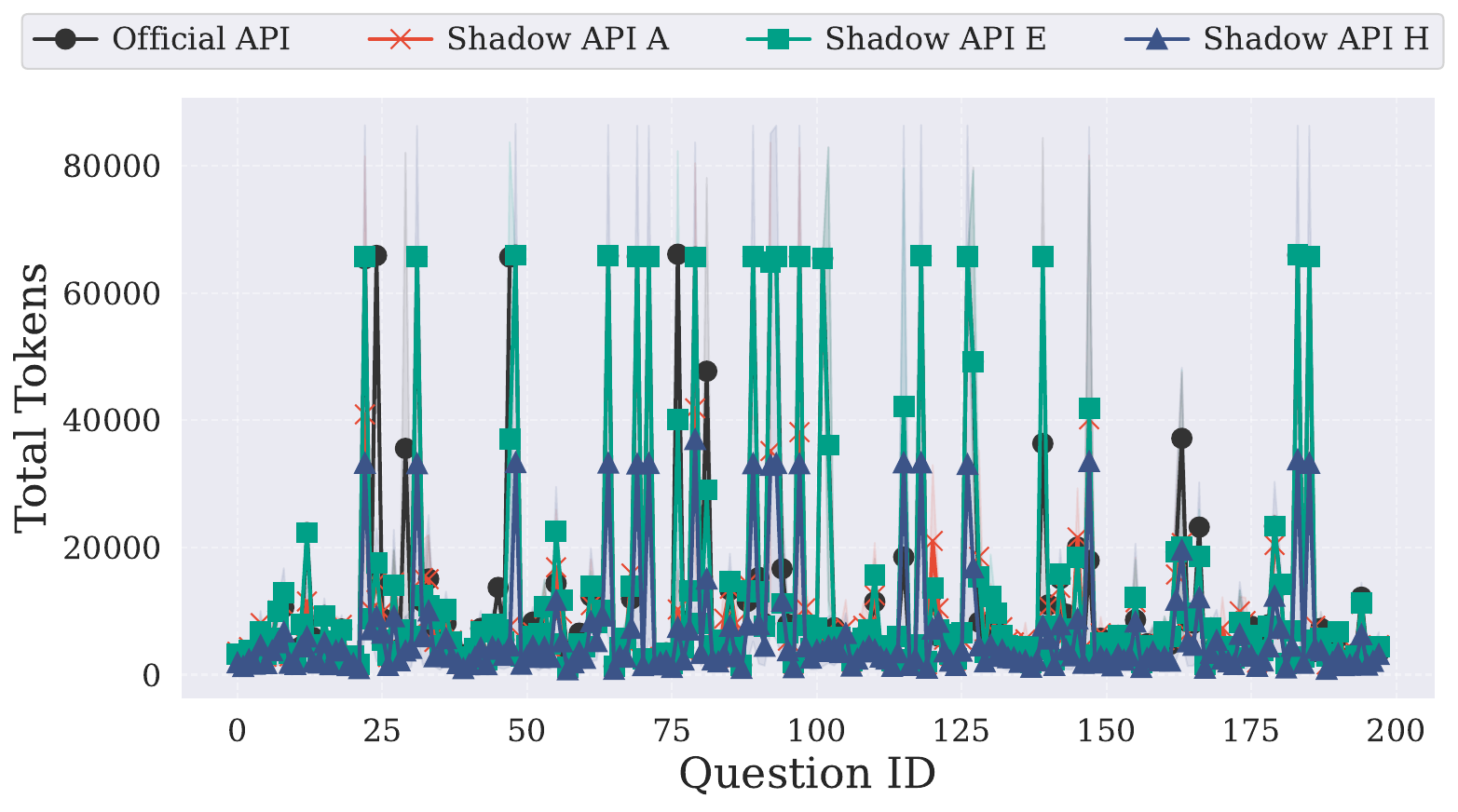}
        \caption{GPQA: Gemini-2.5-flash (Token)}
    \end{subfigure}

    \begin{subfigure}{0.48\linewidth}
        \centering
        \includegraphics[width=\linewidth]{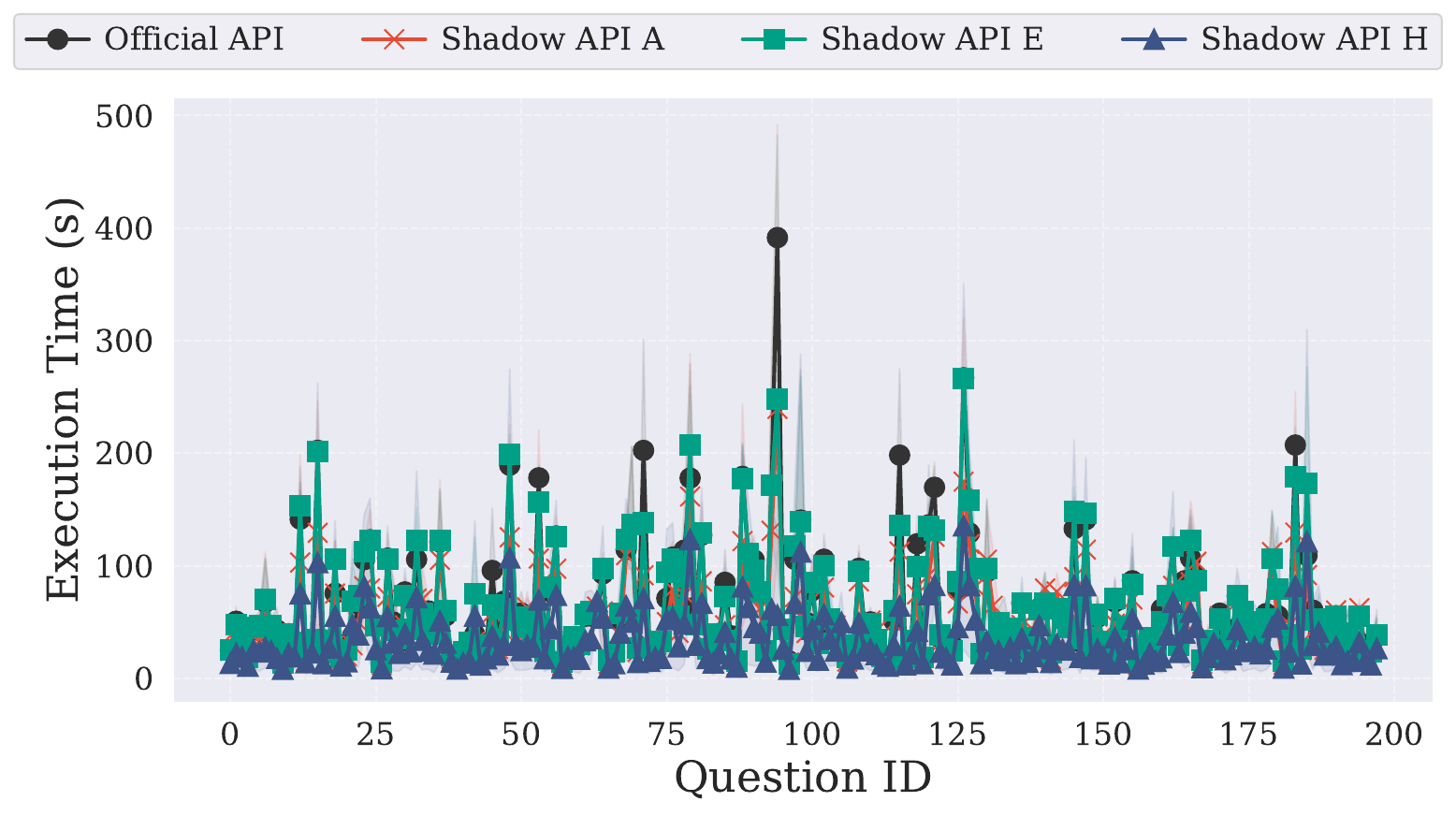}
        \caption{GPQA: Gemini-2.5-pro (Time)}
    \end{subfigure}
    \hfill
    \begin{subfigure}{0.48\linewidth}
        \centering
        \includegraphics[width=\linewidth]{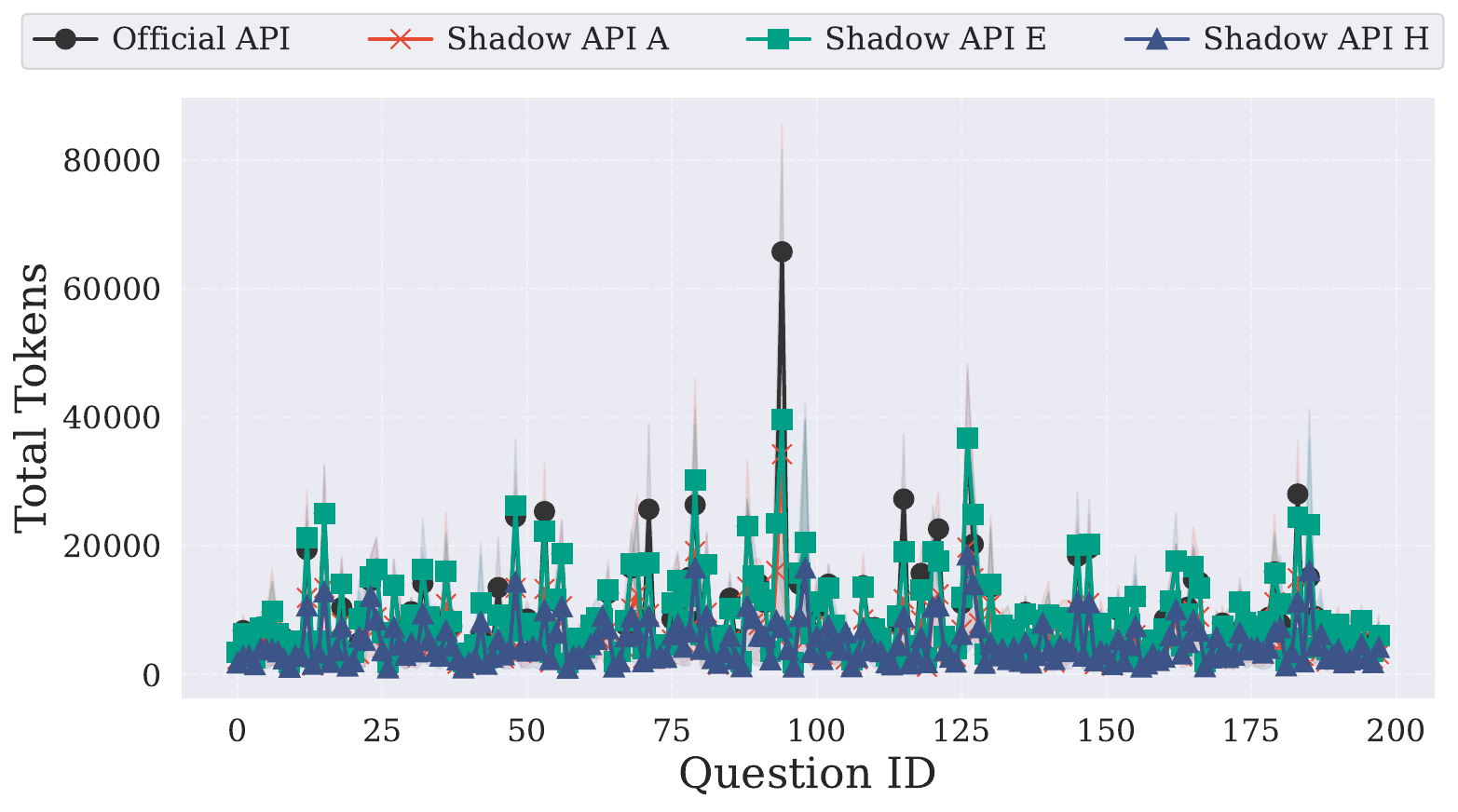}
        \caption{GPQA: Gemini-2.5-pro (Token)}
    \end{subfigure}

    \begin{subfigure}{0.48\linewidth}
        \centering
        \includegraphics[width=\linewidth]{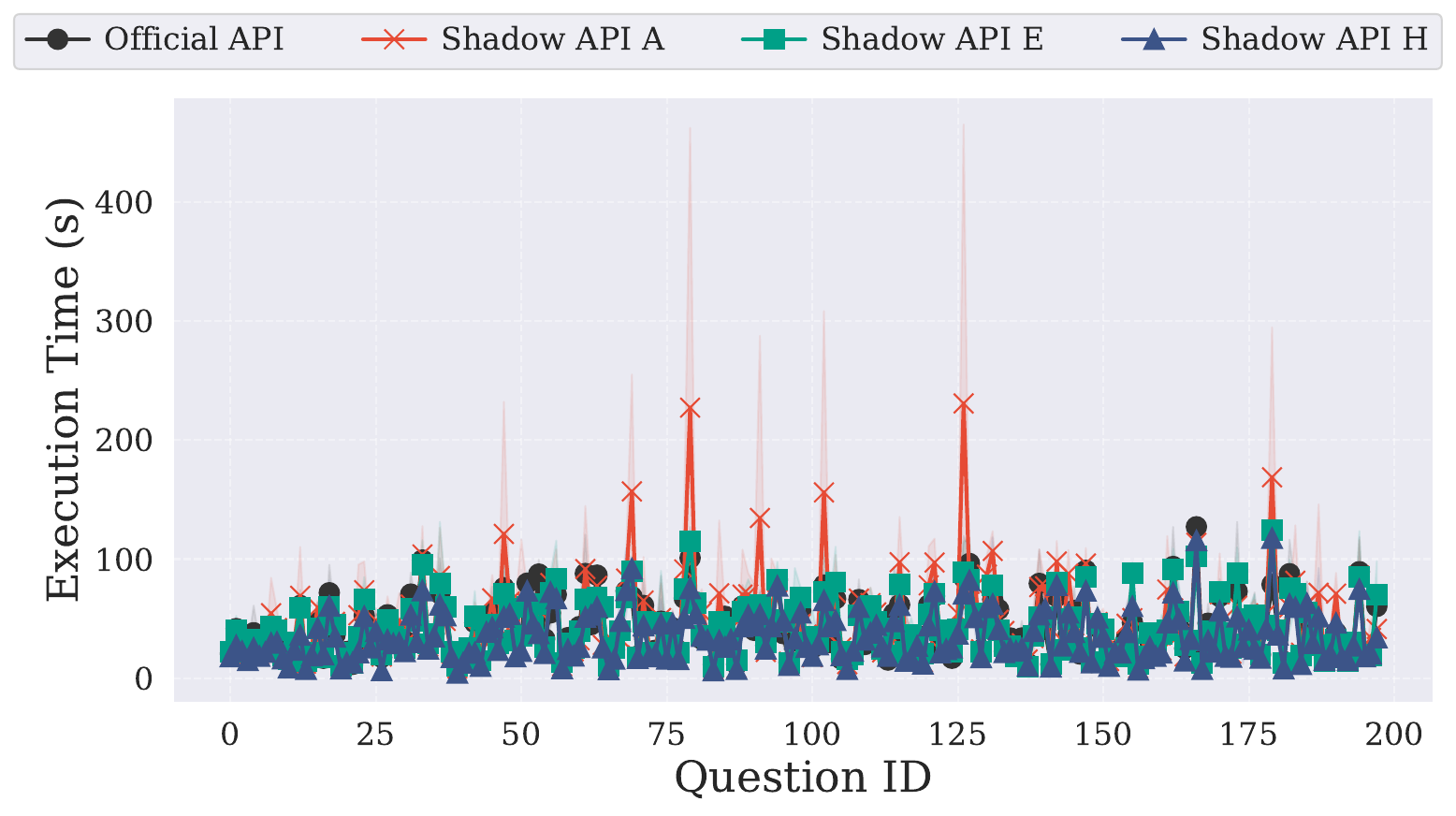}
        \caption{GPQA: DeepSeek-Chat (Time)}
    \end{subfigure}
    \hfill
    \begin{subfigure}{0.48\linewidth}
        \centering
        \includegraphics[width=\linewidth]{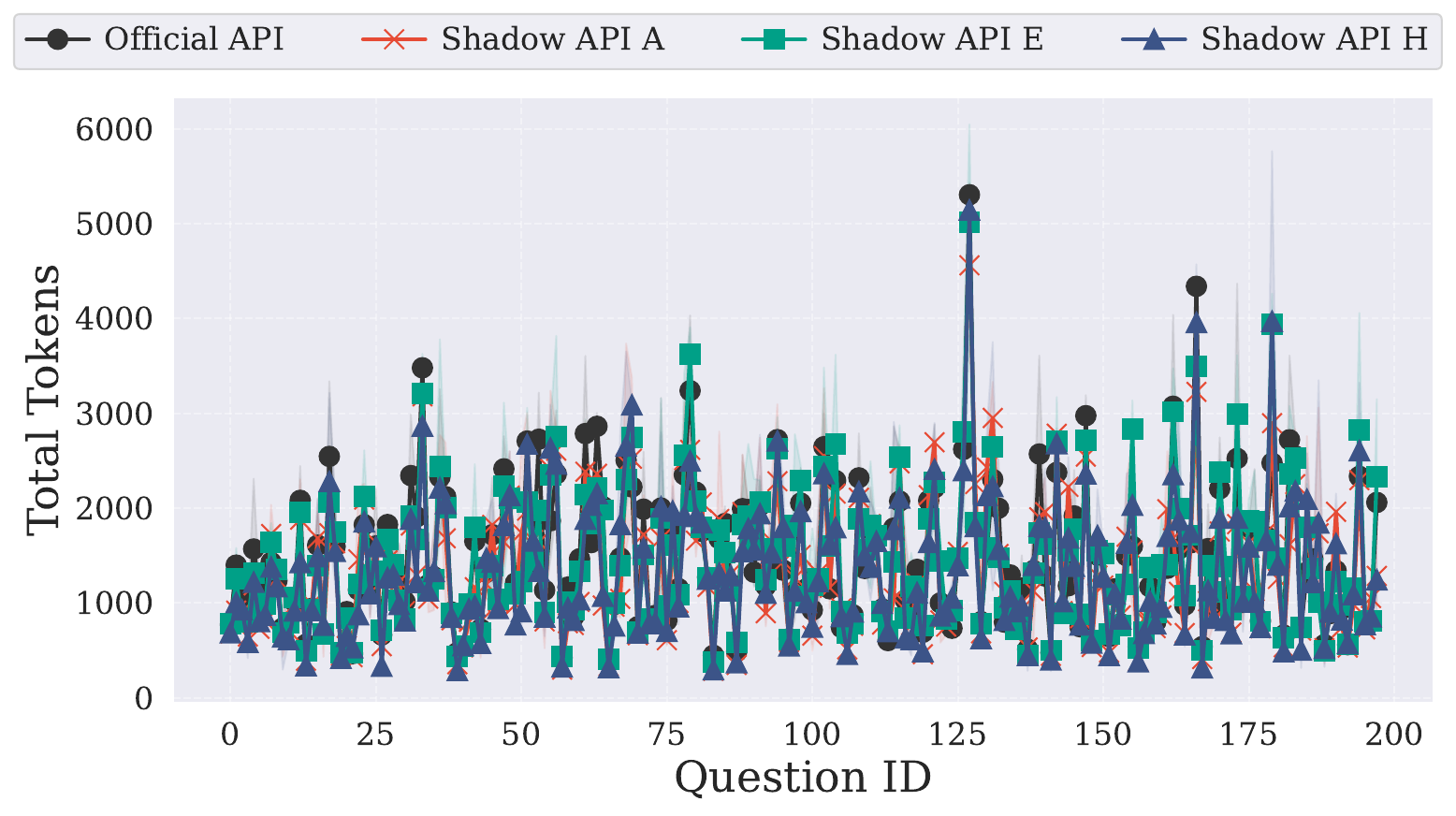}
        \caption{GPQA: DeepSeek-Chat (Token)}
    \end{subfigure}

    \begin{subfigure}{0.48\linewidth}
        \centering
        \includegraphics[width=\linewidth]{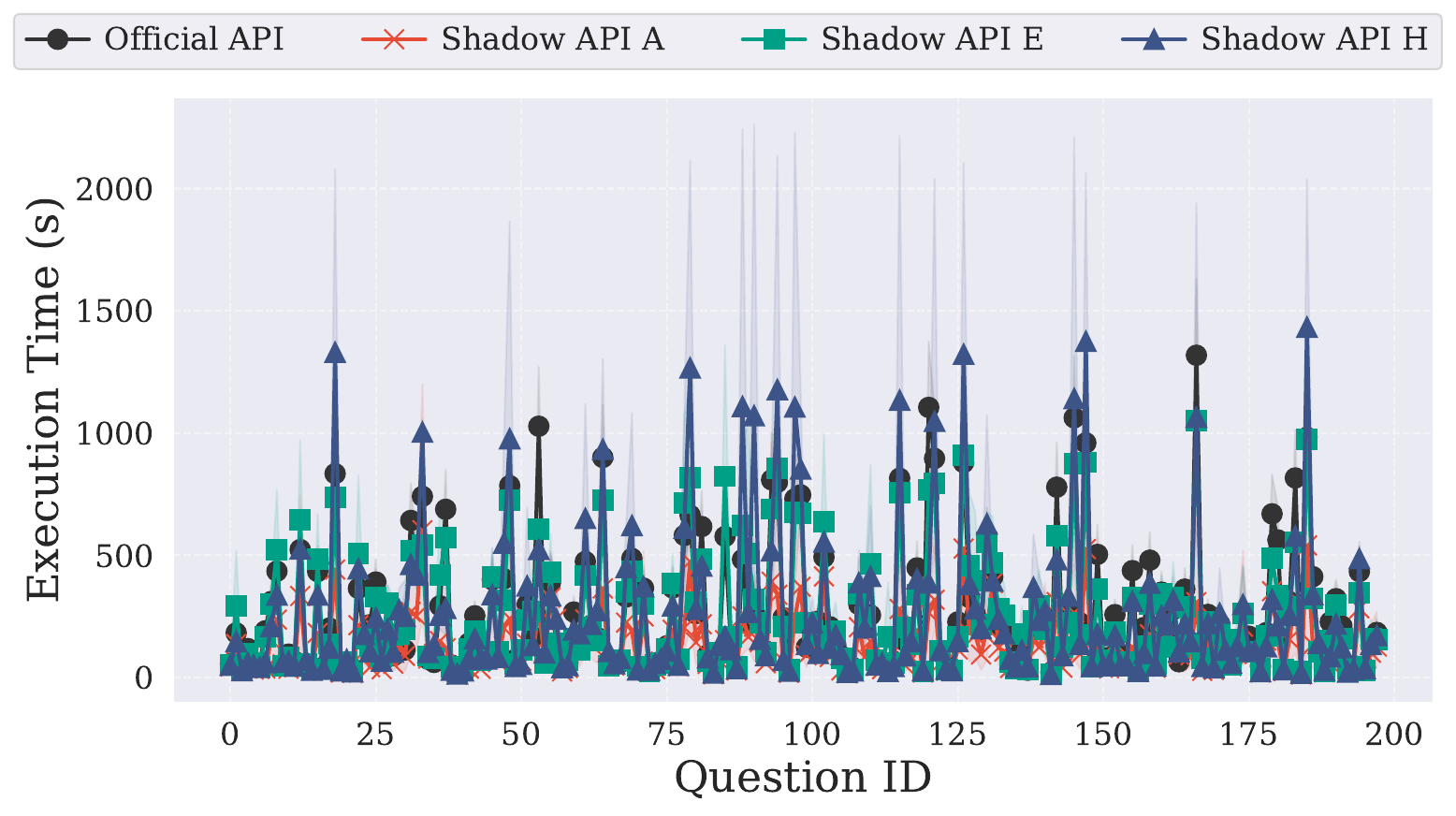}
        \caption{GPQA: DeepSeek-Reasoner (Time)}
    \end{subfigure}
    \hfill
    \begin{subfigure}{0.48\linewidth}
        \centering
        \includegraphics[width=\linewidth]{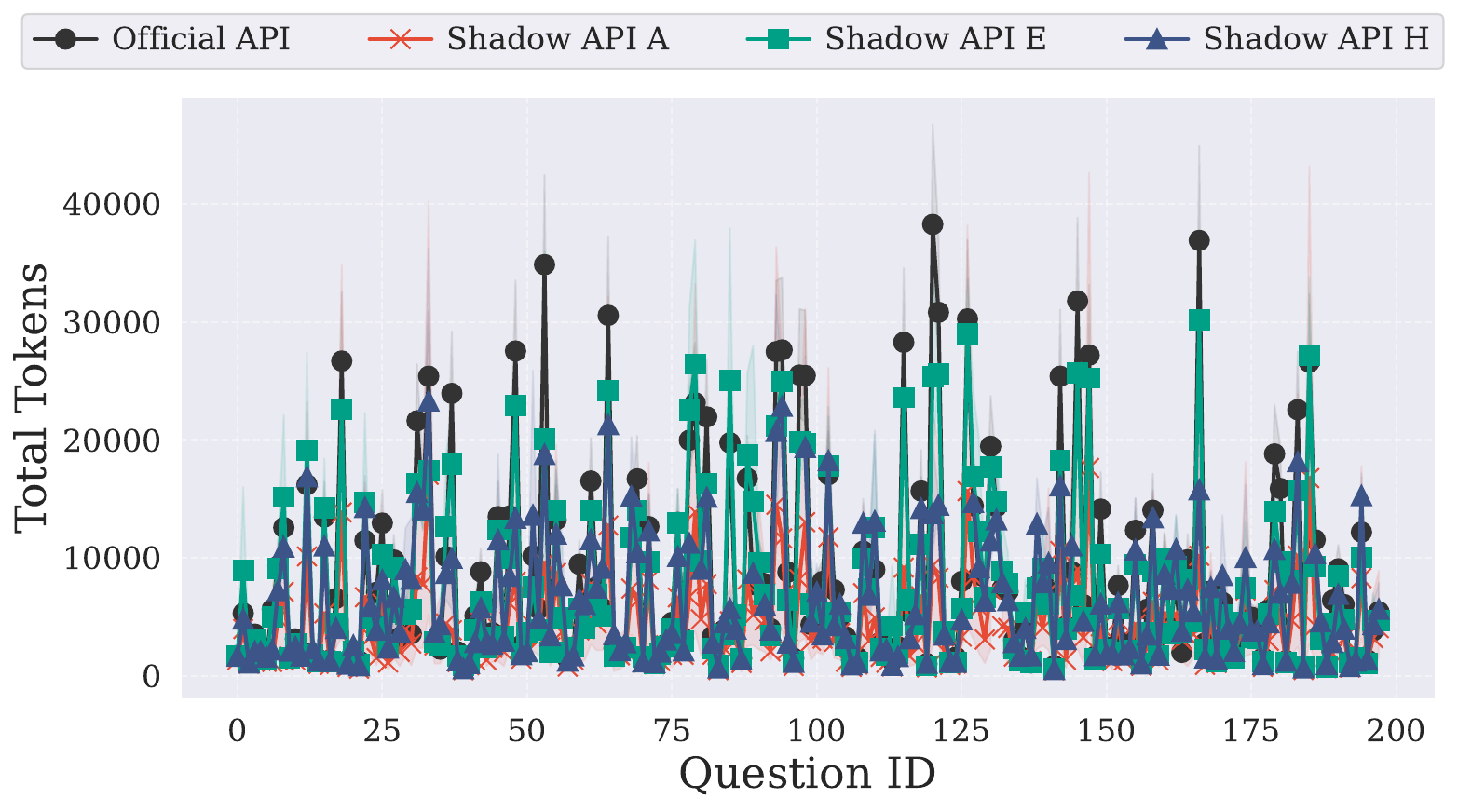}
        \caption{GPQA: DeepSeek-Reasoner (Token)}
    \end{subfigure}

    \caption{Comparison of inference latency time and token counts on the GPQA (Part 2). Solid lines represent mean values, and shaded regions denote the range between the minimum and maximum values across three trials.}
    \label{figure:gpqa_results_part2}
\end{figure*}

%-------------------------------------------------------------------------------
This appendix provides per-query time series that complement the aggregate standard-deviation analysis reported in~\autoref{section:meta}.
\autoref{figure:aime_results_part1} and~\autoref{figure:aime_results_part2} illustrate the inference latency and token counts on AIME 2025.
\autoref{figure:gpqa_part1} and~\autoref{figure:gpqa_results_part2} present the same analysis on GPQA.
Shadow API H exhibits latency and token counts that are sometimes higher and sometimes lower than those of the official API for GPT-4o-mini, indicating inconsistent performance characteristics across runs.

%-------------------------------------------------------------------------------
\section{LLMs Used for Training LLMmap}
\label{appendix:llmmap_model}
%-------------------------------------------------------------------------------

\autoref{table:models} provides a comprehensive inventory of the LLMs utilized in this study, distinguishing between the original baseline models and newly integrated trained models.
All officially trained models are queried directly through the corresponding official APIs.

%-------------------------------------------------------------------------------
\begin{table*}[t]
\centering
\caption{List of LLMs used for training and testing LLMmap. The table is divided into the original baseline models and additionally trained models.}
\scalebox{0.8}{
\begin{tabular}{@{}llcc@{}}
\toprule
\textbf{\#} & \textbf{Version} & \textbf{Provider} & \textbf{Number of Parameters} \\ 
\midrule
\multicolumn{4}{c}{\textit{\textbf{Part I: Original Trained Models}}} \\
\midrule
1 & ChatGPT-3.5 (gpt-3.5-turbo-0125) & OpenAI & / \\ \hline
2 & ChatGPT-4  (gpt-4-turbo-2024-04-09) & OpenAI & / \\ \hline
3 & ChatGPT-4o (gpt-4o-2024-05-13)~\cite{O24} & OpenAI & / \\ \hline
4 & Claude 3 Haiku (claude-3-haiku-20240307) & Anthropic & / \\ \hline
5 & Claude 3 Opus (claude-3-opus-20240229) & Anthropic & / \\ \hline
6 & Claude 3.5 Sonnet (claude-3-5-sonnet-20240620) & Anthropic & / \\ \hline
7 & google/gemma-7b-it & Google & 7B \\ \hline
8 & google/gemma-2b-it & Google & 2B \\ \hline
9 & google/gemma-1.1-2b-it & Google & 2B \\ \hline
10 & google/gemma-1.1-7b-it & Google & 7B \\ \hline
11 & google/gemma-2-9b-it & Google & 9B \\ \hline
12 & google/gemma-2-27b-it & Google & 27B \\ \hline
13 & CohereForAI/aya-23-8B & Cohere & 8B \\ \hline
14 & CohereForAI/aya-23-35B & Cohere  & 35B \\ \hline
15 & Deci/DeciLM-7B-instruct & Deci & 7B \\ \hline
16 & Qwen/Qwen2-1.5B-Instruct & Qwen & 1.5B \\ \hline
17 & Qwen/Qwen2-7B-Instruct & Qwen & 7B \\ \hline
18 & Qwen/Qwen2-72B-Instruct & Qwen & 72B \\ \hline
19 & gradientai/Llama-3-8B-Instruct-Gradient-1048k  & Gradient AI & 8B \\ \hline
20 & meta-llama/Llama-2-7b-chat-hf & Meta & 7B \\ \hline
21 & meta-llama/Meta-Llama-3-8B-Instruct & Meta & 8B \\ \hline
22 & meta-llama/Meta-Llama-3-70B-Instruct & Meta & 70B \\ \hline
23 & meta-llama/Meta-Llama-3.1-8B-Instruct & Meta & 8B \\ \hline
24 & meta-llama/Meta-Llama-3.1-70B-Instruct & Meta & 70B \\ \hline
25 & microsoft/Phi-3-medium-128k-instruct & Microsoft & 14B \\ \hline
26 & microsoft/Phi-3-medium-4k-instruct & Microsoft & 14B \\ \hline
27 & microsoft/Phi-3-mini-128k-instruct & Microsoft & 3.8B \\ \hline
28 & microsoft/Phi-3-mini-4k-instruct & Microsoft & 3.8B \\ \hline
29 & mistralai/Mistral-7B-Instruct-v0.1 & Mistral AI & 7B \\ \hline
30 & mistralai/Mistral-7B-Instruct-v0.2 & Mistral AI & 7B \\ \hline
31 & mistralai/Mistral-7B-Instruct-v0.3 & Mistral AI & 7B \\ \hline
32 & mistralai/Mixtral-8x7B-Instruct-v0.1  & Mistral AI & 8x7B \\ \hline
33 & nvidia/Llama3-ChatQA-1.5-8B & NVIDIA  & 8B \\ \hline
34 & openchat/openchat-3.6-8b-20240522 & OpenChat & 8B \\ \hline
35 & openchat/openchat\_3.5 & OpenChat & 7B \\ \hline
36 & togethercomputer/Llama-2-7B-32K-Instruct & Together AI & 7B \\ \hline
37 & upstage/SOLAR-10.7B-Instruct-v1.0 & Upstage AI & 10.7B \\ \hline
38 & NousResearch/Nous-Hermes-2-Mixtral-8x7B-DPO & Nous Research & 8x7B \\ \hline
39 & abacusai/Smaug-Llama-3-70B-Instruct  & Abacus AI & 70B \\ \hline
40 & microsoft/Phi-3.5-MoE-instruct & Microsoft & 16x3.8B \\ \hline
41 & internlm/internlm2\_5-7b-chat & InternLM & 7B \\ \hline
42 & HuggingFaceH4/zephyr-7b-beta & HuggingFace & 7B \\ 

\midrule
\multicolumn{4}{c}{\textit{\textbf{Part II: Additional Trained Models}}} \\
\midrule

43 & ChatGPT-4o-mini (gpt-4o-mini-2024-07-18) & OpenAI & / \\\hline
44 & ChatGPT-5 (gpt-5-2025-08-07) & OpenAI & / \\\hline
45 & ChatGPT-5-mini (gpt-5-mini-2025-08-07) & OpenAI & / \\\hline
46 & google/gemini-2.0-flash & Google & / \\\hline
47 & google/gemini-2.5-flash & Google & / \\\hline
48 & google/gemini-2.5-pro & Google & / \\\hline
49 & MiniMaxAI/MiniMax-M2 & MiniMaxAI & 230B \\\hline
50 & DeepSeek/deepseek-chat & DeepSeek & / \\\hline
51 & DeepSeek/deepseek-reasoner & DeepSeek & / \\\hline
52 & DeepSeek/deepseek-V3.2-exp-chat & DeepSeek & / \\\hline
53 & DeepSeek/DeepSeek-V3-0324 & DeepSeek & / \\\hline
54 & Qwen/Qwen2.5-7B-Instruct & Qwen & 7B \\\hline
55 & Qwen/Qwen3-VL-32B-Instruct & Qwen & 32B \\\hline
56 & Qwen/Qwen3-8B & Qwen & 8B \\\hline
57 & DeepSeek/DeepSeek-R1-0528-Qwen3-8B & DeepSeek & 8B \\\hline
58 & ZhipuAI/glm-4-9b-chat & ZhipuAI & 9B \\\hline
59 & ZhipuAI/glm-Z1-9B-0414 & ZhipuAI & 9B \\\hline
60 & MoonshotAI/Kimi-K2-Instruct-0905 & MoonshotAI & 1000B \\

\bottomrule
\end{tabular}
}
\label{table:models}
\end{table*}
%-------------------------------------------------------------------------------

%-------------------------------------------------------------------------------
\end{document}